\documentclass[fleqn,10pt]{wlscirep}
\usepackage[utf8]{inputenc}
\usepackage[T1]{fontenc}

\title{Artificial Intelligence Tools Expand Scientists' Impact but Contract Science's Focus\\\textcolor{red}{(Just accepted by \textit{Nature}, to be online soon)}}

\author[1]{Qianyue Hao}
\author[1*]{Fengli Xu}
\author[1,2*]{Yong Li}
\author[3,4*]{James Evans}
\affil[1]{Department of Electronic Engineering, Tsinghua University, Beijing \& 100084, P. R. China.}
\affil[2]{Zhongguancun Academy, Beijing \& 100094, P. R. China.}
\affil[3]{Knowledge Lab and Department of Sociology, University of Chicago, Chicago \& IL 60637, USA.}
\affil[4]{Santa Fe Institute, Santa Fe \& NM 87501, USA.}
\affil[*]{Corresponding authors. Emails: fenglixu@tsinghua.edu.cn, liyong07@tsinghua.edu.cn, jevans@uchicago.edu}

\usepackage{graphicx} 

\usepackage[maxbibnames=150,style=numeric-comp,sorting=none]{biblatex}
\bibliography{references.bib}

\usepackage{url}
\usepackage{comment}

\usepackage{setspace}

\usepackage{multirow}
\usepackage{booktabs}
\usepackage{threeparttable}
\usepackage{csquotes}
\usepackage{rotating}

\date{}

\begin{abstract}
\textbf{Development in Artificial Intelligence (AI) has accelerated scientific discovery~\supercite{wang2023scientific}. Alongside recent AI-oriented Nobel prizes~\supercite{hopfield1982neural,hopfield1984neurons, lecun2015deep, krizhevsky2012imagenet, hinton2006reducing, hinton2002training, kuhlman2003design, jumper2021highly}, these trends establish the role of AI tools in science~\supercite{gao2024quantifying}. This advancement raises questions about the potential influences of AI tools on scientists and science as a whole, and highlights a potential conflict between individual and collective benefits~\supercite{evans2008electronic}. To evaluate, we used a pretrained language model to identify AI-augmented research, with an F1-score of 0.875 in validation against expert-labeled data. Using a dataset of 41.3 million research papers across natural science and covering distinct eras of AI, here we show an accelerated adoption of AI tools among scientists and consistent professional advantages associated with AI usage, but a collective narrowing of scientific focus. Scientists who engage in AI-augmented research publish 3.02 times more papers, receive 4.84 times more citations, and become research project leaders 1.37 years earlier than those who do not. By contrast, AI adoption shrinks the collective volume of scientific topics studied by 4.63\% and decreases scientist's engagement with one another by 22.00\%. Thereby, AI adoption in science presents a seeming paradox---an expansion of individual scientists' impact but a contraction in collective science's reach---as AI-augmented work moves collectively toward areas richest in data. With reduced follow-on engagement, AI tools appear to automate established fields rather than explore new ones, highlighting a tension between personal advancement and collective scientific progress.}
\end{abstract}

\begin{document}

\flushbottom
\maketitle 

\section*{Introduction}
Artificial intelligence (AI) has made significant strides in recent decades, promising to impact myriad aspects of society, including education~\supercite{adiguzel2023revolutionizing,akgun2022artificial}, healthcare~\supercite{mesko2023imperative,loh2022application}, and industry~\supercite{ahmed2022artificial}.
Major investments in predictive and generative AI have catalyzed society-level debates over the future of AI at home and in the workplace.
Perhaps more than any other domain, AI tools have become deeply entwined with the process of knowledge production, yielding findings that attract disproportionate attention in various scientific fields~\supercite{wang2023scientific}.
For example, AlphaFold learns known protein structures to accurately predict the unexplored ones, circumventing the capital and human cost of conventional structural inference and recently granted a 2024 Nobel Prize~\supercite{jumper2021highly,varadi2022alphafold}.
Models improved via deep reinforcement learning have become tuned to contain complex fusion reactions~\supercite{degrave2022magnetic} and discovered new, hardware-optimized forms to matrix multiplication that recursively accelerate deep learning itself~\supercite{fawzi2022discovering}. 
Autonomous laboratory systems driven by ChatGPT have helped some chemists and material scientists upscale the number of adaptive high-throughput experiments~\supercite{boiko2023autonomous,stokel2023chatgpt,gilson2023does}.
Moreover, recent developments in large language models are making them increasingly incorporated in assisting scientific writing~\supercite{salimi2023large,liang2024mapping,hwang2024can,kobak2025delving}, facilitating the distillation of scientific findings, but also raising concerns about weakened confidence in AI-generated content~\supercite{stokel2023chatgpt,gilson2023does, wojtowicz2025undermining}.
AI's increasing capabilities to influence scientific research suggest that it manifests potential to both increase the productivity of individual scientists and raise the visibility of science it supports.

Despite the increasing adoption of AI in science, large-scale empirical measurements of AI’s scientific impact are limited, and a detailed, dynamic understanding of AI’s impact on the entire character of science remains largely unknown. Recent work suggests AI has brought widespread benefits to individual scientists but may lead to demographic disparity resulting from gaps in AI education~\supercite{gao2024quantifying}. Researchers have also identified evolving citation patterns that signal a changing scientific landscape in AI research~\supercite{frank2019evolution}.
Here we seek to explore the impact of AI in scientific research at different scales by posing the question: How does the adoption of AI influence individual scientists’ careers and the collective exploration of science as a whole?

We conduct a large-scale quantitative analysis of the impact of AI on scientists and science, covering 41,298,433 research papers spanning from 1980 to 2025 in the OpenAlex dataset~\supercite{openalex}, with patterns corroborated using the Web of Science~\supercite{WoS,mongeon2016journal}.
Notably, we do not focus on computer science or mathematics, fields that develop AI methodologies directly, but rather on papers that augment research in natural science fields by adopting AI, primarily covering decades involving development and deployment of conventional machine learning algorithms and also extending to a necessarily more preliminary analysis of the latest generative AI techniques.
Specifically, we select six representative disciplines that cover the vast majority of natural science contributions—biology, medicine, chemistry, physics, materials science, and geology.
We then leverage the BERT language model~\supercite{DBLP:conf/naacl/DevlinCLT19,wolf-etal-2020-transformers} to accurately identify such AI-augmented research papers based on their titles and abstracts.

We separate the periods in which AI was predominantly conventional machine learning, deep learning, and most recently generative designs including large language models. With abundant data-based evidence across decades of conventional machine learning and deep learning, we validate these AI-based measurements and use them to reveal that the adoption of AI leads to an amplifying effect on the career of individual scientists, bringing acceleration in the production and visibility of science produced by those scientists who incorporate AI.
Nevertheless, this effect corresponds with a contracted focus within collective science.
Measured with “knowledge extent”, the “diameter” covered by a sampled batch of papers in vector space, AI-driven science spans less topical ground and is associated with a decrease in follow-on scientific engagement, suggesting that AI is currently more likely to focus on existing popular research problems rather than explore new ones.
Meanwhile, analyses using currently available data within the latest era of generative AI including large language models reveal a preliminary consistency with prior periods, providing a starting point for further study as generative AI develops over a longer period.
\section*{Results}

\begin{figure}[ht]
\centering
\includegraphics[width=\textwidth]{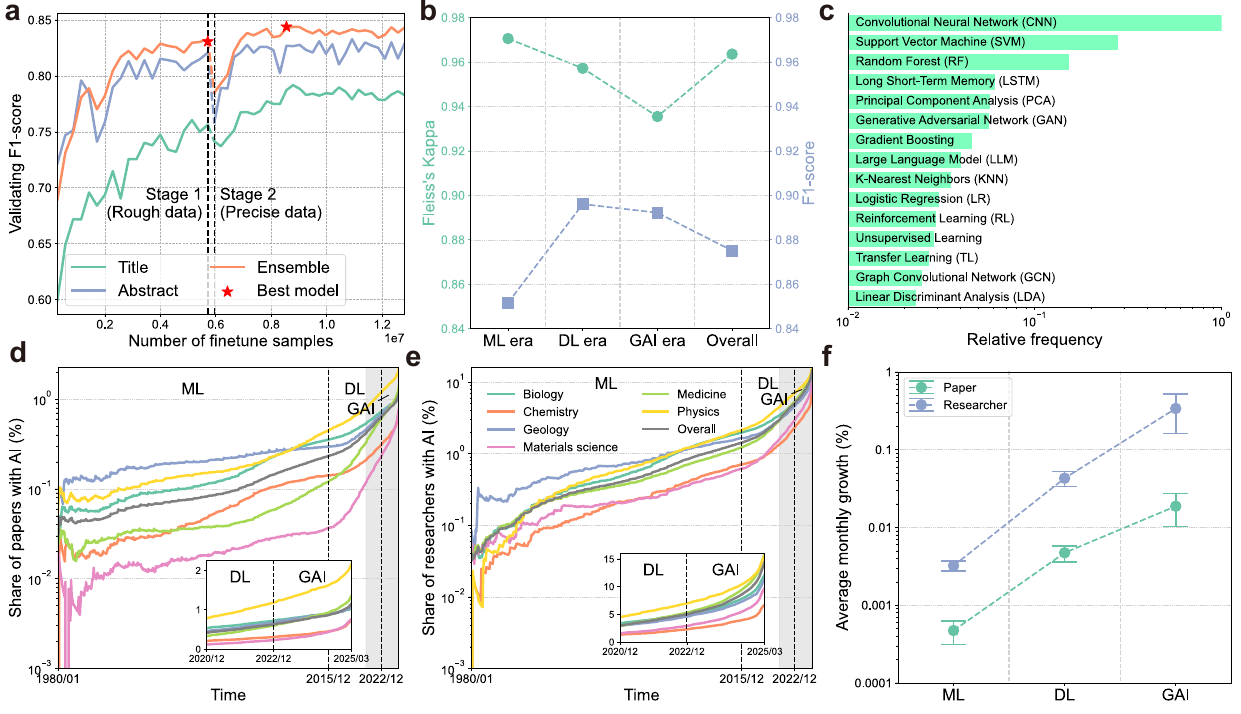}
\caption{
\textbf{Increasing prevalence of AI adoption in science.}
\textbf{(a)} Increasing performance of AI paper identification during the two-stage fine-tuning of the BERT pre-trained models, where we use rough training data in Stage 1 to evolve precise assessments in Stage 2.
We independently train two models based on titles and abstracts, respectively, and then integrate them into an ensemble that selects the optimal models during both stages (red stars) to identify all selected papers.
\textbf{(b)} Accuracy evaluation of our identification results by human experts.
For samples spanning three eras of AI, experts reached consensus with Fleiss’ Kappa ($\kappa$) $\geq 0.93$.
Our model identification results have strong accuracy in validation against expert-labeled data with an F1-score $\geq 0.85$.
\textbf{(c)} Relative adoption frequency of the top 15 AI methods across all disciplines for all selected AI development eras.
\textbf{(d)-(e)} The growth of AI-augmented papers (d, $n=41,298,433$) and AI-adopted researchers (e, $n=5,377,346$) across the eras of ML, DL, and GAI between 1980 and 2025 in selected scientific disciplines. The y-axes are set to log-scale.
\textbf{(f)} The average monthly growth rates for AI papers and researchers across the eras of ML, DL, and GAI across all selected disciplines ($n=543$ month observations), where 99\% CIs are shown as error bars centred at the mean.
}
\label{fig1}
\end{figure}

\subsection*{Increasing prevalence of AI in science}
In this investigation, we focus on research papers using AI methods in various fields of natural science, where we conduct our analysis based on 41,298,433 papers from the OpenAlex dataset~\supercite{openalex}, covering six representative disciplines: biology, chemistry, geology, materials science, medicine, and physics (Methods~M1).
According to the invention of milestone technologies in the trend of AI development, we divide the past decades into three eras, namely machine learning (ML), deep learning (DL), and generative AI (GAI) (Methods~M2).
To identify AI papers in various fields across eras, we fine-tune BERT~\supercite{DBLP:conf/naacl/DevlinCLT19,wolf-etal-2020-transformers}, an established language model~\supercite{DBLP:conf/emnlp/BeltagyLC19,DBLP:conf/acl/CohanFBDW20,DBLP:conf/emnlp/SinghDCDF23}, on articles published in explicitly AI-oriented scientific journals and conferences for automatically extracting and interpreting information from context.
Specifically, we employ a two-stage fine-tuning process to adapt the pre-trained BERT model to the task of AI paper identification.
We first independently train two models based on titles and abstracts of papers, respectively, then ensemble the optimized individual models to identify all selected papers (Fig. 1a, Methods~M3, and Extended Data Fig. 1).
This approach eliminates the need for manual selection of AI-related trigger words, as demonstrated in previous research~\supercite{frank2019evolution}.

To evaluate the accuracy of our identification, we recruited a team of human experts to validate these results (Methods~M4 and Extended Data Fig. 2).
The experts demonstrate strong consensus across their independent annotation of papers sampled at random from the six disciplines mentioned above, achieving an average Fleiss’ Kappa ($\kappa$) of 0.964~\supercite{landis1977measurement,fleiss1971measuring}. The BERT model attains an F1-score of 0.875 in an evaluation that uses the expert labels as ground truth.
Meanwhile, the strong consensus among experts and high quality for identification is consistent across samples from different eras of AI, confirming the reliability of our identification accuracy and laying a robust foundation for our subsequent analysis (Fig. 1b and Supplementary Table~S1-S4).
To provide a rationale and explainability for our identification results, we visualize attention strengths in the BERT model with examples, where the model allocates substantial attention to terms such as “neural network” and “large language model”, illustrating how the model correctly interprets and accurately identifies AI-related contents from papers published in different era of AI development (Supplementary Fig.~S2-S3).

In total, we identify 310,957 AI-augmented papers, comprising 0.75\% of all selected papers.
Semantically, the identified AI-related papers turn out to be around topics combining artificial intelligence and conventional research topics across disciplines (Supplementary Fig.~S4).
Counting all eras and disciplines collectively, the most commonly adopted AI methods in natural science research include support vector machines and principal component analysis from the ML era, and convolutional neural networks and generative adversarial networks from the DL era.
The large language model, which has emerged in recent years, also ranks among the most frequently utilized methods (Fig. 1c and Supplementary Table~S5-S11).
Statistically, despite the overall rise in the number of papers published annually across all disciplines~\supercite{chu2021slowed}, the share of AI-augmented papers surged by 10.70 (geology, $Z=348.60$, $p<0.001$ and $df=1$ in Cochran-Armitage test) to 51.89 (biology, $Z=1388.70$, $p<0.001$ and $df=1$ in Cochran-Armitage test) times from 1980 to 2025 (Fig. 1d).
Similarly, the proportion of researchers adopting AI has grown even more rapidly, from 135.46 times in geology ($Z=546.81$, $p<0.001$ and $df=1$ in Cochran-Armitage test) to 362.16 in physics ($Z=2237.51$, $p<0.001$ and $df=1$ in Cochran-Armitage test) (Fig. 1e).
Meanwhile, growth rates for AI-augmented papers and researchers have accelerated across the three eras (Fig. 1f and Supplementary Fig.~S5-S6).
These findings underscore the increasing prevalence and fast development of AI in science across all disciplines and the importance of understanding AI’s impact on scientific research and progress.

\begin{figure}[ht]
\centering
\includegraphics[width=\textwidth]{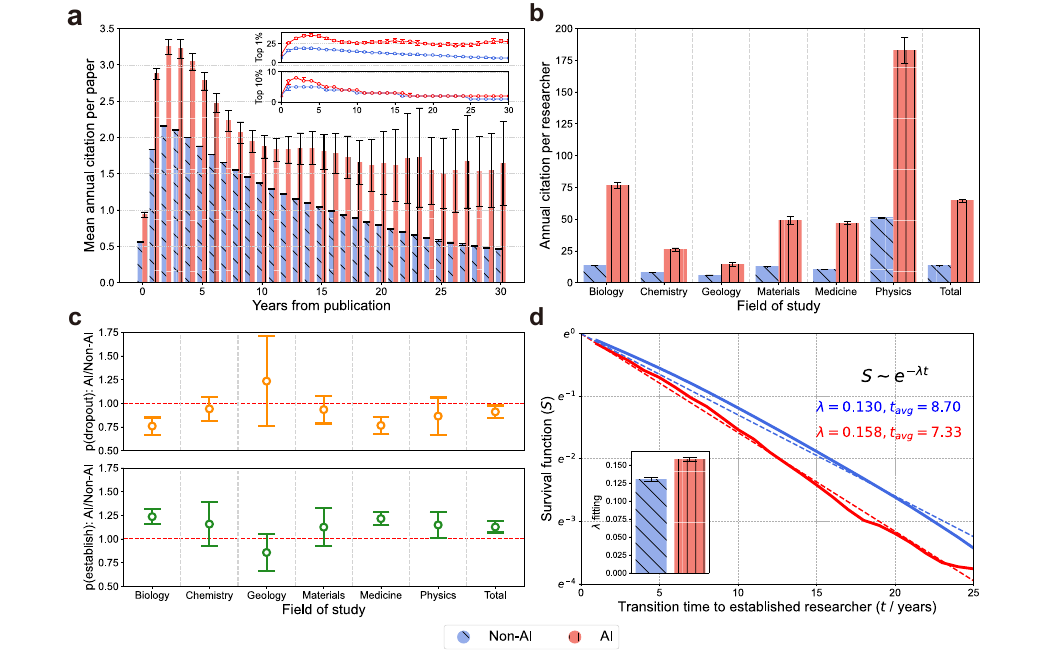}
\caption{
\textbf{AI enlarges paper impact and enhances researcher careers.}
\textbf{(a)} Average (insets: top 1\% and 10\%) annual citations after publication of AI and non-AI papers ($n=27,405,011$), where AI papers attract more citations.
\textbf{(b)} Average annual citations for researchers adopting AI and their counterparts without AI ($p<0.001, n=5,377,346$), where researchers adopting AI garner 4.84 times more citations than their counterparts without AI.
\textbf{(c)} The probability of two role transitions between junior scientists adopting AI and their counterparts without AI ($n=46$ year observations for each field).
Junior scientists adopting AI have a higher probability of becoming established researchers and a lower probability of exiting academia compared with their counterparts without AI.
\textbf{(d)} Survival functions for the transition from junior to established researcher ($p<0.001, n=2,282,029$).
The survival function can be well-fit with exponential distributions, where junior scientists adopting AI become established earlier than their counterparts without AI.
For all panels, 99\% CIs are shown as error bars, with the insets of panel (a) centred at the 1\% and 10\% percentiles and other panels centred at the mean.
All statistical tests use a two-sided t-test.
}
\label{fig2}
\end{figure}

\subsection*{AI enhances individual scientists}
From statistics across 27,405,011 papers with intact reference records in the OpenAlex dataset, we note that from the publication date of each paper until decades later, annual citations to AI papers are 98.70\% higher than non-AI papers on average (Fig. 2a, $t\geq8.33$, $p<0.001$ and $df>10^3$ in t-test on any year).
In addition to higher annual average citations, the higher scientific impact of AI-augmented papers is also reflected by multiple alternative statistical indicators about top and bottom annual citation count (Supplementary Fig.~S8).
Also, AI papers published at different eras consistently receive more citations (Extended Data Fig. 3, $t\geq4.06$, $p<0.001$ and $df>10^3$ in t-test on any era).
Furthermore, we examine the distribution of AI-augmented papers across journals of varying Journal Citation Report (JCR) quantiles~\supercite{jcr} (Supplementary Fig.~S14).
We find that the proportion of AI papers in Q1 journals is 18.60\% higher than non-AI ones in all journals, and in Q2 journals, the AI proportion is 1.59\% higher, while Q3 and Q4 journals hold a relatively lower proportion of papers with AI ($\chi^2=3629.11$, $p<0.001$ and $df=3$ in $\chi^2$-test).
These results indicate a heterogeneous distribution of AI-augmented papers across journals, with a higher prevalence in high-impact journals.
Paralleled by the attention paid to AI papers, the impact of AI researchers also substantially increases.
On average, researchers adopting AI annually publish 3.02 times more papers ($t\geq47.18$, $p<0.001$ and $df>10^3$ in t-test on any discipline) and garner 4.84 times more citations ($t\geq30.32$, $p<0.001$ and $df>10^3$ in t-test on any discipline) compared with those not adopting AI, with consistency across disciplines and robustness for core researchers with multi-year continuous publication records~\supercite{ioannidis2014estimates} (Fig. 2b, Extended Data Fig. 4 and Supplementary Fig.~S17).
Furthermore, when controlling for and comparing scientists with similar early-career positions, the enhanced productivity and impact still hold (Supplementary Fig.~S16).
This suggests that, after accounting for potential selection-biases among researchers with different original achievements that may influence their choice of AI adoption, AI itself contributes to the observed advantages.

To identify the implications of AI adoption on scientist’s career development, we classify the scientists into “junior” and “established”, where junior scientists are defined as newcomers who have not yet led a research project, whereas established scientists refer to those who have led one or more research projects (Methods~M5 and Extended Data Fig. 5).
We extract 2,282,029 career trajectories of scientists from the dataset, each initially identified as a junior scientist (Methods~M6).
The results reveal that AI-augmented research is associated with reduced research team sizes, averaging 1.33 (19.29\%) fewer scientists ($t=20.47$, $p<0.001$ and $df>10^3$ in t-test, Extended Data Fig. 6).
Specifically, the average number of junior scientists decreased from 2.89 in non-AI teams to 1.99 (31.14\%) in AI teams ($t=19.02$, $p<0.001$ and $df>10^3$ in t-test), while the number of established scientists decreased from 4.01 to 3.58 (10.77\%) in AI teams ($t=20.82$, $p<0.001$ and $df>10^3$ in t-test).
This indicates that AI adoption primarily contributes to a reduction in the number of junior scientists in teams, while decrease in the number of established scientists is relatively moderate.
Given the decline in the number of junior scientists, we further calculate the probability of junior scientists becoming established scientists or leaving academia (Fig. 2c).
Across all studied disciplines, the probability for AI-adopted junior scientists to transition to established scientists is 45.00\%, 13.64\% higher than for their counterparts who do not adopt AI ($t\geq1.40$, $p<0.2$ and $df=90$ in t-test on 4 out of 6 disciplines).
This indicates that AI-adopted scientists are associated with increased opportunities to lead research projects and reduced risks of dropping out from academia, thereby experiencing accelerated career transitions from junior to established scientists.

To further quantify this effect, we measure the accelerated career development of junior scientists by employing a birth-death model~\supercite{kendall1960birth} and fitting the model parameter $\lambda$ with scientists’ career trajectories (Fig. 2d and Methods~M7).
We find that the anticipated transition time to becoming established scientists is 1.37 years shorter for AI-adopted junior scientists compared to their counterparts. The expectation of transition time is 7.33 years for junior scientists adopting AI ($R^2=0.995$) and 8.70 years for those without ($R^2=0.987$). 
This demonstrates how AI adoption affords junior scientists opportunities to lead research projects and become established earlier.
Further analysis reveals that this reduction in the transition time for AI-adopted junior scientists to become established ones is universal across examined disciplines (Extended Data Fig. 7).
Moreover, the established scientists involved in AI papers are, on average, 10.77\% younger than those in non-AI papers (Extended Data Fig. 6, $t\geq2.12$, $p<0.05$, and $df>10^3$ in t-test on most year).
Collectively, these findings suggest that AI research receives more attention from academia, and AI-adopting scientists are associated with higher scholarly productivity and impact.
In this way, they become established scientists with higher probability and at earlier ages, experiencing accelerated career development.

\begin{figure}[ht]
\centering
\includegraphics[width=\textwidth]{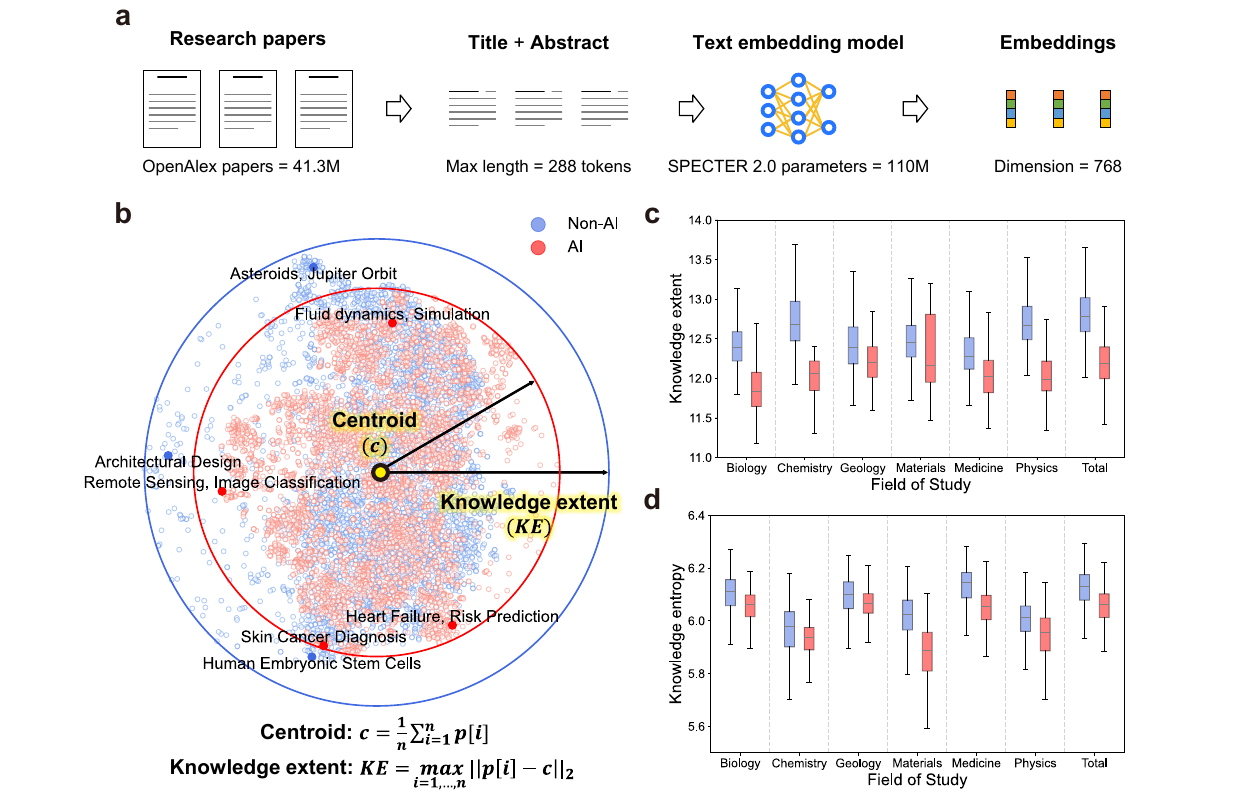}
\caption{
\textbf{AI adoption is associated with a contraction in knowledge extent within and across scientific fields.}
\textbf{(a)} We embed research papers into a 768-dimensional vector space with a pre-trained text embedding model, then measure the knowledge extent of papers within that space.
\textbf{(b)} 
For visualization, we use the t-SNE algorithm to flatten the high-dimensional embeddings of a random batch of 10,000 papers, half of which are AI papers and half are non-AI papers, into a 2-D plot.
As shown by the solid arrows and circular boundaries, the knowledge extent of AI papers (calculated in the unflatted space) is smaller across the entirety of natural science, and AI papers are more clustered in knowledge space, indicating more concentration on specific problems.
\textbf{(c)} Knowledge extent of AI and non-AI papers in each field ($p<0.001, n=1,000$ samples in each field), where AI research focuses on a more contracted knowledge space.
\textbf{(d)} Knowledge entropy of AI and non-AI papers in each field ($p<0.001, n=1,000$ samples in each field), where AI research has a lower entropy.
For panels (c) and (d), boxplots are centred at the median and bounded at the first and third quartile (Q1 and Q3), with 1.5 times of the inter-quartile range (IQR) shown as whiskers from the box.
All statistical tests use a median-test.
}
\label{fig3}
\end{figure}

\subsection*{AI contracts science’s focus}
The accelerating use of AI in science and its impact on individual scientists raises questions about its influences across the entire scientific field.
To evaluate how AI collectively impacts the frontiers of scientific exploration, we design a measurement to characterize the breadth of scholarly attention represented by a collection of research papers.
We employ SPECTER 2.0, a specialized text embedding model pre-trained on a large scientific literature corpus and fine-tuned with citation information~\supercite{DBLP:conf/emnlp/SinghDCDF23}, to project research articles onto this 768-dimensional embedding space of science (Fig. 3a).

Within the high-dimensional embedding space, we design the measurement of knowledge extent (KE), which is the “diameter” of vector space covered by a sampled batch of papers, which allows us to compare the coverage of topical ground between AI and non-AI papers in each given domain~\supercite{fortunato2018science,milojevic2015quantifying} (Fig. 3b and Methods~M8).
Compared to conventional research, AI research is associated with a 4.63\% contracted median collective knowledge extent across science, which is consistent across all six disciplines (Fig. 3c and Extended Data Fig. 8, $\chi^2\geq84.05$, $p<0.001$ and $df=1$ in median-test on any discipline).
Moreover, when dividing these disciplines into more than two hundred sub-fields, the contraction of knowledge extent can be observed in more than 70\% of sub-fields (Extended Data Fig. 9).
When we compare the median entropy of knowledge distribution between AI and non-AI research in each domain (Fig. 3d), results demonstrate that the knowledge distribution of AI research has an lower entropy ($\chi^2\geq79.20$, $p<0.001$ and $df=1$ in median-test on any discipline), indicating an increasingly disproportionate focus on specific problems rather than across entire fields.

Generally, these results highlight an emerging conflict between individual and collective incentives to adopt AI in science, where scientists receive expanded personal reach and impact, but the knowledge extent of entire scientific fields tends to shrink and focus attention on a subset of topical areas.
According to analyses on possible factors that may influence the selectivity of AI adoption across different topics, we find that factors like inherent topicality, original impact, and funding priority, remain almost unrelated to the disproportionate AI adoption (Supplementary Fig.~S22-S24).
In contrast, data availability appears to be a major impacting factor, where areas with an abundance of data are increasingly and disproportionately amenable to AI research, contributing to the observed concentration within knowledge space (Supplementary Fig.~S25).

\begin{figure}[ht]
\centering
\includegraphics[width=\textwidth]{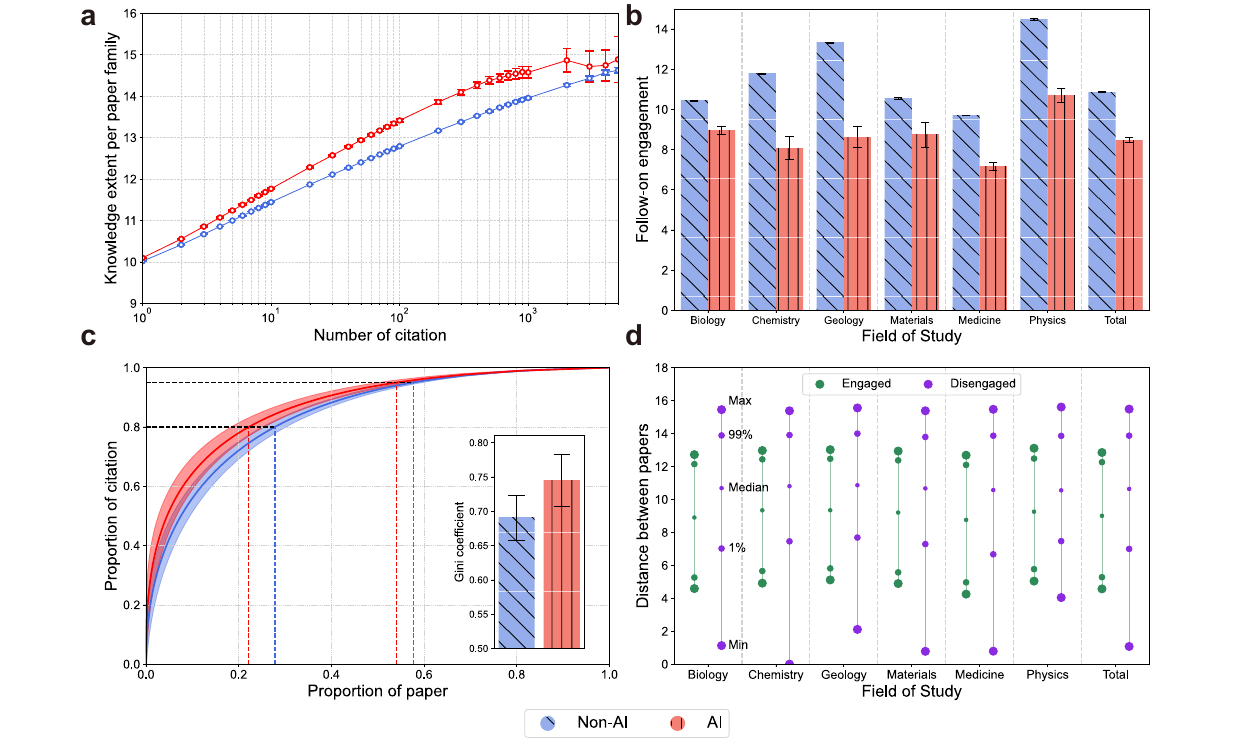}
\caption{
\textbf{Reduced follow-on engagement and more overlapping works in AI research.}
\textbf{(a)} Knowledge extent of individual AI and non-AI paper families, i.e., an original paper and its cumulative citations ($n=27,405,011$), where the knowledge space of individual AI paper families is broader and grows faster.
\textbf{(b)} Engagement among papers that cite AI vs. non-AI papers ($p<0.001, n=23,342,516$), where there are fewer follow-on interactions among papers that cite the same original paper in AI research.
\textbf{(c)} Distribution of citations to AI vs. non-AI papers, where AI papers tend to concentrate more on a smaller number of top papers ($p<0.001, n=100$ sampled paper groups).
\textbf{(d)} Distribution of distances between paper pairs that cite the same prior research, with or without citing one another—engaged versus disengaged ($n=590,325,130$ sampled paper pairs).
Results show that for papers not engaged with each other, the median distance is larger, but the minimum distance is smaller, indicating a higher probability of overlapping in knowledge space.
For all panels, 99\% CIs are shown as error bars or error bands centred at the mean.
All statistical tests use a two-sided t-test.
}
\label{fig4}
\end{figure}

\subsection*{AI reduces scientific engagement}
In order to analyze mechanisms underlying the conflict between the growing influence of individual papers and researchers and the narrowing of domain knowledge within AI research, we examine the relationship between articles that cite AI and non-AI work.
We first examine the knowledge extent of “paper families”, i.e., a focal paper and its follow-on citations, which measures the size of the space covered by research derived from each original paper (Fig. 4a and Methods~M9).
Results show that the knowledge extent of AI papers’ citation families is on average 3.46\% \textit{more} expanded compared to non-AI papers’ ($t\geq1.91$, $p\leq0.1$ and $df>10^3$ in t-test on 30 out of 32 pairs of data).
Therefore, the contraction of knowledge space in AI research is not attributable to the narrowing of knowledge space that can be derived from each original research work.

To further investigate, we examine relationships between papers by measuring the degree of follow-on paper engagement, namely how frequently citations of the same original paper cite each other (Fig. 4b and Methods~M10).
Results demonstrate AI research spawns 22.00\% less follow-on engagement ($t\geq8.10$, $p<0.001$ and $df>10^3$ in t-test on any discipline), suggesting that AI papers tend to only concentrate on the original paper, rather than forming dense interactions among each other, which is the characteristic of emerging fields~\supercite{mcmahan2018ambiguity}.
This results in a star-like structure around specific popular research topics, rather than a network of emergent and interconnected research works.
Further evidence of this concentration is found in the Matthew effect~\supercite{merton1968matthew} among AI paper citations across different fields (Fig. 4c and Extended Data Fig. 10).
In AI research, a small number of superstar papers dominate the field, with 22.20\% of top papers receiving 80\% of the citations and the top 54.14\% receiving 95\% of citations. This unequal distribution leads to a GINI coefficient of 0.754 in citation patterns surrounding AI research, higher than 0.690 for non-AI papers ($t=27.86$, $p<0.001$ and $df=198$ in t-test), signaling a disparity in recognition.

To further analyze the impact of reduced follow-on engagement, we sample 590,325,130 pairs of papers, where each pair cites the same original work.
Among these, 51,723,984 pairs not only cite the same original work but also cite each other (engaged), while the remaining pairs do not cite each other (disengaged).
We examine distances between these pairs of papers within the 768-dimensional vector space (Fig. 4d) and find that median distance between paper pairs disengaged from one another tends to be 18.11\% larger than between paper pairs engaged with each other.
In contrast, the closest disengaged paper pairs are 76.51\% closer to one another than the closest paper pairs engaged with one another.
Taken together, a pair of disengaged papers commonly focus on less related topics and lie farther apart in the embedding space.
Occasionally, however, due to the lack of reciprocal engagement, it is possible that mutually-unaware papers lie very close to each other, which indicates more overlapping research.
These findings suggest that AI in science has become more concentrated around popular research topics that become “lonely crowds” with reduced interaction among papers, linking to more overlapping research and a contraction in knowledge extent and diversity across science.
\section*{Discussion}
In this study, we perform a large-scale empirical measurement of the effect of adopting AI in science on both individual scientists and scientific communities. We identify three waves of AI adoptions in science, corresponding with the dominance of machine learning, deep learning, and large language models. Each wave is marked with an accelerated AI adoption rate in research papers and authors.
We find that individual scientists are increasingly rewarded with expanded academic impact and accelerated career development for incorporating AI assistance in research across these waves and in all natural science research fields we studied.
On average, the use of AI helps individual scientists publish 3.02 times more papers, receive 4.84 times more citations, and become team leaders 1.37 years earlier. This substantial academic benefit may be a driving force behind the accelerated rate of AI adoption.
However, we also find unintended consequences from the increased prevalence of AI-augmented research.
In all fields, AI-augmented research focuses on a narrower scope of scientific topics and reduces the scientific engagement of follow-on research, leading to more overlapping research works that slows the expansion of knowledge.
Further, with a greater concentration of collective attention to the same AI papers, the adoption of AI appears to induce authors to engage in collective hill-climbing\footnote{The metaphor of “collective hill-climbing” describes a situation where researchers act like a group of climbers all scaling the same popular mountain from the same route. Because both the “path” of a known approach and the “peak” of an anticipated solution are constrained, this collective rush leads to “crowding” and may discourage the search for other, potentially higher “mountains” representing new questions and answers.}, catalyzing solutions to known problems rather than creating new ones.

These findings raise critical questions for science policy. What are the topics most likely left behind from AI-augmented research across fields? Those with less available data include critical scientific questions regarding the origins of natural phenomena, where data are necessarily reduced. Accelerating scientific activity “under the lamp post” of highly visible, data-rich phenomena moves science away from many foundational questions and towards operational ones. By driving attention toward the most popular new developments, AI appears to drive problem solution over generation. These issues become particularly concerning in the face of calls to further increase support for AI-augmented science~\supercite{borger2023artificial,lawrence2024accelerating}, coupled with the personal scientific incentives we demonstrate. This could shift collective attention away from new and original questions that lack the data required for AI to demonstrate benefit.
It is true that more overlapping attention and a contracted focus may benefit scientific replication and extension, accelerating the emergence of solid and practical solutions to specific questions.
Insofar as scientific discovery represents a vast and complex landscape, however, concentrating attention on the same developments may increase the likelihood that science becomes fixed on local maxima of scientific explanation and prediction rather than searching in a more broad, decoupled, and diverse way.

While our analysis provides new insight into AI's impact on science, clear limitations remain. Our identification approach, though validated by experts, misses subtle and unmentioned forms of AI use, and our focus on natural sciences excludes important domains where AI adoption patterns may differ. Moreover, despite consistently suggestive evidence, we cannot fully identify the causal linkage between AI adoption and scientific impact. Nevertheless, our findings demonstrate that currently attributed uses of AI in science primarily augment cognitive tasks through data processing and pattern recognition. Looking forward, these findings illuminate a critical and expansive pathway for AI development in science. To preserve collective exploration in an era of AI use, we will need to reimagine AI systems that expand not only cognitive capacity but also sensory and experimental capacity~\supercite{king2009automation,burger2020mobile}, enabling scientists to search, select, and gather new types of data from previously inaccessible domains rather than merely optimizing analysis of standing data. The history of major discoveries has been most consistently linked with new views on nature~\supercite{krauss2024debunking}.
Expanding the scope of AI's deployment in science will be required for sustained scientific research and to stimulate new fields rather than merely automate existing ones.
\section*{Methods}
\subsection*{M1. Dataset and Paper selection}
In this section, we introduce the procedure of selecting the research papers included in our analysis.
In this paper, we conduct our major analyses based on OpenAlex~\supercite{openalex}.
OpenAlex is a scientific research database built upon the foundation of the Microsoft Academic Graph (MAG)~\supercite{mag,oag}.
Supported by non-profit organizations, OpenAlex is continuously updated, providing a sustainable global resource for research information.
As of March 2025, OpenAlex contains 265.7M research papers, along with related data about citation, author, institution, etc.
Among the massive quantity of papers in the OpenAlex dataset, we select 66,117,158 English research papers published in journals and conferences spanning from 1980 to 2025 and filter out those with incomplete titles or abstracts.
We identify the scientific discipline each paper belongs to utilizing the topics contained in OpenAlex, which are extracted using a natural language processing approach that annotates titles and abstracts with Wikipedia article titles as topics sharing textual similarity.
In the raw dataset, these topics form a hierarchical structure and each paper is associated with several.
Adopting the 19 basic scientific disciplines in the Microsoft Academic Graph (MAG)~\supercite{mag,oag}, i.e., art, biology, business, chemistry, computer science, economics, engineering, environmental science, geography, geology, history, materials science, mathematics, medicine, philosophy, physics, political science, psychology, and sociology, we trace along the hierarchy to determine to which disciplines each topic belongs.
We note that because the original topics of one paper may be retraced to different topics, the scientific discipline of each paper may not be unique.
In other words, one paper may span two or more academic disciplines, e.g., chemistry and biology, which reflects the common phenomena of borderline or interdisciplinary research~\supercite{porter2009science}.

In this paper, we emphasize the adoption of AI methods in conventional natural science disciplines and exclude research developing AI methodologies themselves, separating the influence of AI on science from AI’s own invention and refinement.
Therefore, we select biology, medicine, chemistry, physics, materials science, and geology as representatives of natural science disciplines, while we exclude computer science and mathematics, where most works introducing and developing AI methods are published.
We also exclude art, business, economics, history, philosophy, political science, psychology, and sociology, in order to focus on how AI is changing the natural sciences and career trajectories in the sciences.
Our 6 natural science disciplines include the majority of OpenAlex articles, resulting in 41,298,433 papers, containing 18,392,040 in biology, 4,209,771 in chemistry, and 2,380,666 in geology, 4,755,717 in materials science, 24,315,342 in medicine, 5,138,488 in physics.
The selected disciplines cover various dimensions of natural science, representing an broad view of scientific research as a whole.

\subsection*{M2. Divide three stages of AI development}
We divide the history of AI development into three key eras: the traditional machine learning (ML) era (1980–2014), the deep learning (DL) era (2015–2022), and the generative (GAI) era (2023–present).
We consider 1980 as the start of the traditional machine learning era because several landmark researches were published in the 1980s, such as the back-propagating method~\supercite{rumelhart1986learning,lecun1989backpropagation}.
We regard that the deep learning era began in 2015, as marked by breakthroughs including ResNet, which enabled the training of ultra-deep neural networks, revolutionizing fields including computer vision and speech recognition~\supercite{he2016deep}.
Finally, we divide the GAI era to begin in 2023 with the publication of ChatGPT, a representative large language model, in December 2022, which saw the advent of large-scale transformer-based models capable of strong generalized performance across a wide range of tasks, sparking new applications in natural language processing and beyond.
Each of these transitions was driven by advances in algorithms, computational power, and data availability, substantially expanding the capabilities and scope of AI for science.

\subsection*{M3. Design and fine-tune the language model for AI paper identification}
Insofar as both a paper’s title and the abstract contain important information about its content, we independently train two separate models based on paper titles and abstracts, and then integrate the two models into an ensembled one by averaging their outputs.
The structure of our NLP model for paper identification consists of two parts.
The backbone network is a 12-layer BERT model with 12 attention heads in each layer, and the sequence classification head is a linear layer with a 2-dimensional output atop the BERT output.
We normalize the 2-dimensional output with a softmax function and obtain the probability that the paper involves AI-assistance.
We utilize the BERT model named “bert-base-uncased” from Hugging Face~\supercite{HuggingFace}, which is pre-trained with a large-scale general corpus, and set the maximum length of tokenization to be 16 for titles and 256 for abstracts.
We design a two-stage fine-tuning process with training and validation sets, which we extracted from the OpenAlex dataset, to transfer the pre-trained model to our paper identification task.
The construction of positive and negative data is different between the two stages. In both stages, we randomly split the positive and negative data into 90\% and 10\% to correspondingly obtain training and validation sets.
We use the training set for model training and employ the validation set to select the optimal model.
Because the numbers of positive and negative cases are unbalanced, we use the bootstrap sample technique on positive cases to balance its number with negative cases at both stages.

In the first stage, we construct relatively coarse positive data, only considering eight typical AI journals and conferences, i.e., \textit{Nature Machine Intelligence}, \textit{Machine Learning}, \textit{Artificial Intelligence}, \textit{Journal of Machine Learning Research}, \textit{International Conference on Machine Learning (ICML)}, \textit{International Conference on Learning Representations (ICLR)}, \textit{AAAI Conference on Artificial Intelligence}, and \textit{International Joint Conference on Artificial Intelligence (IJCAI)}.
Among the papers belonging to our chosen 6 disciplines, we extract all papers published in these venues as positive cases and randomly sample 1\% of the remaining papers in our six chosen natural science fields as negative cases, resulting in 26,165 positive and 291,035 negative data.
We fine-tune the pre-trained model for 30 epochs on the training set and select the optimal model according to the F1-score on the validation set.

In the second stage, we construct more precise positive data based on the obtained optimal model in the first stage.
We identify papers in the whole OpenAlex dataset and aggregate the results for each venue, obtaining the probability for each venue across OpenAlex to be an AI venue by averaging the AI probability for all papers within it.
We then select the venues with $>80\%$ AI probability and $>100$ papers as AI venues.
We also incorporate venues with “machine learning” or “artificial intelligence” in their names.
In papers belonging to our 6 chosen disciplines, we extract all papers published in the selected AI venues as positive cases and randomly sample 1\% of those remaining as negative cases, resulting in 31,311 positive and 231,258 negative cases.
Then, we fine-tune the obtained optimal model in the first stage for another 30 epochs with the new training set and select the best model according to F1-score on the new validation set.
Finally, we utilize optimal ensemble models during both stages to identify all papers that use AI to support natural science research from the selected representative natural science disciplines.

\subsection*{M4. Scrutinize our identification results by disciplinary experts}
We arbitrarily sample 220 papers (110 papers $\times$ 2 groups) from each of the 6 disciplines, resulting in 12 paper groups in total.
We enlisted 12 experts with abundant AI research experience (Supplementary Table~S1) and assigned 3 different groups of papers to each.
Without revealing the identification results obtained with the BERT model, we queried our experts whether each paper was an AI paper.
In this way, each paper is repeatedly labeled by three distinct experts, and we evaluate the consistency among these experts based on Fleiss’ Kappa coefficient ($\kappa$)~\supercite{landis1977measurement,fleiss1971measuring}, which is an unsupervised measurement for assessing the agreement between independent raters.
Having confirmed consensus among our experts, we draw the final expert label of each paper from the three experts according to the principle of the minority obeying the majority. We regard the expert labels as ground truth and validate the result of our BERT model against it with the F1-Score, which is a supervised accuracy measurement of accuracy.

\subsection*{M5. Determine the project leader of papers}
Here, we define the project leader as the last author of a research paper, in alignment with conventions established by previous studies~\supercite{sekara2018chaperone}.
To ensure that in most papers, the last authors represent the project leader, we examine the fraction of papers that list authors following alphabetical order.
First, we directly traverse all selected papers and obtain that the prevalence of papers listing authors in alphabetical order, which ranges from 14.87\% in materials science to 22.15\% in geology.
Nevertheless, it is difficult to distinguish whether these papers are really intended to list the authors in alphabetical order or the list of authors according to their roles, which just happen to unintentionally fall in alphabetical order. The latter situation is more likely to occur when there are fewer authors, i.e., two or three.
To tackle this analytical challenge, we determine the fraction of “unintended” alphabetical author lists through a Monte Carlo method.
We generate 10 randomly shuffled copies of the author list for each paper and obtained that from 13.82\% (materials science, standard deviation $\sigma=0.02$) to 20.28\% (geology, standard deviation $\sigma=0.03$) papers have alphabetically listed authors among the random author lists.
This indicates the proportion of “unintended” alphabetical author lists, and we can derive the actual fraction of papers with intentionally alphabetical author lists by the difference between the above two sets of statistical results.
The actual fraction obtained illustrates that only 1.58\% of papers across all disciplines intentionally list the authors in alphabetical order (Supplementary Table~S12), and therefore, we can, with negligible interference, assume that we can identify last authors as team leaders.

\subsection*{M6. Detect scientists’ career role transition}
The OpenAlex dataset incorporates a well-designed Author Name Disambiguation (AND) mechanism~\supercite{openalex}, which utilizes an XGBoost model~\supercite{chen2016xgboost} to predict the likelihood that two authors are the same based on features like their institutions, co-authors, and citations, and then applies a custom, ORCID-anchored clustering process to group their works, assigning a unique ID for each author.
Simply utilizing unique IDs, we are able to track a large number of authors at the same time~\supercite{hill2025pivot}, where we depict an individual scientist’s career trajectory using a role transition model (Extended Data Fig. 4a) and extract the role transition trajectories for scientists.

First, we traverse all selected papers in the 6 disciplines and extract all the scientists involved in any of these papers.
Then, for each individual scientist, we extract all papers in which they have been involved and record the time of their first publication in any role, the time of their first publication as team leader (if ever), and the time of their last publication.
Subsequently, we filter out scientists whose publication records span only a single year.
We also filter out those who directly start as established scientists leading research teams without a role transition from junior scientists.
Finally, we detect the time that each scientist abandons academic publishing.
Considering that one scientist may not publish papers continuously every year, we cannot regard them as having left academia based on their absence in the published record for a single year.
Therefore, we follow the settings in previous research~\supercite{milojevic2018changing} to use a threshold of 3 years and regard scientists who have no more publications after 2022 as having exited academia, while those who still publish papers after 2022 are considered to have an unclear ultimate status and are excluded from analysis.
Finally, we obtain 2,282,029 scientists in the 6 disciplines with complete role transition trajectories.
We also classify them into AI and non-AI scientists according to whether they have published AI-augmented papers.

Moreover, by analyzing author contribution statements collected in previous studies~\supercite{xu2022flat,lin2023remote}, we further validate our detection results by examining changes in scientists’ self-reported contributions throughout their careers (Extended Data Fig. 4b).
Results indicate that junior scientists primarily engage in a large proportion of technical tasks, such as conducting experiments and analyzing data, and less in conceptual tasks, such as conceiving ideas and writing papers. Nevertheless, the proportion of conceptual work significantly rises ($p<0.01$ and $df=1$ in Cochran-Armitage test) during their tenure as junior scientists, reaching saturation at a high level (60\% or more) upon transition to becoming established scientists. This finding validates our definition of role-transition by demonstrating a shift in the nature of scientists’ contributions from participating in research projects to leading them.

\subsection*{M7. Estimate the birth-death model for career development of junior scientists}
To obtain a more precise quantification of how much AI accelerates the career development of junior scientists, we employ a general birth-death model~\supercite{kendall1960birth}.
This type of stochastic process model depicts the dynamic evolution of a population as members join and exit.
In our context, it models the role transitions of junior scientists.
Specifically, we use two separate birth-death models for junior scientists who eventually become established and those who leave academia, respectively.
Here, “birth” processes refer to the entry of junior scientists into academia, and “death” processes symbolize their transition out of the junior stage, either by becoming established scientists or quitting academia.
Because the entry and exit of each junior scientist are independent from one another, we use Poisson processes to model “birth” (entry) and “death” (exit) events, respectively.

The Poisson process is a typical stochastic process model for describing the occurrence of random events that are independent of each other~\supercite{kingman1992poisson}.
The mathematical formula of the Poisson process is:
\begin{equation}
    P(N(t_0)=k)=\frac{(\lambda_0 t_0)^k}{k!}e^{-\lambda_0 t_0}, t_0>0, k=0, 1, 2, \dots,
\end{equation}
where $N(t_0)$ denotes the number of random events that happened before time $t_0$, and $\lambda_0$ is the parameter of the Poisson process, depicting the happening rate of random events.
We consider a birth-death model where birth and death dynamics are both Poisson processes, and rate parameters are $\mu$ and $\omega$, respectively.
Through mathematical derivation~\supercite{meisling1958discrete}, we conclude that the duration time $t$ from birth to death follows an exponential distribution with the parameter $\omega-\mu$, where the exact form of the probability density function is:
\begin{equation}
    P(t)=(\omega-\mu)e^{-(\omega-\mu)t}, t>0.
\end{equation}
We consider the difference between the two rate parameters $\omega-\mu$ as a whole and fit it with a single parameter $\lambda$.
Then, the transition time for junior scientists to become established scientists or leave academia follows the exponential distribution:
\begin{equation}
    P(t)=\lambda e^{-\lambda t}, t>0,
\end{equation}
and the corresponding survival function is
\begin{equation}
    S(t)=1-\int_0^t P(u)du=e^{-\lambda t}, t>0.
\end{equation}
Hence, the average transition time is the conditional expectation of the distribution defined as follows:
\begin{equation}
    \Bar{t}=E[t|t>1]=\int_1^{\infty}t\cdot P(t)dt=\int_1^{\infty}t\cdot\lambda e^{-\lambda t}dt=\frac{1}{\lambda}+1.
\end{equation}

We fit the role transition time of the scientists with the aforementioned exponential distribution, thereby determining the respective values of $\lambda$ for AI-adopted junior scientists and their non-AI counterparts.
Guided by the underlying mechanism of junior scientists’ career development incorporated within the birth-death model, expectations from the model offer a more accurate estimate of the average role transition time.

\subsection*{M8. Measure the knowledge extent of papers}
To assess the knowledge extent of a set of research papers within their high-dimensional embeddings
\begin{equation}
\{\mathbf{p[1]},\mathbf{p[2]},\dots,\mathbf{p[n]}\}, \mathbf{p[i]}\in\mathbb{R}^{768},
\end{equation}
we first compute the centroid as the mean of their vector locations:
\begin{equation}
    \mathbf{c} = \frac{1}{n} \sum_{i=1}^{n} \mathbf{p[i]}.
\end{equation}
Next, we compute the Euclidean distance from each embedding to the centroid, where the knowledge extent of the set of papers is defined as the maximum distance or “diameter” of the vector space covered:
\begin{equation}
    KE = \max_{1\leq i\leq n}\|\mathbf{p[i]}-\mathbf{c}\|_2.
\end{equation}
We note that Euclidean distance is highly correlated with the cosine and related angular distances.

In practice, the number of AI and non-AI papers in each domain differ substantially, introducing bias to the measurement of knowledge extent.
To address this issue, we build on prior work~\supercite{milojevic2015quantifying} about cognitive extent\footnote{Cognitive extent is a measure of the breadth of a scientific field’s cognitive territory. It is quantified by the number of unique phrases, as a proxy for scientific concepts, found within a sampled batch of papers with given size.}.
For each domain, we randomly sample 1,000 papers from both AI and non-AI categories, compute their respective knowledge extent, and repeat this process 1,000 times.
By comparing knowledge extent values across these 1,000 random samples, we ensure that the number of AI and non-AI papers is balanced, making our knowledge extent results comparable.

\subsection*{M9. Measuring the knowledge extent of paper families}
To measure how much knowledge space can be derived from each original research, we calculate the knowledge extent of “paper families”, i.e., a focal paper and its follow-on citations.
Focusing on an original research paper $\phi$, which corresponds to a high-dimensional embedding $\mathbf{p_{\phi}}\in\mathbb{R}^{768}$, we extract all $n_{\phi}$ research papers that cite this original paper.
These citing papers are sorted chronologically by publication date, from earliest to most recent.
The corresponding high-dimensional embeddings of these sorted papers are:
\begin{equation}
    \{\mathbf{p_{\phi}[1]},\mathbf{p_{\phi}[2]},\dots,\mathbf{p_{\phi}[n_{\phi}]}\}, \mathbf{p_{\phi}[i]}\in\mathbb{R}^{768}.
\end{equation}
Thereby, we calculate knowledge extent covered by the “paper family” consists of the original paper $\phi$ and the first $n$ follow-on papers citing it ($1 \leq n \leq n_{\phi}$) as:
\begin{equation}
    KE_{\phi}[n] = \max_{1\leq i\leq n\leq n_{\phi}}\|\mathbf{p_{\phi}[i]}-\mathbf{p_{\phi}}\|_2.
\end{equation}

\subsection*{M10. Measure follow-on engagement among papers}
To quantify how frequently citations of the same original paper interact with each other, we design a metric called follow-on engagement building on prior work~\supercite{mcmahan2018ambiguity}.
For an original paper with $n$ citations, there are at most $\frac{n(n-1)}{2}$ possible citations among these $n$ citing papers if everyone cites all papers published earlier than their own.
We then count how many times these $n$ citing papers actually cite one another, denoted as $k$.
Our metric for follow-on engagement is calculated as the ratio of actual to maximum possible citations:
\begin{equation}
    EG = \frac{k}{\frac{n(n-1)}{2}} = \frac{2k}{n(n-1)} = \frac{2k}{n(n-1)} \times 100 (\%). 
\end{equation}
This metric helps quantify the degree of interactions and collaboration among papers that cite the same original work. Prior work has demonstrated a positive association between the ambiguity of a focal work and follow-on engagement~\supercite{mcmahan2018ambiguity}.
\clearpage
\newpage
\section*{Extend data figures}

\setcounter{figure}{0}

\begin{figure}[ht]
\centering
\includegraphics[width=\textwidth]{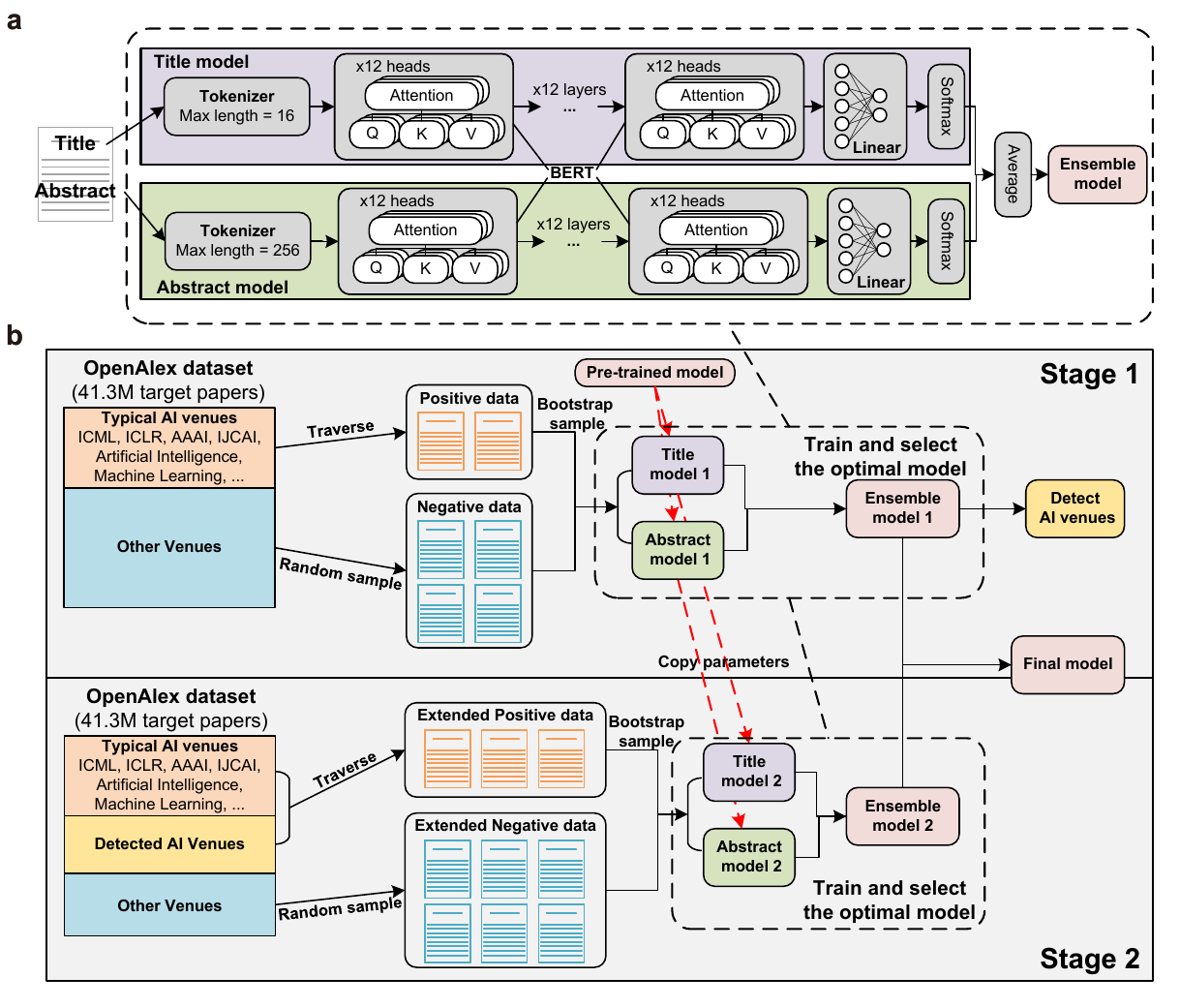}
\caption{
\textbf{Illustration for the method of identifying AI usage in research papers with fine-tuned language models.}
\textbf{(a)} Structure of our deployed language model, which consists of the tokenizer, the core BERT model, and the linear layer.
\textbf{(b)} Procedure of the two-stage model fine-tuning process, where we design specific approaches for constructing positive and negative data at each stage.
}
\label{figSD1}
\end{figure}

\clearpage
\begin{figure}[ht]
\centering
\includegraphics[width=\textwidth]{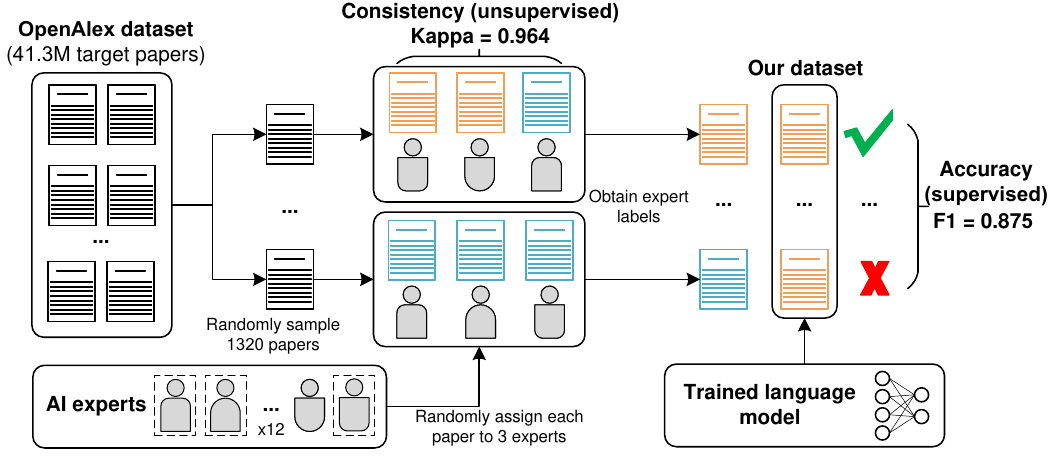}
\caption{
\textbf{Procedure of accuracy evaluation via expert evaluation.}
We randomly sample 1320 papers and delegate three experts to scrutinize the identification results for each paper.
We then draw the final expert label of each paper from the three experts according to the principle of the minority obeying the majority and validate the result of the language model with it.
Results indicate strong consistency among experts and high accuracy with our identification results.
}
\label{figSD2}
\end{figure}

\clearpage
\newpage
\begin{figure}[ht]
\centering
\includegraphics[width=\textwidth]{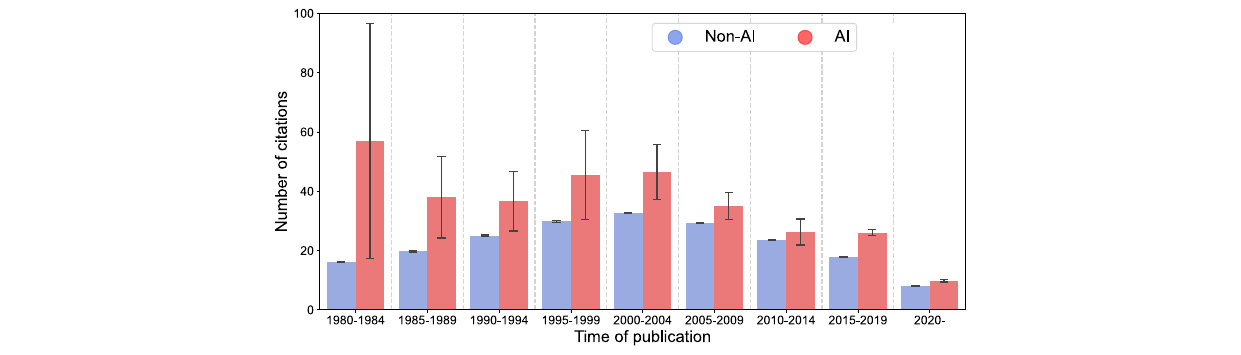}
\caption{
\textbf{Comparison of the total citations of AI and non-AI papers published in different eras.}
The results show that AI papers consistently attract more citations over different eras ($p < 0.001, n=27,405,011$), indicating a higher academic impact than non-AI papers.
99\% CIs are shown as error bars centred at the mean, and the statistical tests use a two-sided t-test.
}
\label{figs2}
\end{figure}

\clearpage
\newpage
\begin{figure}[ht]
\centering
\includegraphics[width=\textwidth]{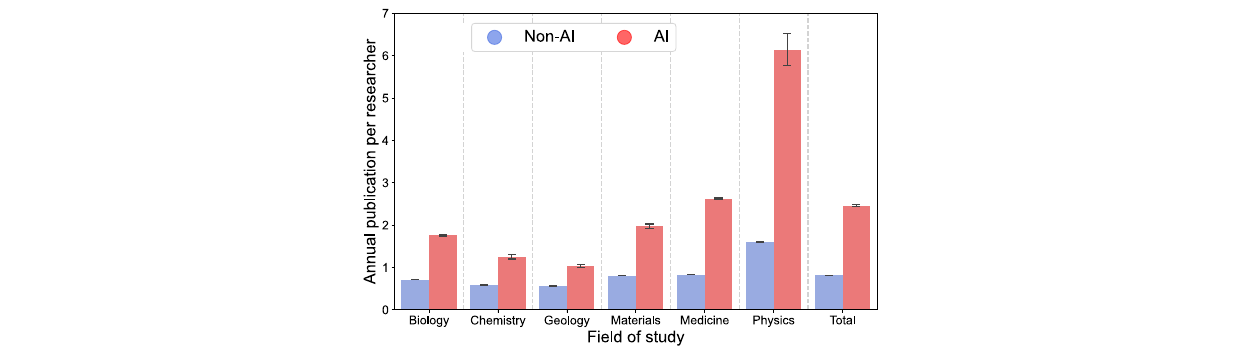}
\caption{
\textbf{Annual publications of researchers adopting AI and their counterparts without AI.}
Results show that in all 6 scientific disciplines, researchers adopting AI are more productive than their counterparts without AI ($p < 0.001, n=5,377,346$).
On average, researchers adopting AI annually publish 3.02 times more papers compared with those not using AI.
99\% CIs are shown as error bars centred at the mean, and the statistical tests use a two-sided t-test.
}
\label{figs3}
\end{figure}

\clearpage
\newpage
\begin{figure}[ht]
\centering
\includegraphics[width=\textwidth]{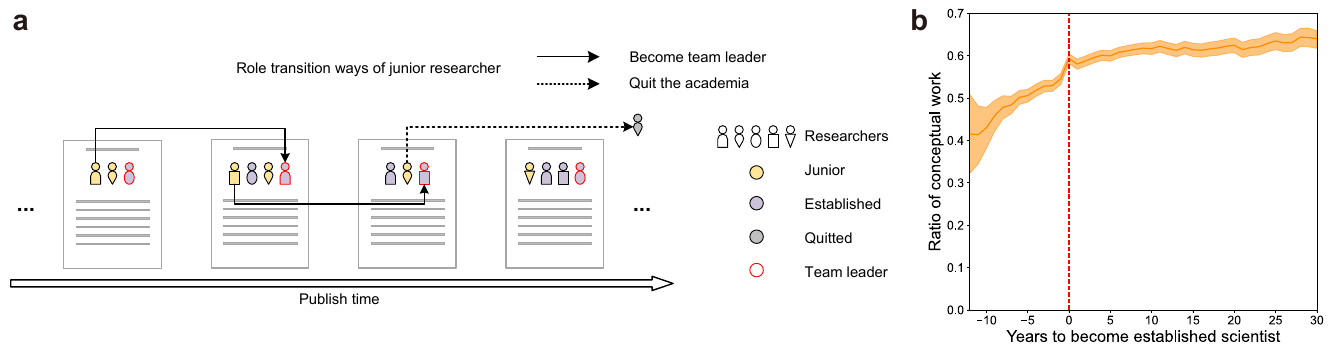}
\caption{
\textbf{Scientists’ career role transition.}
\textbf{(a)} The career role transition of researchers.
We consider the last author of each paper as research project leader and researchers who have been research project leaders as established researchers.
Researchers who have yet to lead a research project are junior researchers, and they have two potential role transition pathways in the future: (1) become established researchers (solid arrow), and (2) abandoning academia (dashed arrow).
\textbf{(b)} Change in the ratio of conceptual work across the research career, before and after becoming an established researcher.
The ratio increases rapidly before the role transition to established researchers, while it remains stable and high after that transition.
99\% CIs are shown as error bands centred at the mean.
}
\label{figs4}
\end{figure}

\clearpage
\newpage
\begin{figure}[ht]
\centering
\includegraphics[width=\textwidth]{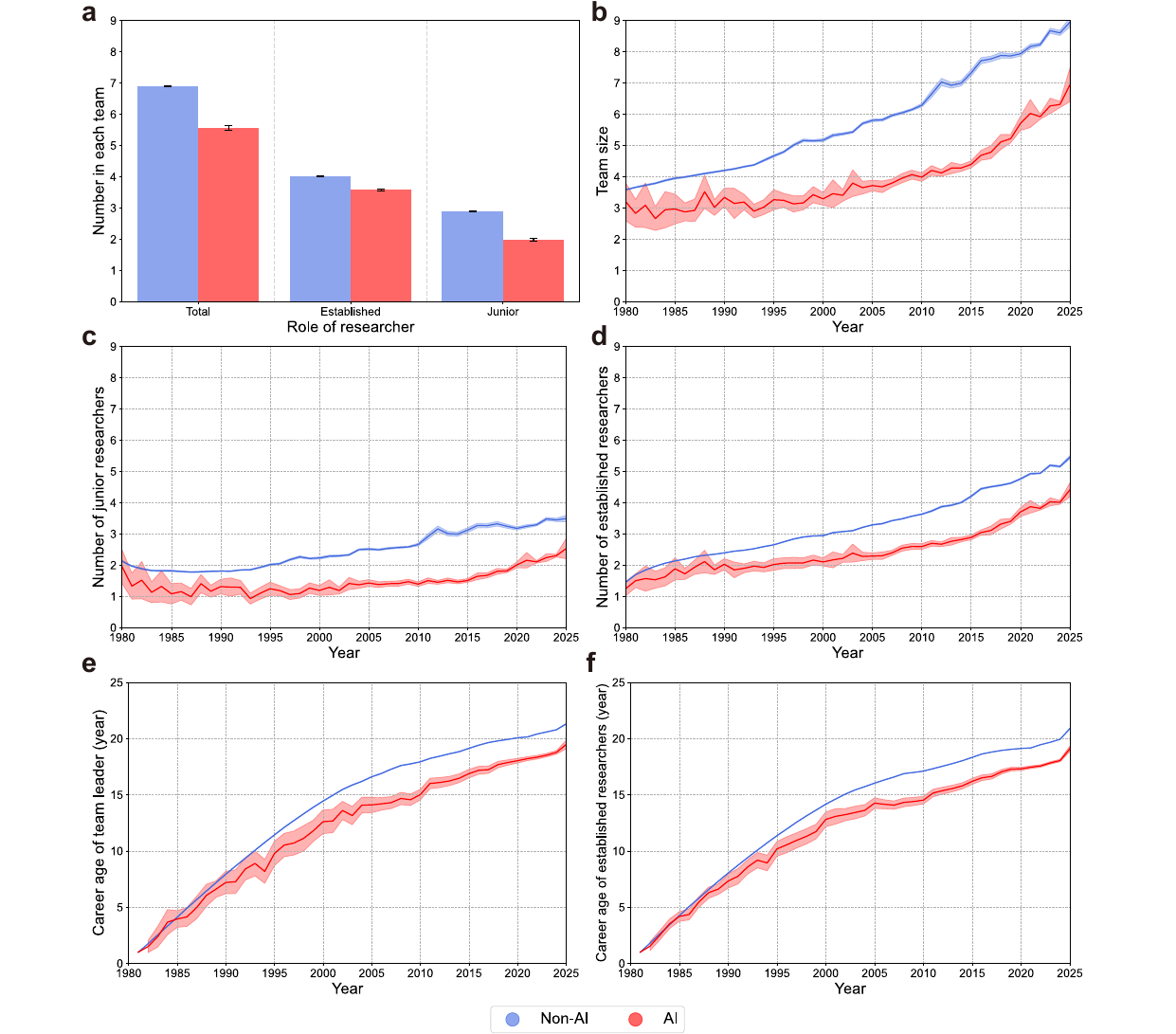}
\caption{
\textbf{Team composition of AI and non-AI papers.}
\textbf{(a)} AI research is associated with reduced research team sizes, averaging 1.33 fewer scientists ($p<0.001, n=33,528,469$).
Specifically, the average number of junior scientists decreased from 2.89 in non-AI teams to 1.99 in AI teams (31.14\%), while the number of established scientists decreased from 4.01 to 3.58 (10.77\%).
\textbf{(b)-(d)} Change in team size, average number of junior researchers, and average number of established researchers.
These findings indicate that within the overall trend of increasing size of scientific research teams, AI adoption primarily contributes to a reduction in the number of junior scientists in teams, while a decrease in the number of established scientists is more moderate.
\textbf{(e)} The average career age of team leaders in AI and non-AI papers.
\textbf{(f)} The average career age of all involved established researchers in AI and non-AI papers.
Results indicate that AI accelerates the transition from junior to established scientists, enabling AI-adopted researchers to become established at a younger age than those without AI.
For all panels, 99\% CIs are shown as error bars or error bands centred at the mean.
All statistical tests use a two-sided t-test.
}
\label{figs5}
\end{figure}

\clearpage
\newpage
\begin{figure}[ht]
\centering
\includegraphics[width=\textwidth]{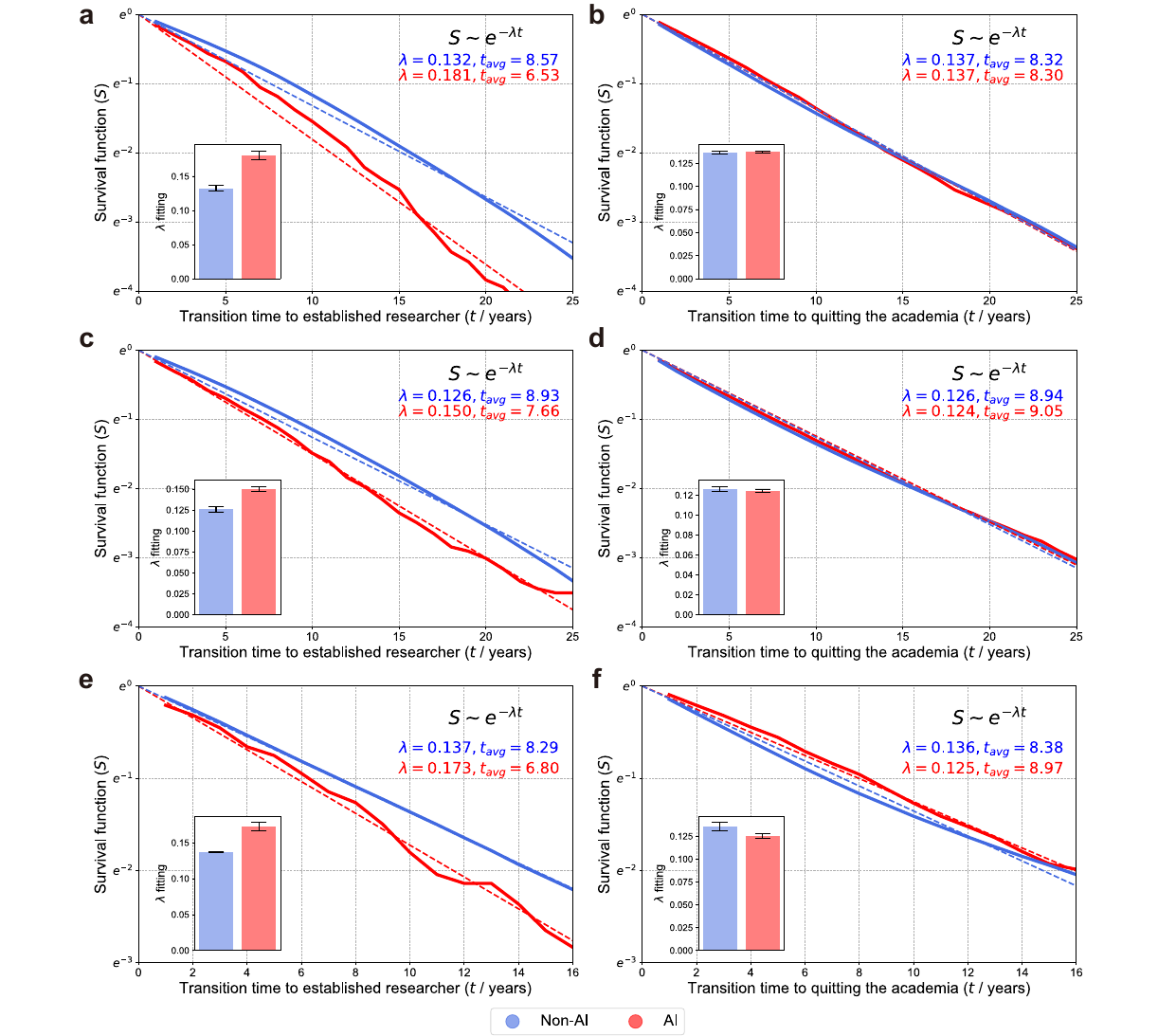}
\caption{
\textbf{Model fitting the role transition time of junior scientists.}
\textbf{(a) (c) (e)} Survival functions for the transition from junior to established researcher in biology ($n=625,093$), medicine ($n=1,137,076$), and physics ($n=120,366$).
\textbf{(b) (d) (f)} Survival functions for the transition from junior researcher to leave academia in biology ($n=625,093$), medicine ($n=1,137,076$), and physics ($n=120,366$).
All survival function can be well-fit with exponential distributions, where the expected time for junior scientists to become established is shorter for those who adopt AI ($p<0.001$), while the expected time for junior scientists to abandon academia is similar or slightly longer for those who adopt AI.
Results indicate that AI not only provides junior scientists opportunities to become established scientists at a younger age, but also reduces the risk of their exiting academia early.
For all panels, 99\% CIs are shown as error bars centred at the mean.
All statistical tests use a two-sided t-test.
}
\label{figs6}
\end{figure}

\clearpage
\newpage
\begin{figure}[ht]
\centering
\includegraphics[width=0.8\textwidth]{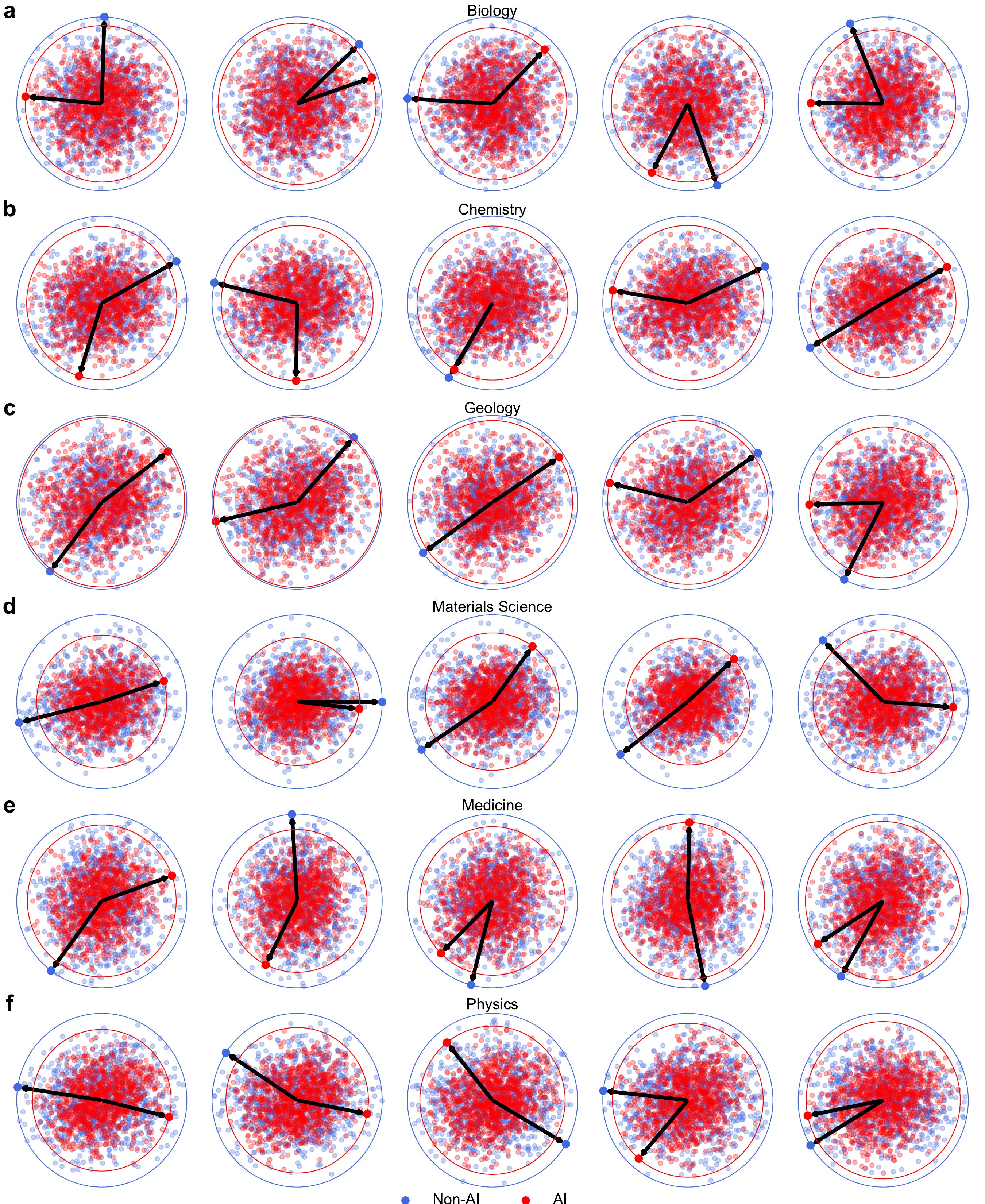}
\caption{
\textbf{The knowledge extent of AI and non-AI papers.}
Here we visualize the embeddings of a small random sample of 2,000 papers, half of which are AI papers and half are non-AI papers.
To eliminate randomness introduced by the T-SNE algorithm, here we simply pick out the first two dimensions of the high-dimensional embeddings to flatten them into a 2-D plot, and we provide 5 different random batches for each field to ensure robustness.
As shown by the solid arrows and circular boundaries, the knowledge extent of AI papers is smaller than that of a comparable sample of non-AI papers, which is consistent across the fields studied in our analysis.
}
\label{figs8}
\end{figure}

\clearpage
\newpage
\begin{figure}[ht]
\centering
\includegraphics[width=\textwidth]{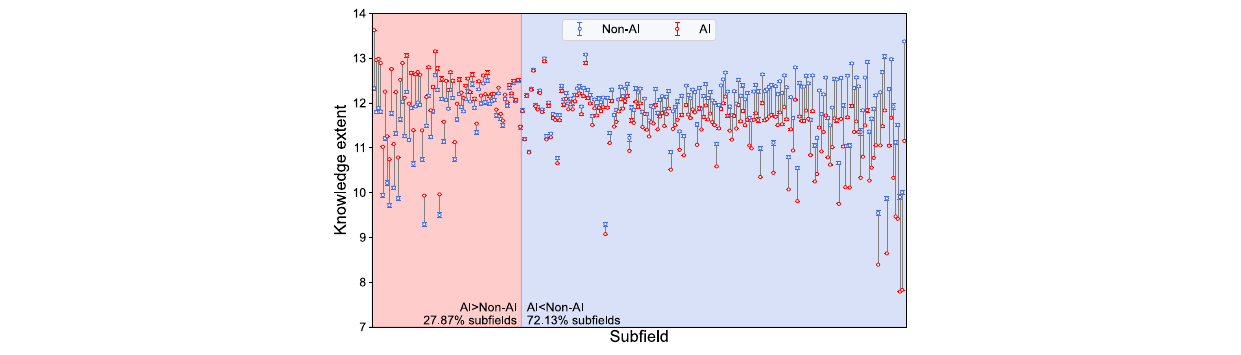}
\caption{
\textbf{The knowledge extent of AI and non-AI papers in each subfield.}
Compared with conventional research, AI research is associated with a shrinkage in the collective knowledge extent of science, where the contraction of knowledge extent can be observed in more than 70\% of over two hundred sub-fields ($n=1,000$ samples in each subfield).
For all subfields, 99\% CIs are shown as error bars centred at the mean.
}
\label{figs9}
\end{figure}

\clearpage
\newpage
\begin{figure}[ht]
\centering
\includegraphics[width=\textwidth]{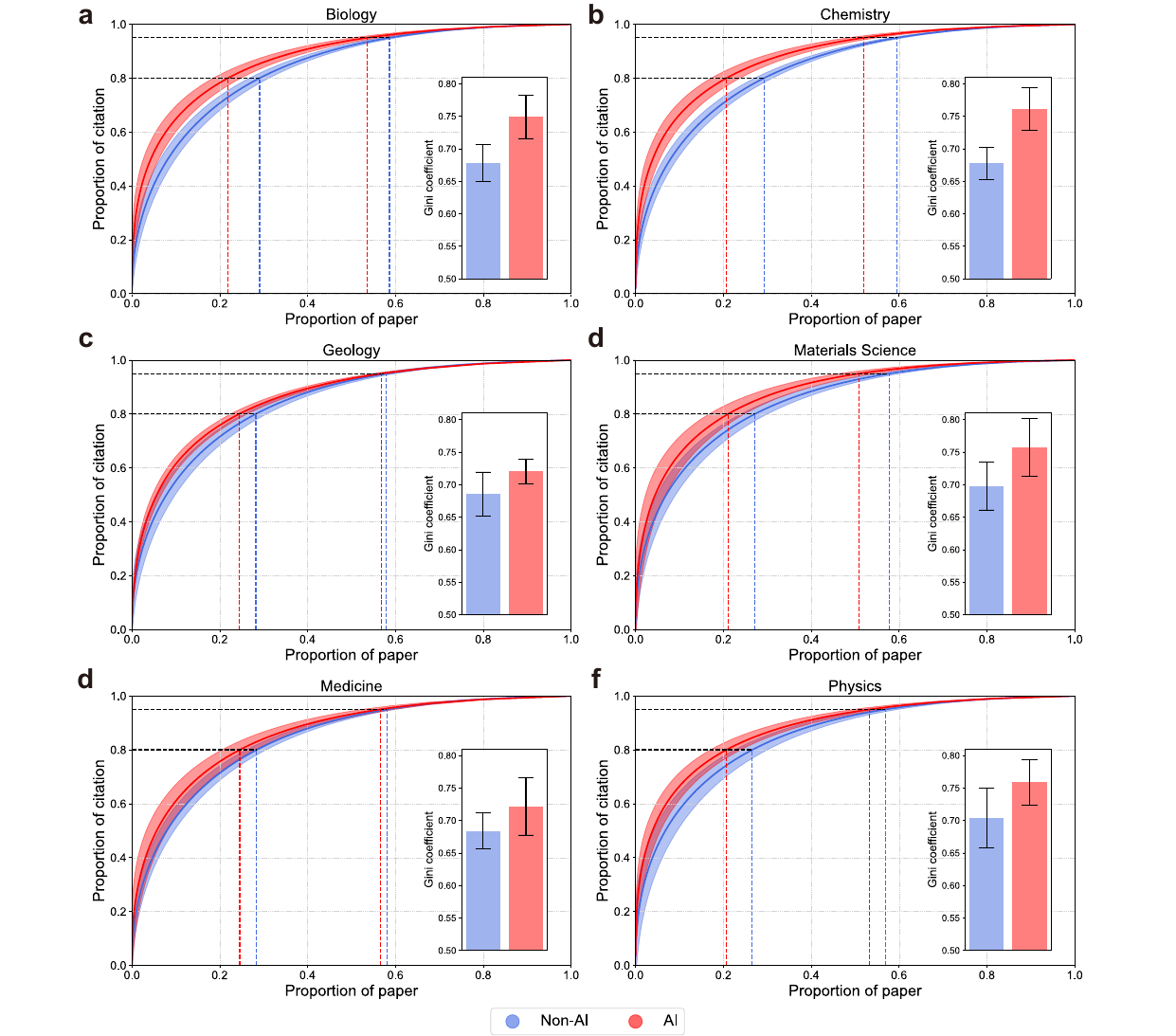}
\caption{
\textbf{The Matthew effect in citations to AI and non-AI papers.}
In AI research, a small number of superstar papers dominate the field, with approximately 20\% of top papers receiving 80\% of citations and 50\% receiving 95\%.
This unequal distribution leads to a higher GINI coefficient in citation patterns surrounding AI research ($p<0.001, n=100$ sampled paper groups for each discipline).
Such disparity in the recognition of AI papers is consistent across all fields examined.
For all panels, 99\% CIs are shown as error bars or error bands centred at the mean.
All statistical tests use a two-sided t-test.
}
\label{figs10}
\end{figure}
\clearpage
\newpage
\setcounter{figure}{0}

\newpage
\section*{Supplementary Notes}
\section{Identifying artificial intelligence in scientific research}
To thoroughly investigate how the adoption of AI impacts scientific research, the very rudimentary step is to identify the AI papers from massive research papers published over the past decades.
In this section, we discuss the details of AI paper identification, covering the model and method design (Section~\ref{Identification of AI papers}), the accuracy evaluation (Section~\ref{Identification accuracy evaluation}), the statistical (Section~\ref{Increasing prevalence of AI in science}) and semantic (Section~\ref{Profile of the identified AI papers}) features of identified AI papers along the trend of scientific research over decades.

\subsection{Identification model and method}
\label{Identification of AI papers}
As described in Methods~M2 of the main text, we design a two-stage fine-tuning process to leverage the pre-trained BERT model to identify papers that use AI to support natural science research.
In the first stage, we construct coarse positive data with typical AI journals and conferences to fine-tune the pre-trained model.
In the second stage, we identify papers in the whole dataset with the obtained optimal model in the first stage and aggregate the results for each venue.
We then select the venues with $>80\%$ AI probability and $>100$ papers as AI venues for positive data, and we also incorporate venues with “machine learning” or “artificial intelligence” in their names.
Finally, we utilize optimal ensemble models during both stages to identify all papers that use AI to support natural science research from the selected representative natural science disciplines.
To illustrate the final identification results, we show the proportion and number of identified AI papers in each venue in Supplementary Fig.~S\ref{figs1}.


\subsection{Identification accuracy evaluation}
\label{Identification accuracy evaluation}
To evaluate the accuracy of our identification, we recruited a team of human experts with abundant AI research experience (Supplementary Table~S\ref{tab1}) to validate the results.
We sample 2 groups of papers at random from each of the 6 disciplines, resulting in 12 paper groups in total, where samples in each group span all three eras of AI (Supplementary Table~S\ref{tab2}).
We assign three different experts to independently label each sampled paper and measure the consistency among experts based on Fleiss’ Kappa coefficient~\supercite{landis1977measurement}.
Results exhibit an overall Fleiss’ Kappa of 0.964, and each of the three eras of AI has a Fleiss’ Kappa greater than 0.93 (Supplementary Table~S\ref{tab3}), indicating strong consensus across the independent annotation of distinct experts.
Taking the expert labels as ground truth and validating the identification results of our BERT model against it, our model reaches an overall F1-score of 0.875, and the F1-scores of the three eras of AI are all greater than 0.85 (Supplementary Table~S\ref{tab4}).
This indicates the consistent high quality of our AI paper identification, which lays a robust foundation for our subsequent analysis.

To provide rationale and explainability for our identification results, we offer multiple identification examples from different eras of AI and visualize the average attention strengths within our title and abstract BERT models.
Our example for the ML era is a medicine paper with AI published in \textit{PNAS} 2006~\supercite{damoiseaux2006consistent}, where our model allocates substantial attention to terms such as “independent component analysis” (Supplementary Fig.~S\ref{figsR1S14a}a).
In the DL era, our first example is a chemistry paper with AI published in \textit{Nature} 2018~\supercite{segler2018planning}, and our second is a biology paper with AI published in \textit{Nature} 2021~\supercite{jumper2021highly}, where our model allocates substantial attention to terms such as “neural network” (Supplementary Fig.~S\ref{figsR1S14a}bc).
Our example of the generative AI era is a biology paper using AI and published in \textit{Science} 2023~\supercite{lin2023evolutionary}, where the model allocates substantial attention to terms such as “large language model” (Supplementary Fig.~S\ref{figsR1S14a}d).
These illustrate how our identification model correctly interprets and accurately identifies various AI-related contents from papers published in different eras of AI.

Furthermore, to gain a deeper understanding of our identification results, we present an example in which the model makes a mistake (Supplementary Fig.~S\ref{figsR1S14b}).
The sample paper, published in \textit{Géoscience}~\supercite{kandlikar2005representing}, is incorrectly classified by the model as AI-related, but human experts confirm that the article does not involve AI methodologies.
As we illustrate, the model assigns substantial attention to terms like “representing” and “deep”, which are commonly associated with AI research.
In the context of this article, however, “representing” refers to general expression or depiction, which is unrelated to representation learning in AI, and “deep” is used to describe the extent of uncertainty rather than indicating a deep neural network or similar AI concepts.
This example illustrates an edge case where the identification model may produce misclassifications.
Nevertheless, given the high F1-score observed in our evaluation of identification accuracy, such cases appear to be rare, and the effect on our subsequent analyses minimal.


\subsection{Profile of the identified AI papers}
\label{Profile of the identified AI papers}
To better illustrate how AI has been applied in natural science research across different disciplines and AI development eras, we present a semantic overview of the major research topics and the primary AI methods employed.
For research topics, AI-related studies tend to cluster around areas that integrate artificial intelligence with conventional disciplinary subjects (Supplementary Fig.~S\ref{figs3}).
For example, in the field of medicine, the topics “Radiology, Nuclear Medicine and Imaging” and “Computer Vision and Pattern Recognition” rank among the most prominent. This reflects the widespread application of computer vision techniques to enhance research in medical imaging.

For adopted AI methods, we extracted phrases from the abstracts of identified AI papers and calculated the frequency of phrases related to specific AI techniques. We identified and listed the top 10 most frequently used AI methods for each discipline and AI era (Supplementary Tables~S\ref{tab5}–S\ref{tab11}).
It is worth noting that in earlier years, due to the limited number of AI papers, the number of commonly used AI methods may be fewer than ten.
The results show that during the ML era, the most frequently applied AI methods in natural science research included Artificial Neural Networks (ANN), Principal Component Analysis (PCA), and Support Vector Machines (SVM).
In the DL era, Convolutional Neural Networks (CNN), a landmark of the time, dominated the landscape, while traditional methods such as SVM continued to see widespread adoption.
Since the beginning of the generative AI era in 2023, Large Language Model (LLM) has increasingly appeared among the most commonly used AI methods, with their frequency rankings steadily rising across disciplines.
The evolution of AI methods in natural science research reflects the development and transformation of AI technologies over the past decades.


\subsection{Increasing prevalence of AI adoption in science}
\label{Increasing prevalence of AI in science}
In the main text, the proportion of papers and researchers adopting AI exhibit exponential growth over the past decades.
To further illustrate and analyze this upward trend, we separately plot the proportion of papers and researchers adopting AI in each discipline with a log-scale y-axis and estimate the growth rate with exponential fitting in each era (Supplementary Fig.~S\ref{figsR1S12a}-S\ref{figsR1S12b}).
As results show, the proportion of papers and researchers adopting AI fit to straight regions in the log-scale plot, indicating an exponential growth trend.
Meanwhile, the growth rate of AI papers and AI researchers in all disciplines increased progressively from the ML to the DL and generative AI eras, as estimated by the exponential fitting, which underscores the increasing prevalence and fast development of AI in science.

Beyond the general upward trend of AI adoption in science, we further investigate how the recent emergence of generative AI influences this trajectory.
We calculate the monthly increase in the proportion of AI papers and researchers across disciplines in recent years (Supplementary Fig.~S\ref{figsR1S13}).
Results show that following the release of ChatGPT in December 2022, which we regard as the beginning of the generative AI era, growth rates in the proportion of papers and researchers initially remain consistent with prior trends.
A period of time later, however, these growth rates begin to exhibit marked acceleration across disciplines, providing evidence for the impact of generative AI advancement.
Meanwhile, it also aligns with intuitive expectation: the mass adoption of generative AI in natural science research requires time, and there is an inherent delay for generative AI-based research to pass through peer review and achieve publication.


\section{Extended analyses on how AI impacts individual science}
In the main text, we illustrate that AI research attracts more attention from academia, and AI-adopting scientists are more likely to achieve higher scholarly
productivity and impact.
In this section, we discuss in more detail how AI impacts individual science, covering the effect of AI on individual papers (Section~\ref{The effect of AI on individual papers}) and individual scientists’ careers (Section~\ref{The effect of AI on individual scientists’ careers}).

\subsection{The effect of AI on individual papers}
\label{The effect of AI on individual papers}
As demonstrated in the main text, AI-related papers, on average, receive higher annual citation counts than non-AI papers.
Given that citation distributions are empirically right-skewed~\supercite{redner1998popular,seglen1992skewness}, we adopt multiple statistical indicators beyond the average annual citation counts to ensure more robust conclusions (Supplementary Fig.~S\ref{figsR1S4}-S\ref{figsR2S4}).
For the right tail of the distribution, namely papers with high citation counts, we observe that from year of publication and across subsequent decades, citations for the top 1st and 5th-percentile of AI papers exceed those of non-AI papers by 152.39\% and 48.27\%, respectively.
For the left tail of the distribution, namely papers with low citation counts, we find that, over the same time period, the proportion of AI papers receiving fewer than three and fewer than five citations each year is 2.49\% and 2.46\% lower, respectively, than non-AI papers.
Taken together, these findings suggest that AI papers not only tend to receive higher citations but are also less likely to become low-impact publications, indicating the enhanced visibility and academic attention garnered by AI research.

In addition to the “velocity” of citation, shown above as the number of citations received per year after publication, we also compare the “acceleration” of citations, namely the year-over-year change in annual citation counts, between AI and non-AI papers (Supplementary Fig.~S\ref{figsR1S3}).
Results reveal that during the initial post-publication years when annual citations are generally increasing, AI papers exhibit a faster acceleration in citation growth.
In the subsequent period, after reaching peak citation “velocity”, AI papers also exhibit a more rapid deceleration in annual citations. Over the longer term, the changing patterns of annual citations to AI and non-AI papers converge, with no significant differences observed.
Nevertheless, throughout the entire citation lifecycle, the citation “velocity” of AI papers consistently remains higher than that of non-AI papers.

Empirically, review articles, editorial pieces, and other types of special publications tend to exhibit markedly different citation patterns compared to original research papers.
To verify the robustness of our finding that AI papers garner higher academic impact, we distinguish these special publications from those reporting original research and replicate our analyses exclusively on the latter (Supplementary Fig.~S\ref{figsR1S5a}-S\ref{figsR1S5b}).
Specifically, we filter publications labeled as “article” in the OpenAlex dataset~\supercite{openalex}, thereby excluding journals dedicated to reviews as well as publication types such as letters, editorials, and erratum.
To further eliminate a small number of review articles that may still be included in non-review journals, we additionally filter all papers with “review” or “survey” in their titles.
As a result, these special pieces account for 9.26\% of the 27,405,011 publications with intact reference records from our original analysis.
After excluding them, we obtained a refined dataset of 24,867,012 original research papers.
On this subset, AI papers still receive higher visibility and academic attention than their non-AI counterparts across the previously introduced statistical indicators.

In our analysis of annual citations for AI and non-AI papers, we find that both types of papers continue to be cited even 20 or more years after publication.
Given that fewer papers, especially AI-related ones, were published several decades ago and that most scholarly works cite “historical” papers to situate their research within a broader context, it is important to understand whether referenced papers are being cited as foundational citations or simply “throwaway” citations that the scholarly crowd uses more colloquially.
To address this, we follow an established methodology for distinguishing between core and superficial citations~\supercite{hao2024HLM}, and we compute the times AI and non-AI papers are cited as core citations in each year following publication (Supplementary Fig.~S\ref{figsR1S8}).
Results show that the proportion of AI papers to be cited as core citations tends to decrease over time, which aligns with the intuitive expectation that older papers are more likely to be cited for historical context rather than as active intellectual foundations.
Nevertheless, foundational influence can still be observed in decades-old papers.
For example, two recent studies published in 2020 and 2021, which used AI techniques to predict diabetes~\supercite{hasan2020diabetes,kumari2021ensemble}, both cite a 1988 paper on the ADAP learning algorithm~\supercite{smith1988using} as a foundational reference.
Both newer papers build upon the prior algorithm to design more advanced ensemble models.
Moreover, in terms of absolute counts of being core citations, AI papers still surpass non-AI papers by 98.70\%, reflecting the heightened academic attention attracted by AI research.

AI represents a technology that started small and then became the most prominent method in multiple fields. As such, AI papers exhibit a distinctive citation pattern compared to non-AI papers.
To investigate whether this citation pattern is unique to AI or also characteristic of other emerging technologies that started small and then became widely appreciated and used, we conduct a comparative analysis using Nanotechnology as case study (Supplementary Fig.~S\ref{figsR1S9}).
We identified Nanotechnology-related research by detecting the presence of the word “nano” in the titles of all publications.
As shown, Nanotechnology was sparsely studied in its early stages but began to gain widespread attention across disciplines such as biology, chemistry, and physics beginning in the 1990s.
We observe that Nanotechnology exhibits a citation pattern partially similar to AI: Nanotechnology papers are still cited as core citations many years following publication, although the proportion of being such core citations declines over time.
In contrast to AI, however, both total number of citations and the frequency of being core citations for Nanotechnology papers eventually decline to levels indistinguishable from non-Nanotechnology papers, suggesting that nanotechnology does not exhibit the same sustained higher academic influence as AI.

To provide a more comprehensive assessment of the impact of AI papers, we analyze additional indicators beyond citation-related statistics.
First, we examine the distribution of AI papers across journals of varying Journal Citation Report (JCR) quantiles~\supercite{jcr} (Supplementary Fig.~S\ref{figs2}).
We find that the proportion of AI papers in Q1 journals is 18.60\% higher than non-AI ones in all journals, and in Q2 journals, the AI proportion is a scant 1.59\% higher, while Q3 and Q4 journals hold a relatively lower proportion of papers with AI.
These results indicate a heterogeneous distribution of AI-augmented papers across journals, with a higher prevalence in high-impact journals.
Second, we examine the influence of AI and non-AI papers from the perspective of disruption~\supercite{wu2019large,park2023papers} (Supplementary Fig.~S\ref{figsR1S7}).
Despite higher citation counts for AI papers, we find that their disruption scores are lower.
This suggests that while AI papers tend to influence a larger number of subsequent studies, the higher impact is primarily developmental rather than disruptive, which means that they contribute to advancing (and completing) existing fields rather than initiating new ones.
This observation is consistent with our later finding that AI contracts science’s focus.


\subsection{The effect of AI on individual scientists’ careers}
\label{The effect of AI on individual scientists’ careers}
As shown in the main text, scientists who adopt AI tend to achieve higher scholarly productivity and impact.
However, an alternative explanation for this observation is whether these advantages are driven by AI adoption itself, or whether scientists who are already on a trajectory toward greater productivity and impact are simply more likely to adopt AI.
To investigate this, we conducted a match-and-comparison analysis of scientists with similar early-career productivity and impact trajectories, but who differ in their subsequent AI adoption behavior (Supplementary Fig.~S\ref{figsR1S11}).
Specifically, we filter 11,019 scientists who began adopting AI in the third year of their careers and match them with 1,926 scientists who exhibited comparable annual citation counts during their first three years but never adopted AI.
The comparison reveals that starting from the third year, when the former group began adopting AI, a divergence in annual citation counts emerges between the two groups.
By the tenth year of their careers, the scientists who adopted AI in their third year have 24.45\% higher annual citations than their non-AI-adopting counterparts with similar early trajectories.
Similarly, their annual productivity in the tenth year is 9.61\% higher than the matched group of non-AI-adopters with comparable early-career productivity.
The same pattern holds when analyzing a second cohort: 9,837 scientists who adopted AI in the fifth year of their careers were compared with peers who had similar citation and productivity levels in their first five years but never adopted AI.
Taken together, these findings suggest that for scientists with comparable early-career positions, adopting AI itself contributes to their subsequent advantages in productivity and impact.

In our main analysis, we include 2.3 million scientists with complete career trajectories and show that those adopting AI are more productive and receive more citations.
Prior research suggests that only a relatively small subset of scientists continue publishing over long periods, however, forming what has been recognized as the global core of active researchers~\supercite{ioannidis2014estimates}.
To examine whether our findings hold for this persistent group, we identified 1,495,265 researchers with uninterrupted publication records over at least five consecutive years and 525,716 researchers over at least ten consecutive years, accounting for 52.30\% and 18.39\% of all researchers in our dataset, respectively (Supplementary Fig.~S\ref{figsR1S6}).
We then compared the productivity and citation impact of AI-adopting versus non-AI-adopting researchers within these subsets.
Among those with at least five consecutive years of publications, researchers adopting AI published 2.40 times more papers and received 3.88 times more citations per year than their non-AI counterparts, with consistent patterns observed across disciplines.
Similarly, for researchers with at least ten consecutive years of publications, who are more productive and receive more citations than researchers with five consecutive years of publications, AI adopters published 2.41 times more papers and received 4.31 times more citations annually.
These results further confirm the positive impact of AI adoption on the career progression of both the continuously publishing core scientists and broader population of normal scientists.

In our career transition model of researchers, we set a threshold of 3 years and regard scientists who have no more publications after 2022 as having exited academia, where the threshold setting is consistent with previous research~\supercite{milojevic2018changing}.
To ensure the robustness of our findings, we switch to different thresholds in detecting the dropout of researchers and replicate our results (Supplementary Fig.~S\ref{figsR1S2b}).
As we illustrate, when using different dropout thresholds of 2 or 4 years, the probability for AI-adopted junior scientists to transition to established scientists is consistently higher than for their counterparts who do not adopt AI, and the anticipated transition time to becoming established scientists is shorter for AI-adopted junior scientists compared to their counterparts.
These robust results underscore the role of AI in bringing about increased opportunities for junior scientists to lead research teams and reducing the risks of their leaving academia.
To assess the appropriateness of our threshold for detecting researcher dropout, we analyzed the distribution of gap year durations in researcher careers (Supplementary Fig.~S\ref{figsR1S2a}).
Here, a period of gap years refers to a temporary interruption in a researcher’s publication activity, followed by a subsequent resumption of publishing. Results show that 44.67\% of all gap periods lasted only one year, and 76.94\% lasted no more than three years.
Therefore, when choosing a three-year threshold for identifying dropout events, it can appropriately capture the majority of cases where researchers temporarily paused and then resumed publication activity.
Additionally, as we illustrate, both shorter and longer thresholds yield similar overall results.
Nevertheless, a threshold that is too short would misclassify many researchers with temporary gaps as dropouts, while a threshold too long would result in a large number of researchers being classified as having uncertain status and excluded from analysis, thereby reducing the sample size.
Taking these considerations into account, we adopt a three-year threshold as a balanced and robust choice.

To better understand whether the adoption of AI affects all groups of scientists uniformly, we provide demographic information on the people who are leaving the field (Supplementary Fig.~S\ref{figsR1S10}).
First, we utilized the “institution” field in the OpenAlex dataset to determine researchers’ affiliations and categorized them into demographic groups based on institutional attributes.
On the one hand, we categorize institutions by type, such as companies, educational institutions, government agencies, etc.
Researchers affiliated with educational institutions exhibit the lowest dropout rates, although differences in dropout rates across institution types are relatively modest.
Notably, across all institutional types, researchers who adopt AI show similarly reduced dropout probabilities.
On the other hand, we categorize institutions by geographic region, including Africa, Europe, Asia-Pacific, etc. We find that researchers based in Africa and South America have higher dropout probabilities compared with those in Europe, Asia-Pacific, and North America.
Furthermore, while AI adoption is associated with reduced dropout rates for researchers in regions such as Asia-Pacific and North America, the benefit is far less pronounced for those in Africa and South America.
Second, we follow established methods in prior work and infer gender and ethnicity information based on author names~\supercite{lockhart2023name}.
The results show that White and API (Asian or Pacific Islander) researchers have lower dropout probabilities compared to their Hispanic and Black counterparts.
Meanwhile, AI adoption is associated with reduced dropout rates for White and API researchers, while the benefit is much less pronounced for Black researchers.
We also observe heterogeneity in the effect between male and female researchers.
Male researchers tend to have lower original dropout probabilities and receive greater benefit from AI adoption.
In contrast, female researchers exhibit higher original dropout probabilities and gain relatively less from adopting AI.
These findings provide preliminary evidence regarding the heterogeneous impact of AI on scientific careers across different demographic groups, suggesting that the benefits of AI are unequally distributed, but the causes and consequences of this heterogeneity merit further investigation.


\section{Extended analyses on how AI impacts the overall science}
In the main text, we illustrate the conflict between individual and collective incentives to adopt AI in science, where individual scientists receive expanded personal reach and impact, but the knowledge extent of entire scientific fields shrinks to a narrower focus.
In this section, we discuss in more detail how AI impacts the entire knowledge extent.
First, we illustrate the benefit of topic diversity (Section~\ref{The benefit of topic diversity}).
Second, we discuss multiple factors that may impact the selectivity of AI adoption in different fields and thereby lead to the observed outcome of narrowed focus of AI-augmented research (Section~\ref{Selectivity of AI adoption on different factors}).

\subsection{The benefit of topic diversity}
\label{The benefit of topic diversity}
Given our observation of the contracted knowledge extent of AI-augmented research, a natural question arises: Is a broader knowledge space, namely a greater diversity of research topics, indeed beneficial to the overall advancement of scientific knowledge?
To examine this, we categorize our selected papers into 252 distinct sub-fields and compute, for each sub-field, the average citation count and disruption score~\supercite{wu2019large,park2023papers} across different eras of AI (Supplementary Fig.~S\ref{figsR1S1}).
Results show that, across different sub-fields and eras of AI, neither citation (as a proxy for impact) nor disruption (as a proxy for novelty) is obviously correlated with the sub-field’s knowledge extent.
Specifically, the absolute values of Pearson’s $r$ are below 0.1 in almost all cases.
This indicates that higher topic diversity within a subfield does not dissipate the sub-field’s intellectual energy and reduce its scholarly impact or innovative capacity.
Therefore, maintaining a broader diversity of research topics does not hinder a sub-field’s performance; rather, it likely offers more opportunities for advancing the scientific frontier and provides researchers with a wider array of investigation choices, ultimately benefiting the state of collective knowledge.


\subsection{Selectivity of AI adoption across different topics}
\label{Selectivity of AI adoption on different factors}
To gain a deeper understanding of our main findings that “AI in science has become more concentrated around some popular research topics” and that “AI-augmented research focuses on a narrower scope”, we examine whether external factors might influence the selectivity of AI adoption across different topics, potentially leading to their disproportionate representation.

\textbf{Topicality.} Certain topics may inherently lend themselves more to AI applications, potentially making them more likely to be represented in AI-assist research.
To investigate this possibility, we analyzed the topical correlations between AI and non-AI research (Supplementary Fig.~S\ref{figsR1S15a}).
Based on citation relationships between AI and non-AI papers, we found that AI research across various fields is more frequently cited by non-AI studies than by AI studies themselves.
This suggests that topics in AI research influence non-AI research rather than forming isolated, self-referential clusters of AI literature.
There is no obvious difference in topic selection between AI and non-AI research, indicating that inherent topicality does not account for the disproportionate prevalence of AI-augmented research in different fields.

\textbf{Original impact.} Another possibility concerns whether the topics being squeezed out by AI are likely to be marginalized anyway.
That is, does AI tend to concentrate on more fruitful areas rather than topics with lower prior and potential impact?
To evaluate this, we categorize our selected papers into 1,883 topics according to the OpenAlex taxonomy. We calculate the average citation count and disruption score~\supercite{wu2019large,park2023papers} of non-AI papers within each topic, obtaining proxies for the topic’s original influence and novelty. We then examine, across topics, the correlation between the degree of AI penetration and the original influence and novelty across different AI development eras, where the absolute values of Pearson’s $r$ are below 0.1 in all cases (Supplementary Fig.~S\ref{figsR1S15b}).
The results show that across topics and AI eras, neither original citation nor original disruption is obviously correlated with the proportion of AI adoption in that topic.
This suggests that AI adoption does not selectively favor topics based on their original impact.
Instead, AI homogeneously penetrates into topics with both high and low original influence.

\textbf{Funding priority.} Another potential factor that might influence the selectivity of AI adoption across different topics is funding priority, which means that funding agencies may tend to directly support AI research more in some specific research topics.
To evaluate this, we merge information from the “grants” field in the OpenAlex dataset and acknowledgment texts in the Web of Science (WOS) dataset~\supercite{WoS,mongeon2016journal}, obtaining 31,115,808 coded funding data acknowledgments from 32,437 funding agencies.
By categorizing our selected papers into topics as mentioned above, we calculate the deviation of the probability that AI papers are funded in each topic from the average level and obtain the indicator of funding priority to the topics.
We then examined, across topics, correlation between the degree of AI penetration and funding priority across different AI development eras, where the absolute values of Pearson’s $r$ are below 0.1 in all cases (Supplementary Fig.~S\ref{figsR1S15c}).
Results show that across topics and AI eras, there are heterogeneous funding priorities, but funding priority is not obviously correlated with proportion of AI adoption, suggesting that policy-makers’ choices do not obviously affect AI adoption on selective topics.

\textbf{Data abundance.} As discussed in the main text, the knowledge extent of entire scientific fields tends to shrink to focus attention on areas most amenable to AI research, such as those with an abundance of data.
To directly evaluate the impact of data abundance on the selectivity of AI adoption across topics, we extract and quantify the appearance frequency of data-related terms, such as “data” and “dataset”, in titles and abstracts of papers within each topic.
We then normalize these frequencies to serve as an indicator of data abundance for each topic.
We then examine, across topics, correlation between the degree of AI penetration and data abundance across different AI development eras (Supplementary Fig.~S\ref{figsR1S15d}).
Results show that, as expected, data abundance is diverse across topics, while across AI eras, data abundance is positively and significantly ($p<0.001$ for all eras) correlated with the proportion of AI adoption, suggesting that AI is more likely to be adopted in topics with abundant data resources. Notably, the correlation of AI-use with frequency of mentioned data resources rises with the size and intensity of AI-models from 0.24 in the machine learning era to 0.36 in the deep learning era to 0.43 in the large model era. 

In general, data abundance is a major external factor influencing the selectivity of AI adoption across different topics, contributing to the observed disproportionate representation within knowledge space and the contraction of scientific focus, whereas the other external variables discussed above appear to be largely unrelated.
This finding elucidates factors underlying the heterogeneous adoption of AI across topics and provides valuable insights into how AI can be better leveraged to promote sustainable and comprehensive scientific development.

Given the selectivity of AI across topics, we further examined the consistency of our findings regarding the contracted knowledge extent of AI-augmented research in different topics.
We conducted stratified analyses to control for the effects of the external factors discussed above, which may impact the finding as confounding variables (Supplementary Fig.~S\ref{figsR1S16}).
Results show that when we control each external factor by partitioning them into 10 tiers based on magnitude, contraction in knowledge extent for AI-augmented research remains evident within each tier. Regardless of whether a topic is more or less likely to be selected for AI adoption, our findings hold consistently: existing AI-augmented research within the topic tends to cover a contracted knowledge space compared to non-AI research.


\section{Robustness analyses}
\subsection{Robustness of results in the generative AI era}
Because generative AI has not been around for very long and there remains a lead time before publications using them appear, the available data regarding generative AI is necessarily limited for analysis.
To verify the robustness of our general findings over decades of AI development to the specific era of generative AI, we replicated our findings with only the subset of papers published during the generative AI era.
Of the 41,298,433 publications in our OpenAlex and Web of Science analyses that cover six disciplines, 4,797,614 are published in the generative AI era.
As we show in Supplementary Fig.~S\ref{figLLMera2}-S\ref{figLLMera4}, all results in the generative AI era are consistent with those from prior decades of AI development.

This separate analysis of AI from the generative AI era could exhibit limitations due to the lack of data caused by the relatively short history of generative AI, including LLMs.
One major point is that our method of detecting scientists’ career role transitions requires scientists to maintain a certain length of publication history, which is not feasible separately in the short generative AI era.
Therefore, we are not able to replicate the analysis regarding the career development of individual scientists.
Although reaching quantitatively consistent conclusions, with the limited data currently available, results in the generative AI era may appear less significant than corresponding results in our general findings and serve as a starting point for future research as these new models develop and are deployed in new ways.
As large, generative models evolve over longer periods of time and produce richer data, additional evaluations should be pursued, further revealing how generative AI impacts scientific development consistent with traditional machine learning techniques or representing new directions.

\subsection{Robustness of results on the Web of Science dataset}
To verify the robustness of our findings on a more restricted dataset of high-quality papers, we replicated our analysis using the subset of articles from the OpenAlex that also appear in the Web of Science (WOS) dataset~\supercite{WoS,mongeon2016journal}.
To extract the publications in WOS, we subsetted the OpenAlex publications that could be linked to the WOS dataset based on a shared DOI or a PubMed Identifier (PMID).
Of the 41,298,433 publications in our OpenAlex analyses that cover six disciplines, 23,576,370 could be linked to a WOS publication.
As we show in Supplementary Fig.~S\ref{figWoS2}-S\ref{figWoS4}, all of the results on the WOS dataset are consistent with those from the OpenAlex dataset in the main text.

\subsection{Robustness of distance calculations in the high-dimensional paper embedding space}
Because multiple important analyses in the main text are presented based on the distance calculated in the paper embedding space, it is crucial to ensure reliability of our distance calculations for high-dimensional vectors.
To evaluate this, we utilize a principal components analysis (PCA) approach. We first identified the number of principal components required to explain 90\% of the total variance in the original 768-dimensional embeddings, which we determined to be 135.
We then projected the original embeddings onto these principal components and recalculated distances in this reduced-dimensional space to replicate our analysis. 
As we show in Supplementary Fig.~S\ref{figsR2S1a}-S\ref{figsR2S1b}, all results in the reduced-dimensional space are consistent with those from the original embedding space reported in the main text.

In addition to replicating our analysis in the reduced-dimensional space, we further tested the sensitivity of our distance calculations.
Using the same PCA method, we reduce the original embeddings to 86, 135, and 197 dimensions, which correspond to the principal components that explain 80\%, 90\%, and 95\% of the total variance, respectively.
We then randomly sample 1,000 NonAI-NonAI, 1,000 AI-AI, and 1,000 NonAI-AI paper pairs and calculate the distance for each pair within each of the reduced-dimensional spaces.
As we show in Supplementary Fig.~S\ref{figsR2S2}, distances in the various reduced-dimensional spaces are highly correlated with those in the original space, with Pearson’s $r$ greater than 0.95 ($p<0.001$ for all groups).

Besides the sensitivity to embedding dimension, we also check the robustness regarding different distance measurements.
We substitute all Euclidean distance into Cosine distance and replicate our analyses.
As we show in Supplementary Fig.~S\ref{figsR2S5}, all results with Cosine distance are consistent with the original ones with Euclidean distance in the main text.
These results demonstrate the robustness of our distance calculations based on paper embeddings, ensuring the reliability of our distance-based analyses and suggesting the potential for future research to leverage these distance metrics for further analyses.
\clearpage
\section*{Supplementary Figures}
\begin{figure}[ht]
\centering
\includegraphics[width=\textwidth]{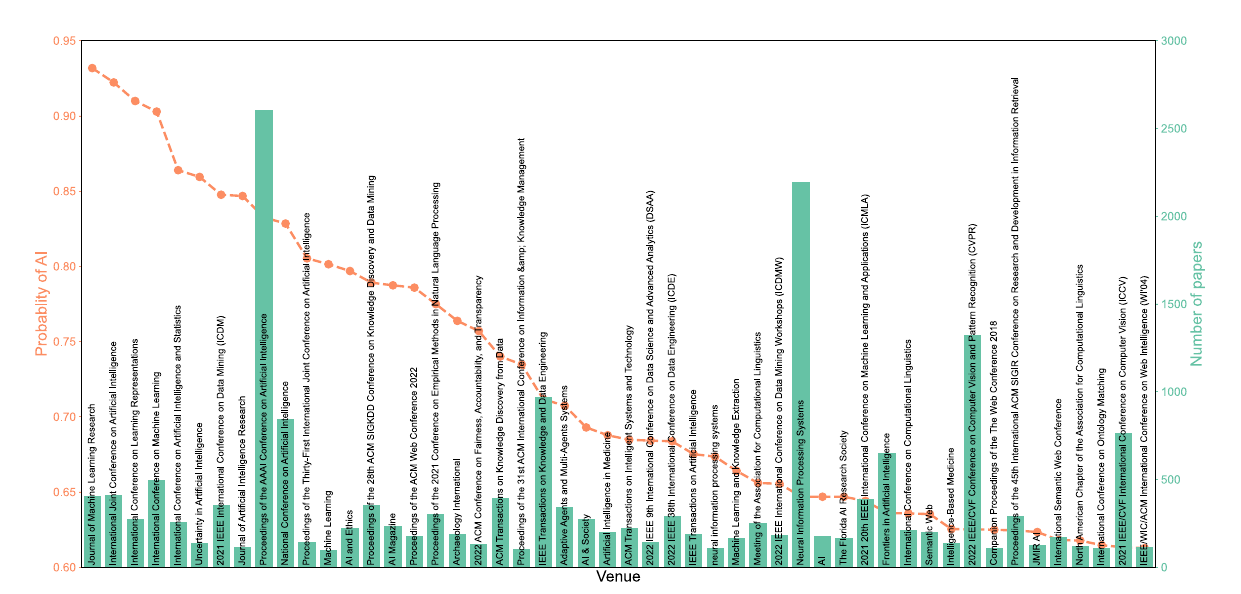}
\caption{
\textbf{Probability of AI (orange) and number of papers (green) for selected venues.} 
Final identification results combine the best models in both stages of fine-tuning. Venues are ordered according to the probability of AI papers within them.
}
\label{figs1}
\end{figure}

\clearpage
\newpage
\begin{figure}[ht]
\centering
\includegraphics[width=\textwidth]{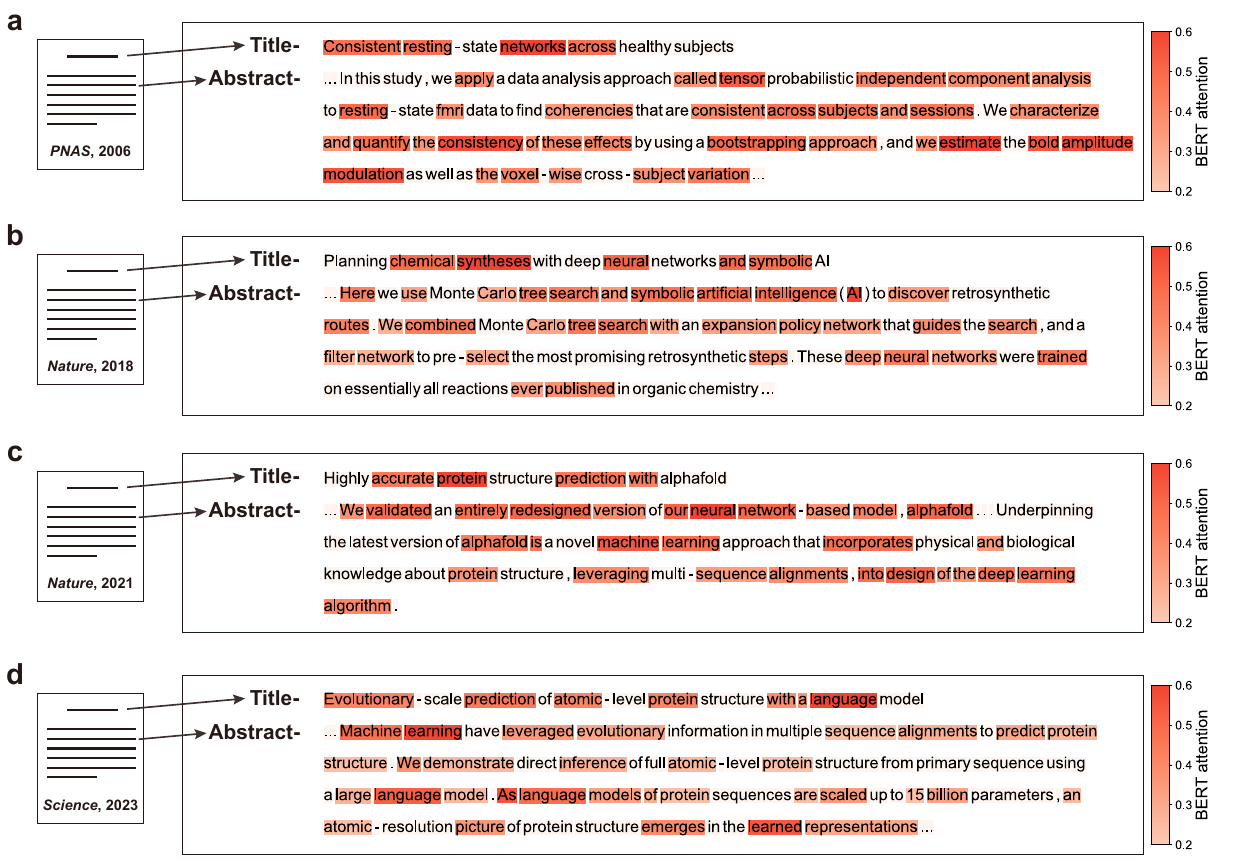}
\caption{\textbf{Correct examples of AI paper identification across different eras of AI development.}
\textbf{(a)} Example from the ML era, a medicine paper with AI published in \textit{PNAS} 2006.
\textbf{(b)} Example from the DL era, a chemistry paper with AI published in \textit{Nature} 2018.
\textbf{(c)} Example from the DL era, a biology paper with AI published in \textit{Nature} 2021.
\textbf{(d)} Example from the GAI era, a biology paper with AI published in \textit{Science} 2023.
}
\label{figsR1S14a}
\end{figure}

\clearpage
\newpage
\begin{figure}[ht]
\centering
\includegraphics[width=\textwidth]{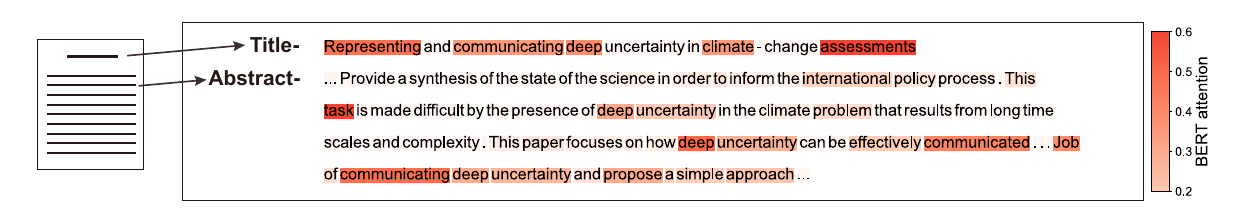}
\caption{
\textbf{An incorrect example of AI paper identification.}
The example is a geology paper without AI published in 2006, which is mistaken to be AI-related by the identification model.
}
\label{figsR1S14b}
\end{figure}

\clearpage
\newpage
\begin{figure}[ht]
\centering
\includegraphics[width=\textwidth]{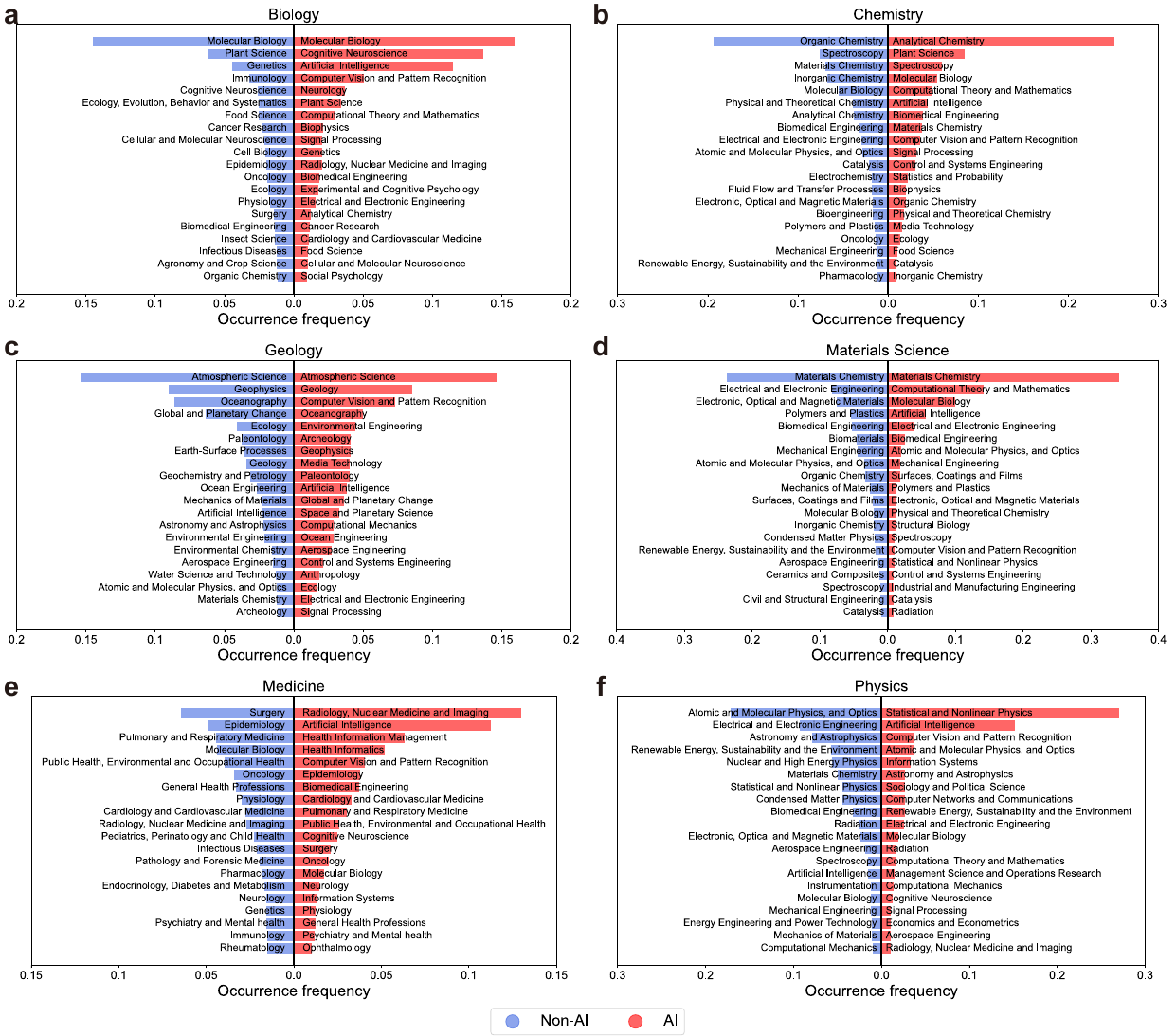}
\caption{
\textbf{Topics with top occurrence frequency in AI and non-AI papers.}
Results show that AI’s primary contribution to conventional research fields is around computer science and machine learning algorithms.
Across disciplines, identified AI papers turn out to be selected combinations of conventional research topics and AI-related techniques, including “Artificial Intelligence”, “Computer Vision and Pattern Recognition”, and “Computational Theory and Mathematics”. 
}
\label{figs3}
\end{figure}

\clearpage
\newpage
\begin{figure}[ht]
\centering
\includegraphics[width=\textwidth]{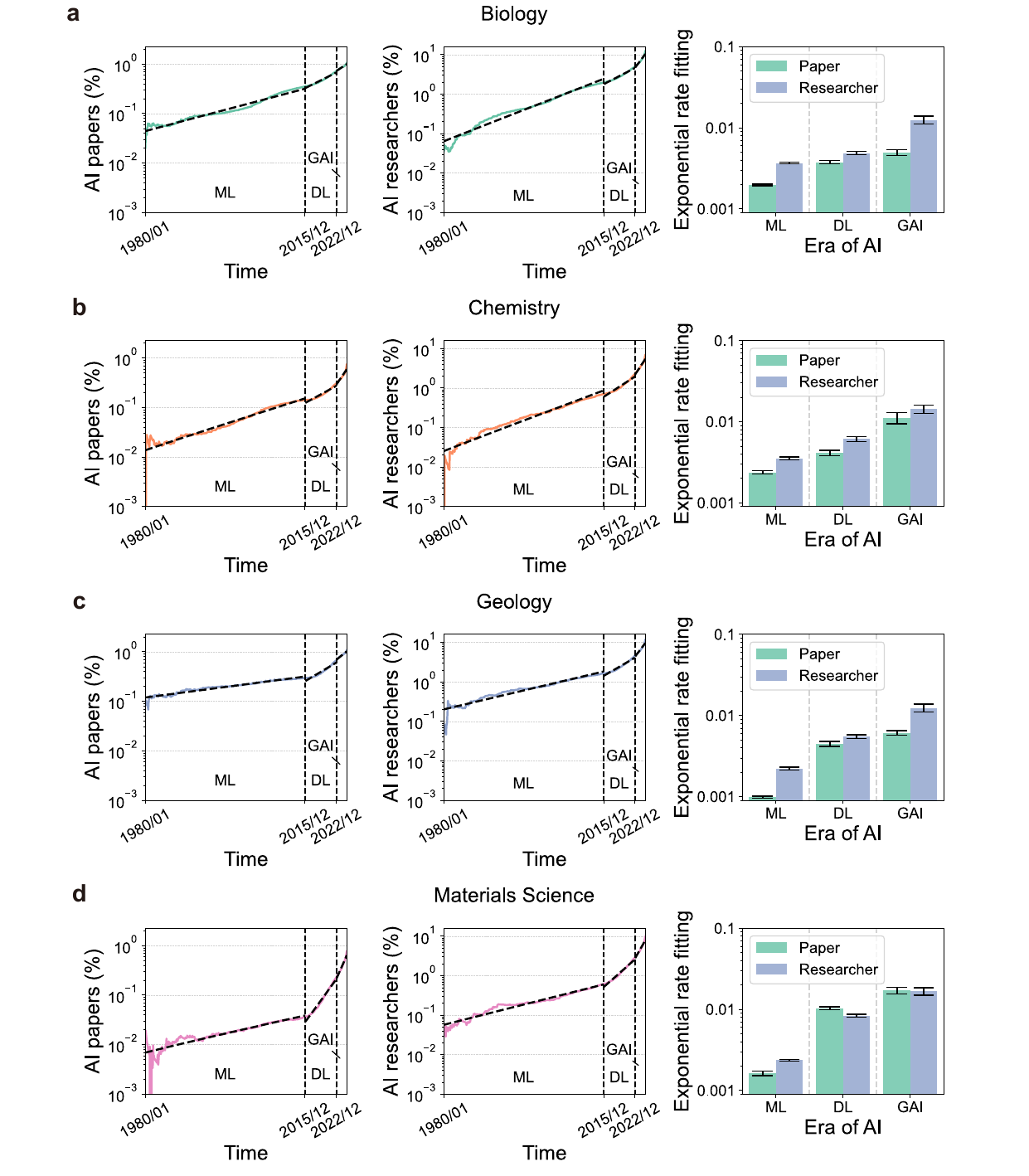}
\caption{
\textbf{Proportion of papers and researchers adopting AI in each discipline with log-scale y-axis and estimated growth rate with exponential fitting for each era.}
Proportion of papers and researchers adopting AI fit to straight regions in the log-scale plot, indicating high precision of the exponential fitting.
Meanwhile, estimated by the exponential fitting, the growth rate of AI papers and AI researchers in all disciplines increased progressively from the ML era to the DL and GAI eras.
For the estimated growth rates ($n=543$ month observations), 99\% CIs are shown as error bars centred at the mean.
}
\label{figsR1S12a}
\end{figure}

\clearpage
\newpage
\begin{figure}[ht]
\centering
\includegraphics[width=\textwidth]{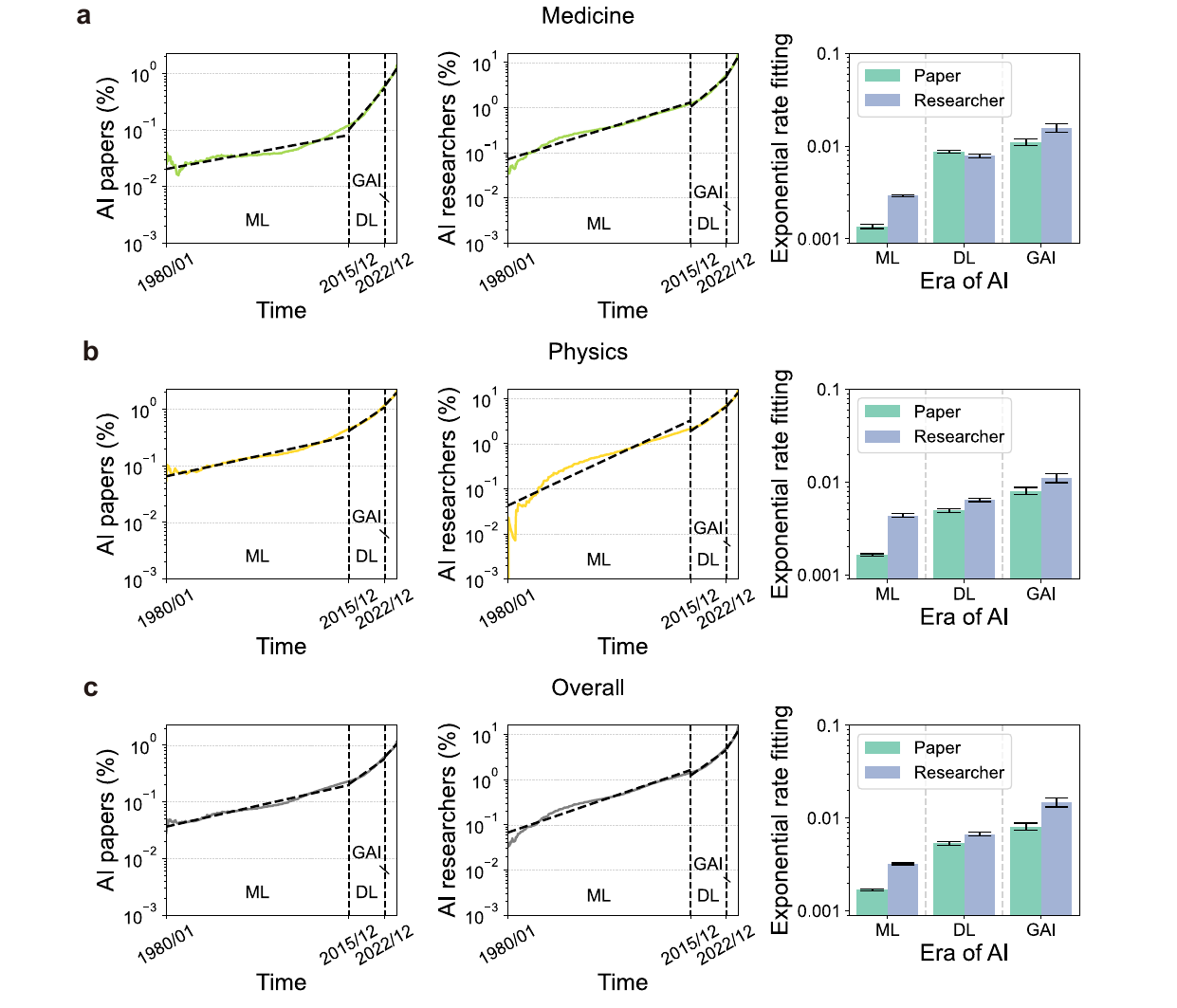}
\caption{
\textbf{(Continued from Supplementary Fig.~S\ref{figsR1S12a}) The proportion of papers and researchers adopting AI in each discipline with log-scale y-axis and estimate the growth rate with exponential fitting in each era.}
The proportion of papers and researchers adopting AI fit to straight regions in the log-scale plot, indicating high precision of the exponential fitting.
Meanwhile, estimated by the exponential fitting, the growth rate of AI papers and AI researchers in all disciplines increased progressively from the ML era to the DL and GAI eras.
For the estimated growth rates ($n=543$ month observations), 99\% CIs are shown as error bars centred at the mean.
}
\label{figsR1S12b}
\end{figure}

\clearpage
\newpage
\begin{figure}[ht]
\centering
\includegraphics[width=\textwidth]{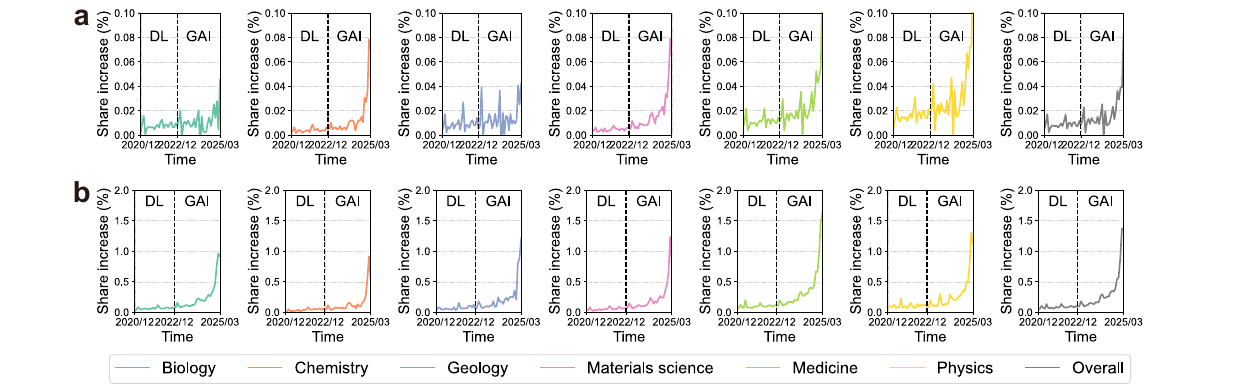}
\caption{
\textbf{The monthly increase in the proportion of AI (a) papers and (b) researchers across disciplines in recent years.}
Following the release of ChatGPT in December 2022, growth rates in the proportion of papers and researchers initially remain consistent with prior trends.
A period of time later, these growth rates begin to exhibit a marked acceleration across disciplines, providing evidence for the impact of generative AI advancement and use.
}
\label{figsR1S13}
\end{figure}

\clearpage
\newpage
\begin{figure}[ht]
\centering
\includegraphics[width=\textwidth]{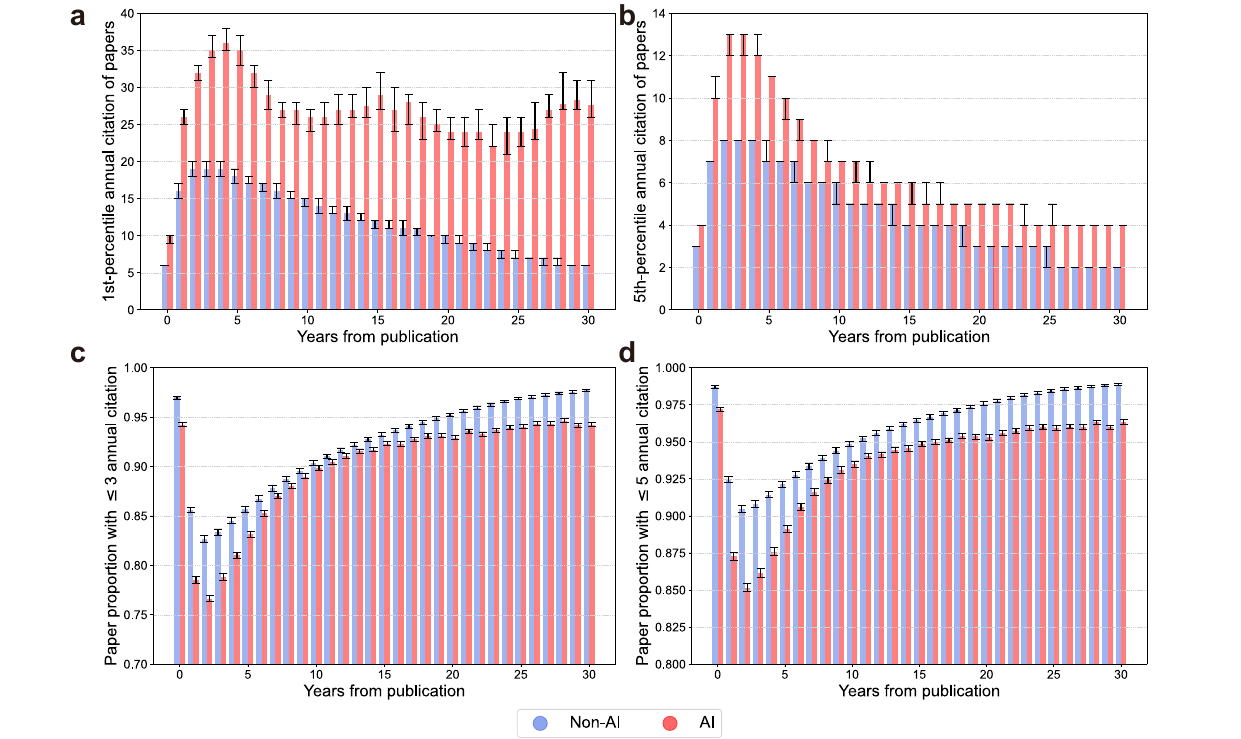}
\caption{
\textbf{Alternative statistical indicators for the annual citation comparison between AI and non-AI papers.}
\textbf{(a)} Top 1st-percentile annual citation for AI and non-AI papers from year of publication ($n=27,405,011$).
\textbf{(b)} Top 5th-percentile annual citation for AI and non-AI papers from year of publication ($n=27,405,011$).
\textbf{(c)} Proportion of AI and non-AI papers receiving fewer than three citations each year from year of publication ($n=27,405,011$).
\textbf{(d)} Proportion of AI and non-AI papers receiving fewer than five citations each year from year of publication ($n=27,405,011$).
For all panels, 99\% CIs are shown as error bars centred at the corresponding percentiles or proportions.
}
\label{figsR1S4}
\end{figure}

\clearpage
\newpage
\begin{figure}[ht]
\centering
\includegraphics[width=\textwidth]{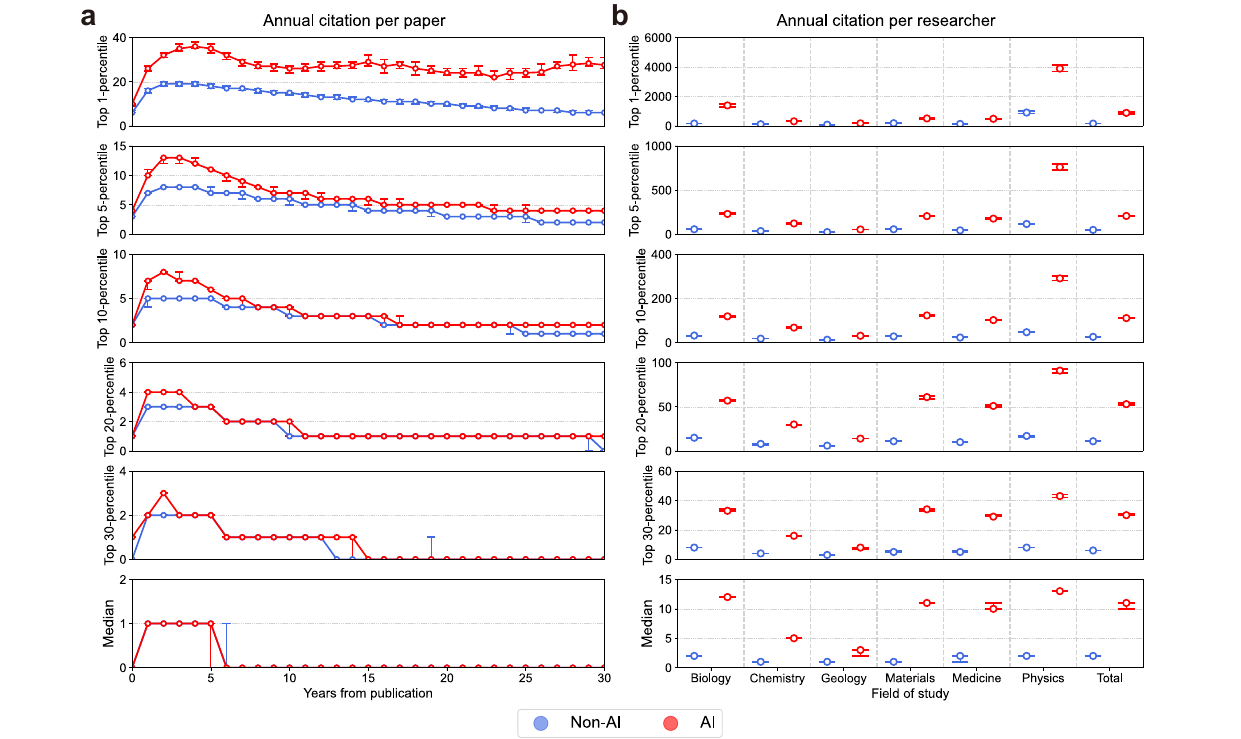}
\caption{
\textbf{Different percentile statistics for the annual citation comparison between AI and non-AI papers.}
\textbf{(a)} Annual citations after publication of AI and non-AI papers ($n=27,405,011$).
\textbf{(b)} Annual citations for researchers adopting AI and their counterparts without AI ($n=5,377,346$).
Consistently indicated by different percentile statistics, AI papers and researchers attract more citations.
For all panels, 99\% CIs are shown as error bars centred at the corresponding percentiles.
}
\label{figsR2S4}
\end{figure}

\clearpage
\newpage
\begin{figure}[ht]
\centering
\includegraphics[width=\textwidth]{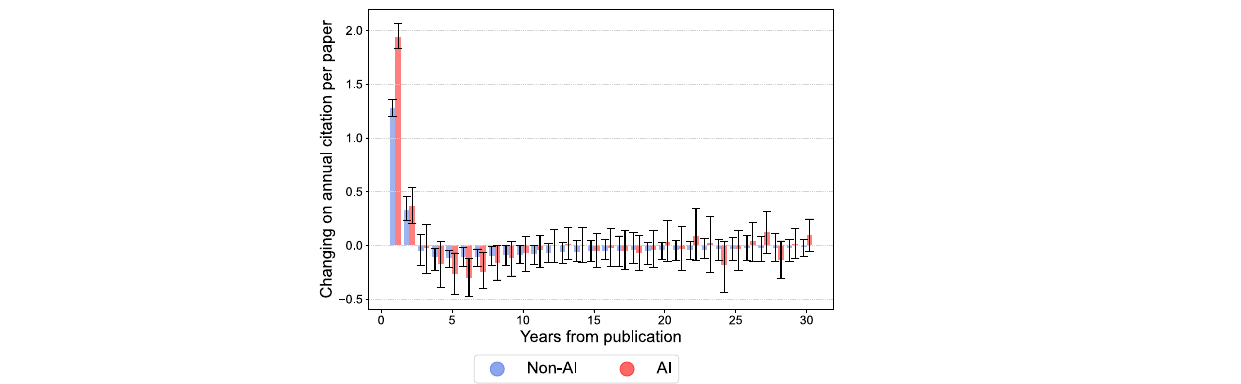}
\caption{
\textbf{Comparison of “acceleration” of citation, namely year-over-year change in annual citation counts, between AI and non-AI papers.}
AI papers experience a faster acceleration in citation growth during the initial post-publication years, and exhibit a more rapid deceleration in annual citations after reaching peak annual citations ($n=27,405,011$).
Throughout the entire citation lifecycle, the annual citation of AI papers consistently remains higher than that of non-AI papers.
99\% CIs are shown as error bars centred at the mean.
}
\label{figsR1S3}
\end{figure}

\clearpage
\newpage
\begin{figure}[ht]
\centering
\includegraphics[width=\textwidth]{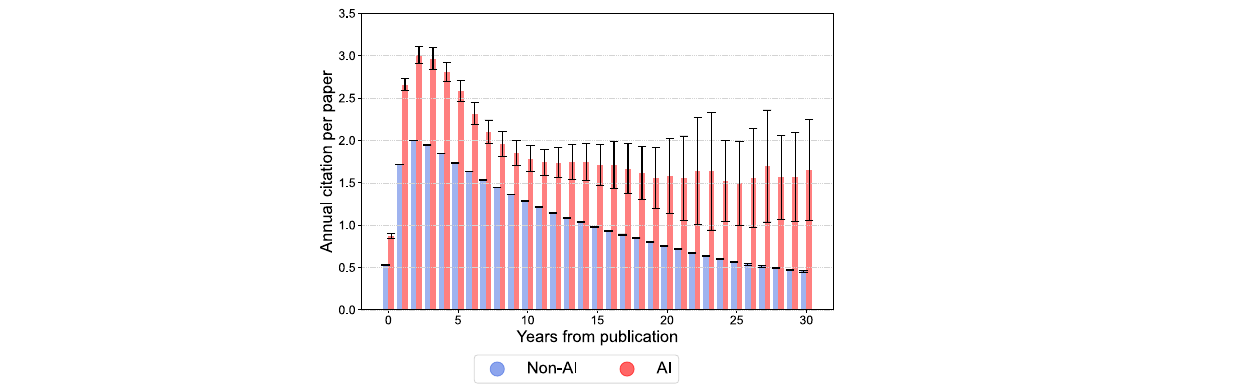}
\caption{
\textbf{Annual citations after the publication of AI and non-AI original research papers, excluding review articles, editorial pieces and other special publications.}
Results show that AI papers attract more citations, indicating higher academic impact than papers without AI ($n=24,867,012$).
Results are consistent with the overall statistics in the main text.
99\% CIs are shown as error bars centred at the mean.
}
\label{figsR1S5a}
\end{figure}

\begin{figure}[ht]
\centering
\includegraphics[width=\textwidth]{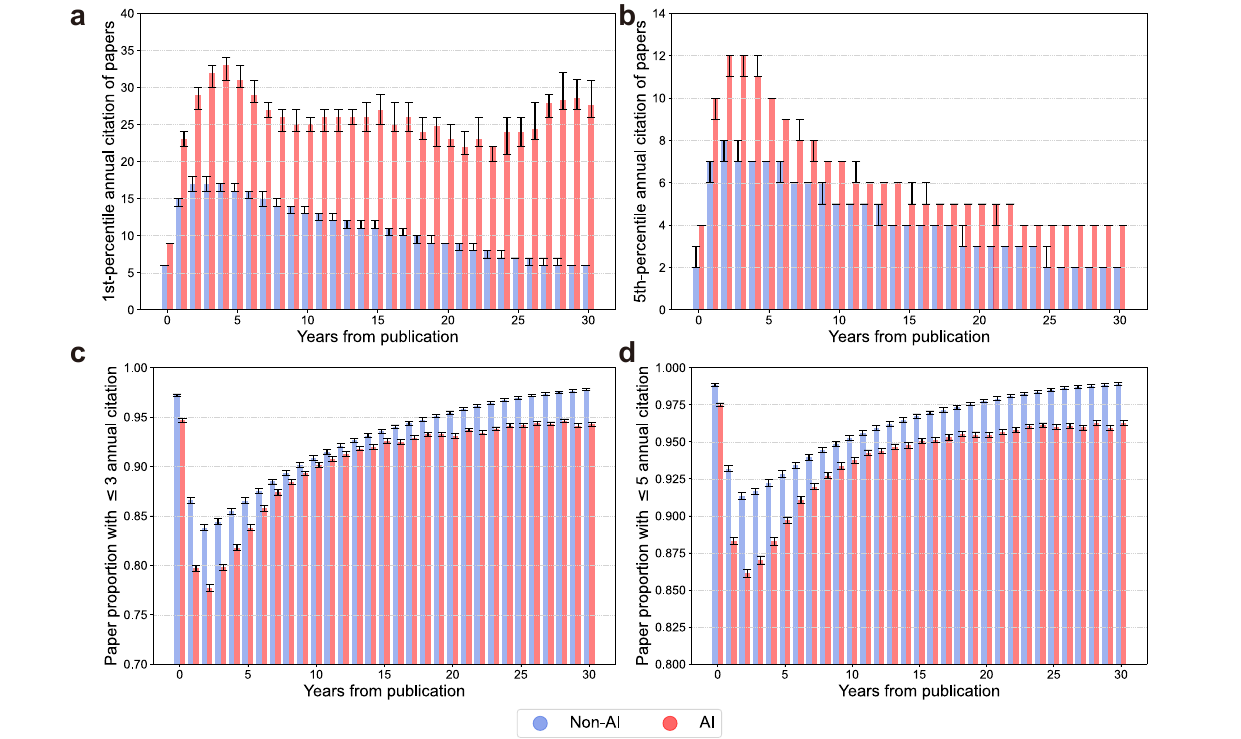}
\caption{
\textbf{Alternative statistical indicators for the annual citation comparison between AI and non-AI original research papers, excluding review articles, editorial pieces and other special publications.}
\textbf{(a)} Top 1st-percentile annual citation for AI and non-AI papers from year of publication ($n=24,867,012$).
\textbf{(b)} Top 5th-percentile annual citation for AI and non-AI papers from year of publication ($n=24,867,012$).
\textbf{(c)} Proportion of AI and non-AI papers receiving fewer than three citations each year from year of publication ($n=24,867,012$).
\textbf{(d)} Proportion of AI and non-AI papers receiving fewer than five citations each year from year of publication ($n=24,867,012$).
Results are consistent with the overall statistics in the main text.
For all panels, 99\% CIs are shown as error bars centred at the mean.
}
\label{figsR1S5b}
\end{figure}

\clearpage
\newpage
\begin{figure}[ht]
\centering
\includegraphics[width=\textwidth]{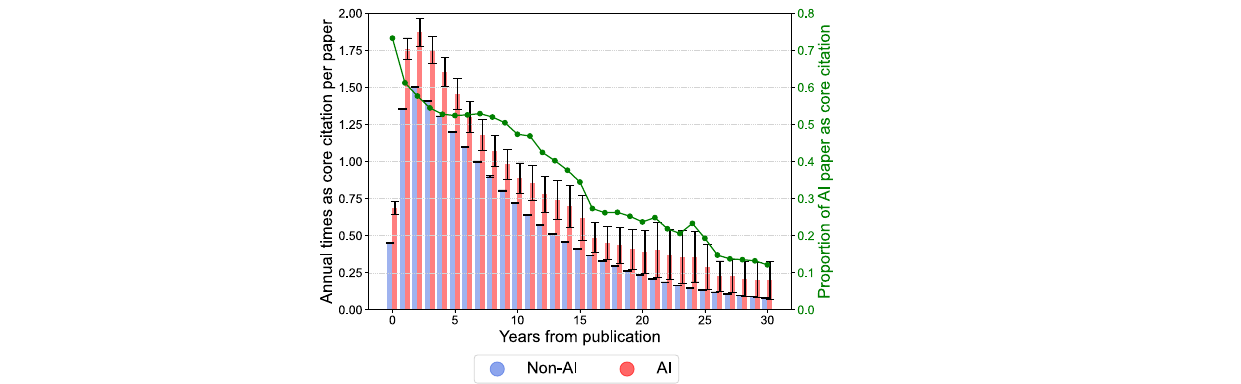}
\caption{
\textbf{Distinguishing between core and superficial citations.}
Results show that the proportion of AI papers cited as core citations tends to decrease over time (green line), while foundational influence can still be observed in decades-old papers ($n=27,405,011$).
In terms of the absolute count of instances being among the core citations in future papers, AI papers still outperform non-AI papers (red and blue bars), reflecting their enhanced academic impact. 
99\% CIs are shown as error bars centred at the mean.
}
\label{figsR1S8}
\end{figure}

\clearpage
\newpage
\begin{figure}[ht]
\centering
\includegraphics[width=\textwidth]{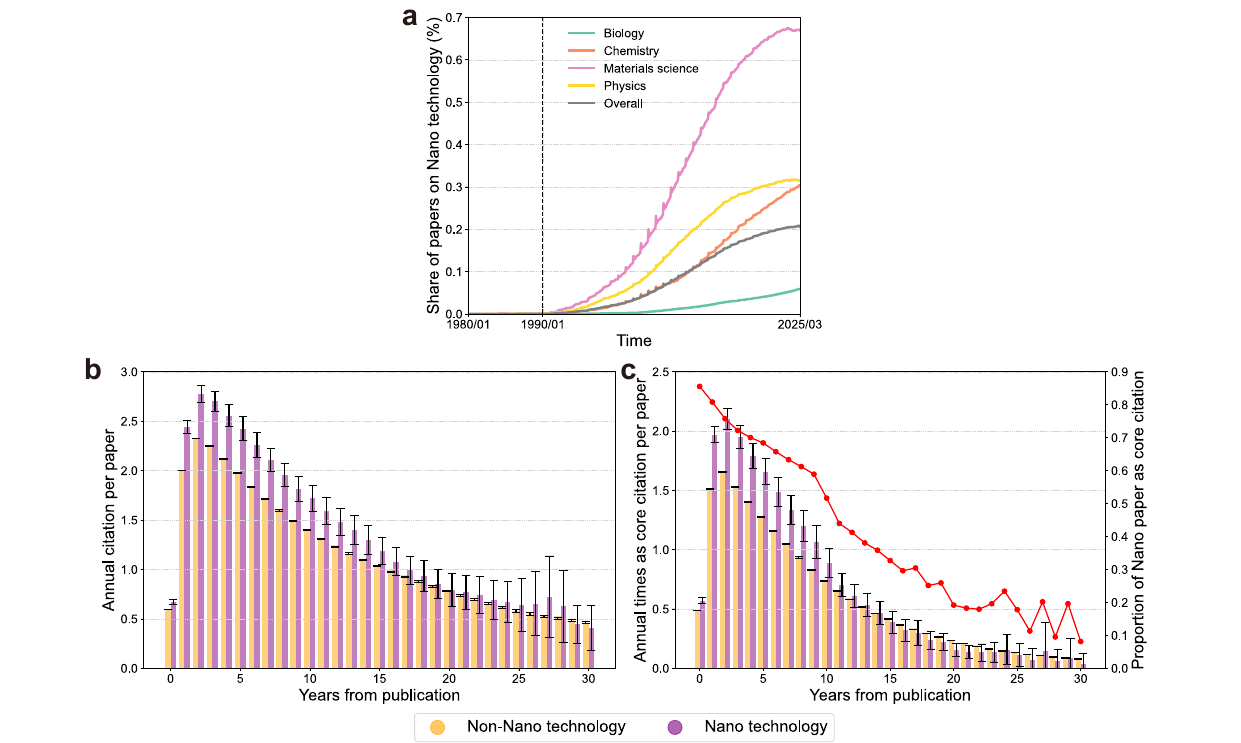}
\caption{
\textbf{Comparison between citation patterns of Nanotechnology and non-Nanotechnology papers.}
\textbf{(a)} Growth of Nanotechnology over time in different disciplines.
\textbf{(b)} Annual citations after publication of Nanotechnology and non-Nanotechnology papers ($n=17,976,303$).
\textbf{(c)} Annual times of being core citation following publication of Nanotechnology and non-Nanotechnology papers ($n=17,976,303$).
Nanotechnology papers are still cited as core citations many years after publication, with the proportion of being among the core citations declining over time.
In contrast to AI, both total number of citations and the frequency of being core citations for Nanotechnology papers eventually decline to levels indistinguishable from non-Nanotechnology papers.
For panels (b) and (c), 99\% CIs are shown as error bars centred at the mean.
}
\label{figsR1S9}
\end{figure}

\clearpage
\newpage
\begin{figure}[ht]
\centering
\includegraphics[width=\textwidth]{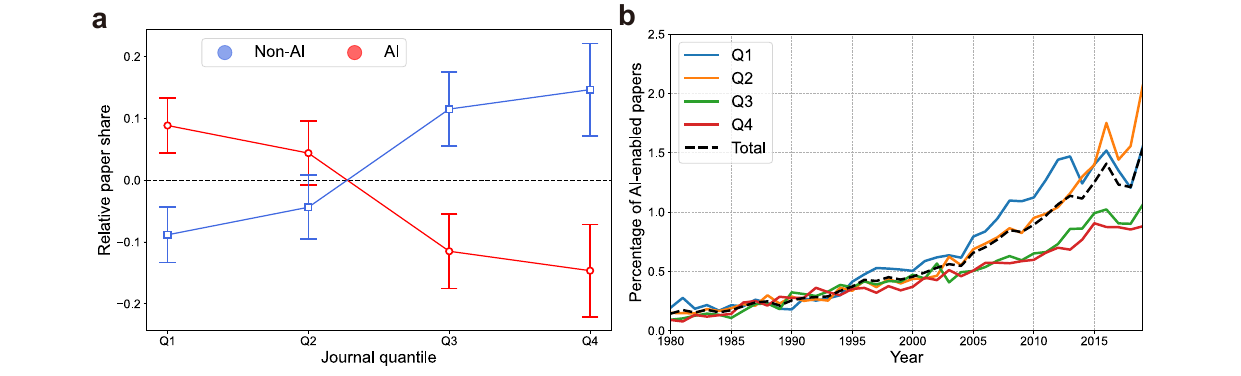}
\caption{
\textbf{Distribution of AI papers across journals of varying Journal Citation Report (JCR) quantiles.}
\textbf{(a)}
Comparison of the relative share of AI and non-AI papers published in different journals ($n=11,098$), where 99\% CIs are shown as error bars centred at the mean.
\textbf{(b)}
The change in the percentage of AI papers in journals with different JCR quantiles. The percentages of AI papers rise in all journals, and the percentages of AI papers in Q1 (blue) and Q2 (orange) journals are higher than the total percentage in all journals (black).
These results indicate that AI-augmented papers are more likely to be published in high-impact journals (Q1 and Q2) than papers without AI.
These statistics are obtained based on papers published before 2021, comparable with the 2021 JCR quantile data we used. 
}
\label{figs2}
\end{figure}

\clearpage
\newpage
\begin{figure}[ht]
\centering
\includegraphics[width=\textwidth]{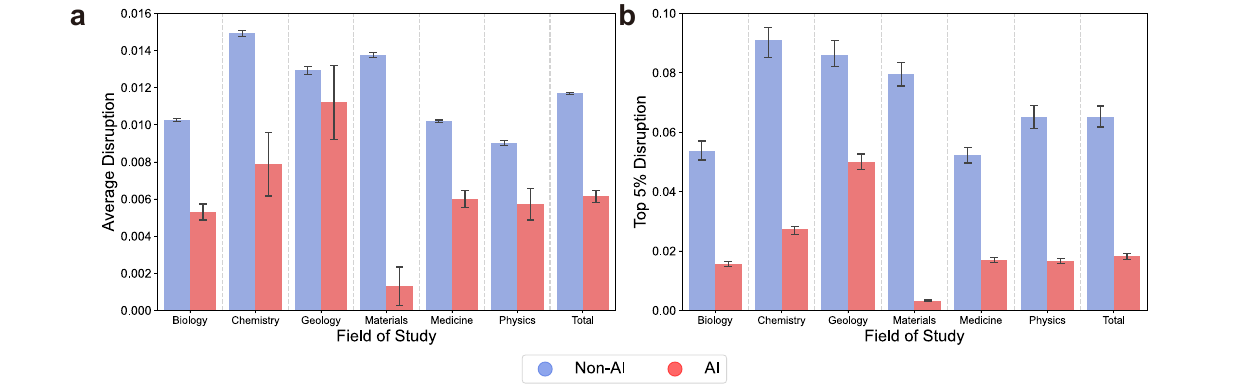}
\caption{
\textbf{Comparison of disruption between AI and non-AI papers.}
\textbf{(a)} Average disruption of AI and non-AI papers in each field ($p<0.001, n=23,199,583$).
\textbf{(b)} Top 5th-percentile disruption of AI and non-AI papers in each field ($p<0.001, n=23,199,583$).
99\% CIs are shown as error bars centred at the mean or the 5\% percentile.
All statistical tests use a two-sided t-test.
}
\label{figsR1S7}
\end{figure}

\clearpage
\newpage
\begin{figure}[ht]
\centering
\includegraphics[width=\textwidth]{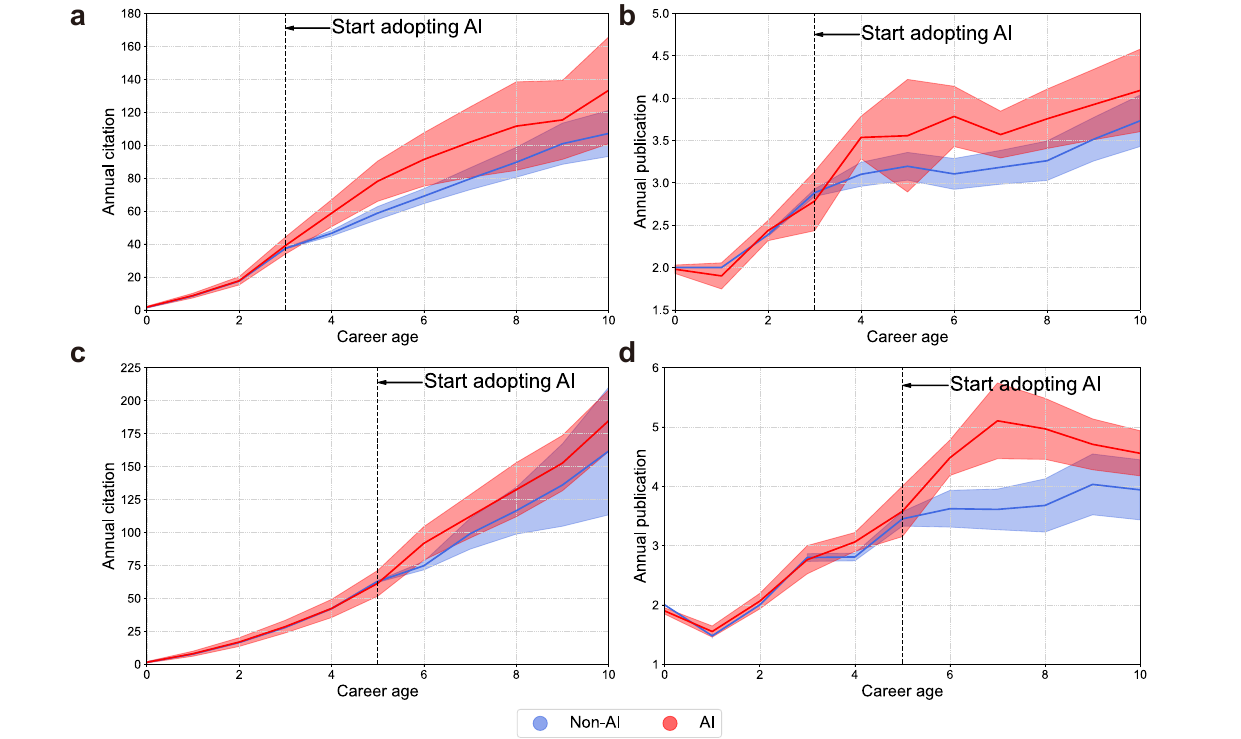}
\caption{
\textbf{Match-and-comparison analysis on the role of AI in driving the advantages of scientific in productivity and scholarly impact.}
\textbf{(a)} Comparison between scientists who began adopting AI in the third year of their careers and matched counterparts who exhibited comparable annual citation counts during their first three years but never adopted AI.
\textbf{(b)} Comparison between scientists who began adopting AI in the third year of their careers and matched counterparts who exhibited comparable annual productivity during their first three years but never adopted AI.
\textbf{(c)} Comparison between scientists who began adopting AI in the fifth year of their careers and matched counterparts who exhibited comparable annual citation counts during their first three years but never adopted AI.
\textbf{(d)} Comparison between scientists who began adopting AI in the fifth year of their careers and matched counterparts who exhibited comparable annual productivity during their first three years but never adopted AI.
The results suggest that for scientists with comparable early-career positions, adopting AI itself contributes to their subsequent advantages in productivity and scholarly impact.
For all panels, 99\% CIs are shown as error bands centred at the mean.
}
\label{figsR1S11}
\end{figure}

\clearpage
\newpage
\begin{figure}[ht]
\centering
\includegraphics[width=\textwidth]{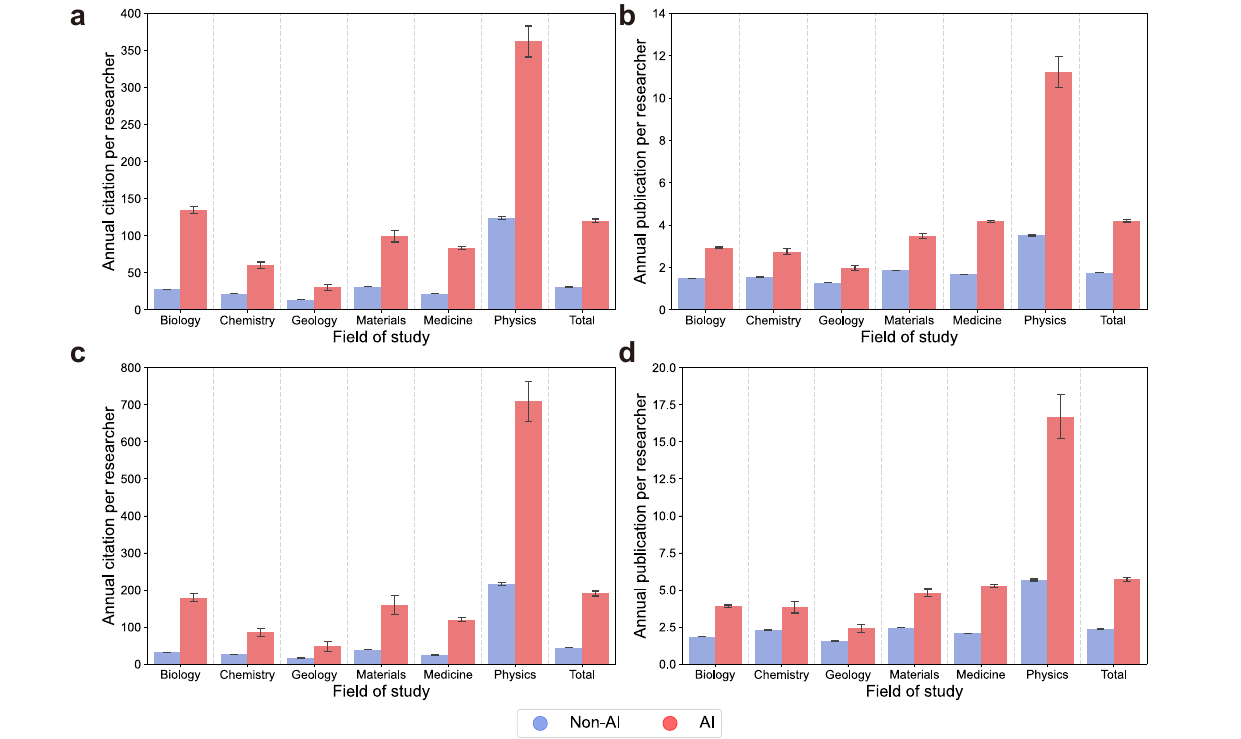}
\caption{
\textbf{The impact of AI on the productivity and citation of researchers with multi-year continuous publication records.}
\textbf{(a)} Comparison of annual citations between researchers adopting AI and their counterparts without AI among those with at least five consecutive years of publications ($p<0.001, n=1,495,265$).
\textbf{(b)} Comparison of annual publications between researchers adopting AI and their counterparts without AI among those with at least five consecutive years of publications ($p<0.001, n=1,495,265$).
\textbf{(c)} Comparison of annual citations between researchers adopting AI and their counterparts without AI among those with at least ten consecutive years of publications ($p<0.001, n=525,716$).
\textbf{(d)} Comparison of annual publications between researchers adopting AI and their counterparts without AI among those with at least ten consecutive years of publications ($p<0.001, n=525,716$).
These results confirm the positive impact of AI adoption on the career progression of continuously publishing core scientists.
For all panels, 99\% CIs are shown as error bars centred at the mean.
All statistical tests use a two-sided t-test.
}
\label{figsR1S6}
\end{figure}

\clearpage
\newpage
\begin{figure}[ht]
\centering
\includegraphics[width=\textwidth]{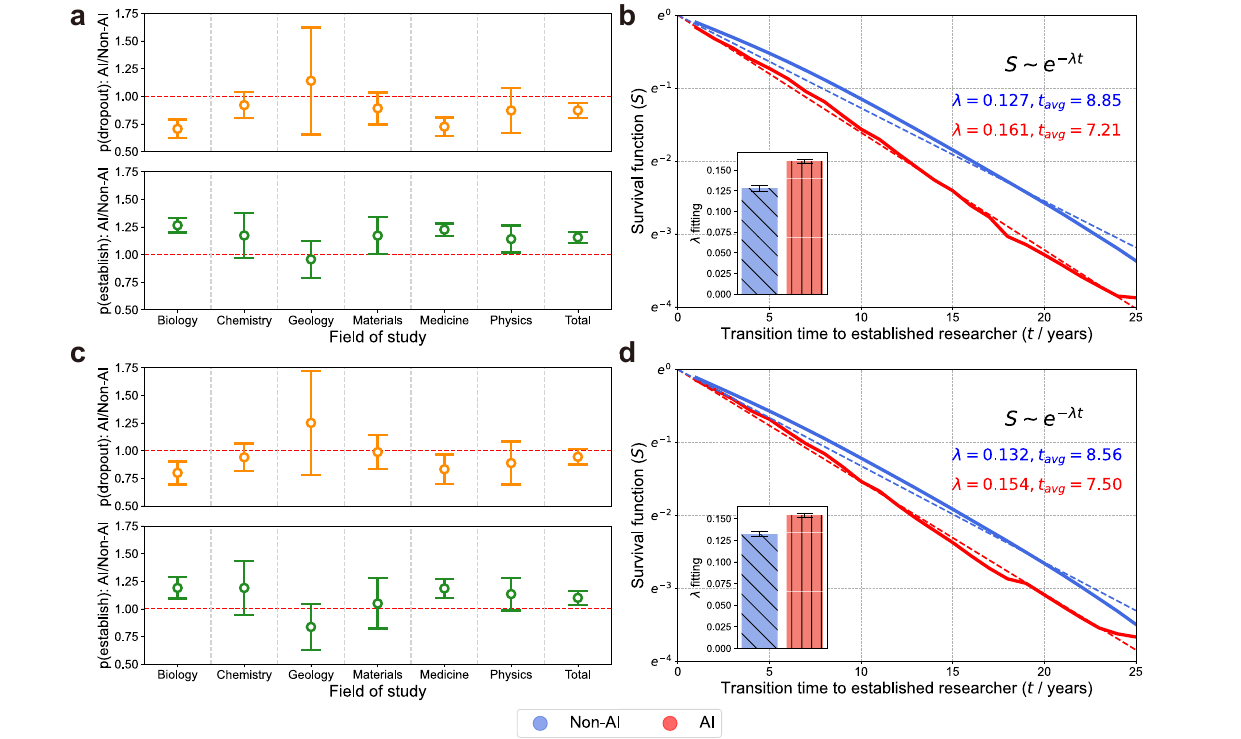}
\caption{
\textbf{Impact of AI adoption on researchers’ careers with different thresholds in detecting the dropout of researchers.}
\textbf{(a) (c)} The probability of two role transitions between junior scientists adopting AI and their counterparts without AI ($n=46$ year observations for each field).
\textbf{(b) (d)} Survival functions for the transition from junior to established researcher (b, $p<0.001, n=2,858,901$; d, $p<0.001, n=1,947,315$).
In (a) and (b), the threshold for detecting researcher dropout is 2 years.
In (c) and (d), the threshold for detecting researcher dropout is 4 years.
Results are consistent for both shorter and longer thresholds.
For all panels, 99\% CIs are shown as error bars centred at the mean.
All statistical tests use a two-sided t-test.
}
\label{figsR1S2b}
\end{figure}

\clearpage
\newpage
\begin{figure}[ht]
\centering
\includegraphics[width=\textwidth]{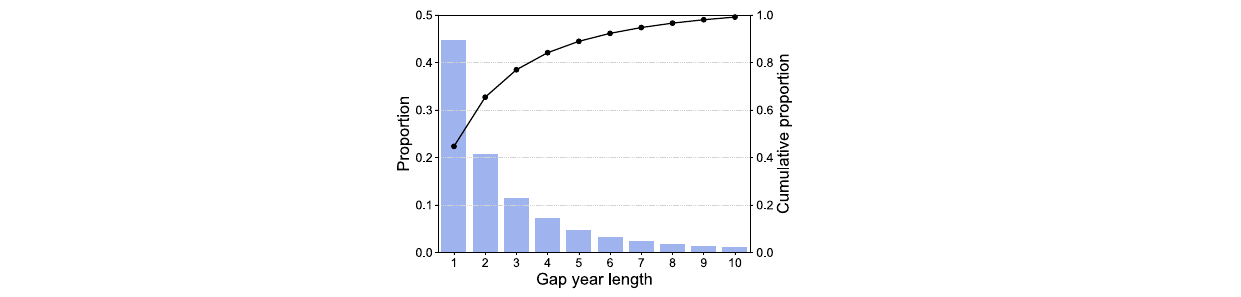}
\caption{
\textbf{The distribution of gap year durations in researcher’s careers.}
A period of gap years refers to a temporary interruption in a researcher’s publication activity, followed by a subsequent resumption of publishing. Results show that 44.67\% of all gap periods lasted only one year, and 76.94\% lasted no more than three years.
}
\label{figsR1S2a}
\end{figure}

\clearpage
\newpage
\begin{figure}[ht]
\centering
\includegraphics[width=\textwidth]{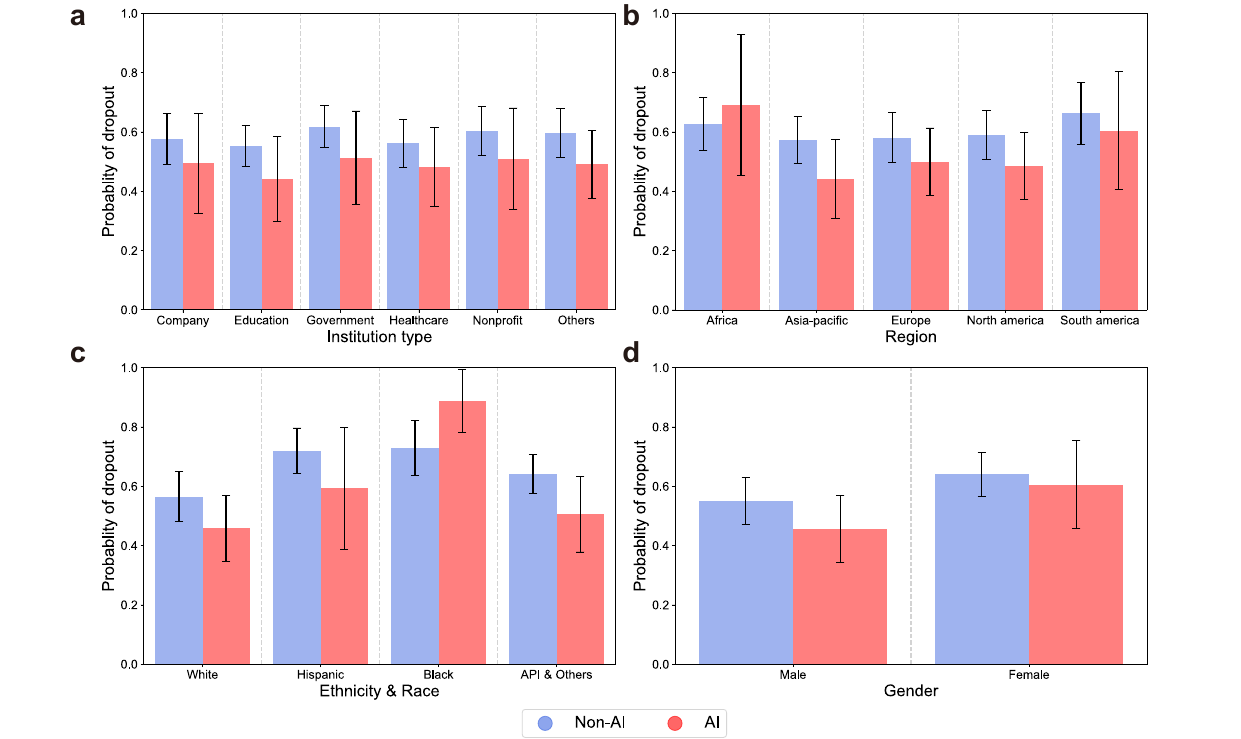}
\caption{
\textbf{Impact of AI adoption on research careers across different demographic groups.}
\textbf{(a)} Researchers affiliated with institutions of different types ($n=2,140,845$).
\textbf{(b)} Researchers affiliated with institutions within different geographic regions ($n=2,140,845$).
\textbf{(c)} Researchers of different ethnicity and race ($n=1,438,544$).
\textbf{(d)} Researchers of different genders ($n=502,336$).
Results show the heterogeneous impact of AI on scientific careers across different demographic groups. 
For all panels, 99\% CIs are shown as error bars centred at the mean.
}
\label{figsR1S10}
\end{figure}

\clearpage
\newpage
\begin{figure}[ht]
\centering
\includegraphics[width=\textwidth]{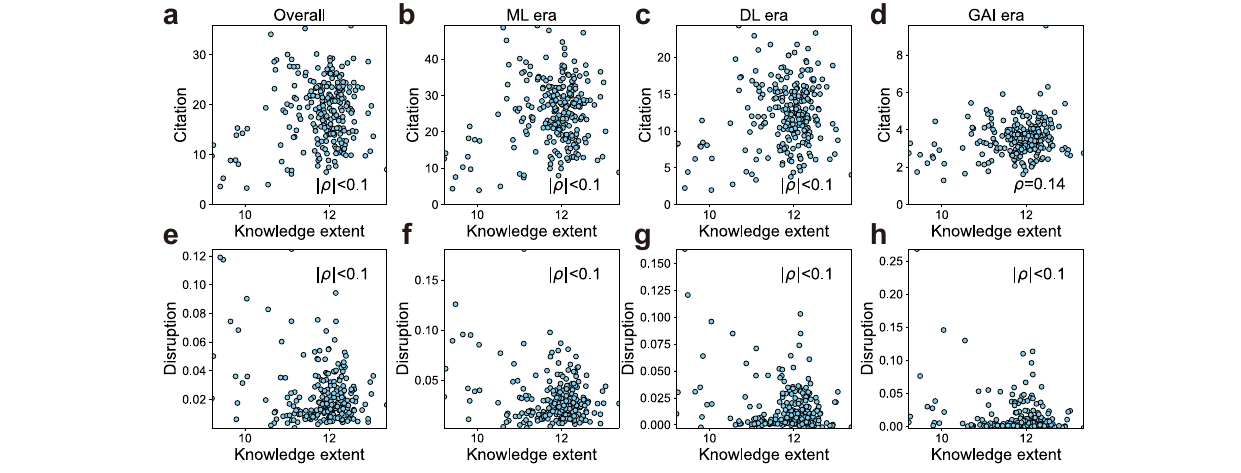}
\caption{
\textbf{Correlation analysis between knowledge extent and citation or disruption of sub-fields.}
For all panels, each scatter represents a sub-field, and the corresponding Pearson’s $r$ is indicated correspondingly.
Results show that, across sub-fields and eras of AI development, neither citation impact (as a proxy for impact) nor disruption (as a proxy for novelty) is obviously correlated with the sub-field’s knowledge extent.
}
\label{figsR1S1}
\end{figure}

\clearpage
\newpage
\begin{figure}[ht]
\centering
\includegraphics[width=\textwidth]{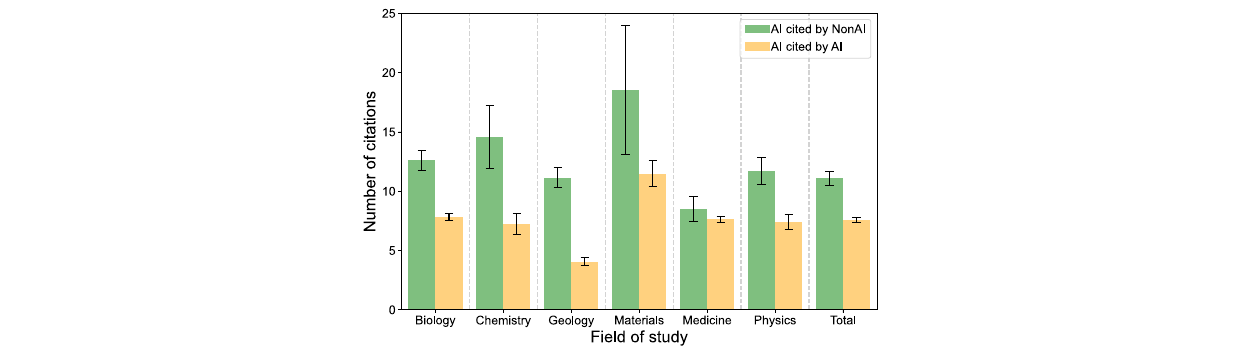}
\caption{
\textbf{Selectivity of AI adoption in different topics regarding topicality itself.}
The citation relationships between AI and non-AI papers show that AI research across various fields is more frequently cited by non-AI studies than by AI studies themselves ($p<0.001, n=27,405,011$).
This suggests that topics in AI research influence non-AI research rather than forming isolated, self-referential clusters of AI literature.
There tends to be no obvious difference in topic selection between AI and non-AI research, indicating that inherent topicality does not account for the observed more narrow focus of AI-augmented research.
99\% CIs are shown as error bars centred at the mean, and the statistical tests use a two-sided t-test.
}
\label{figsR1S15a}
\end{figure}

\clearpage
\newpage
\begin{figure}[ht]
\centering
\includegraphics[width=\textwidth]{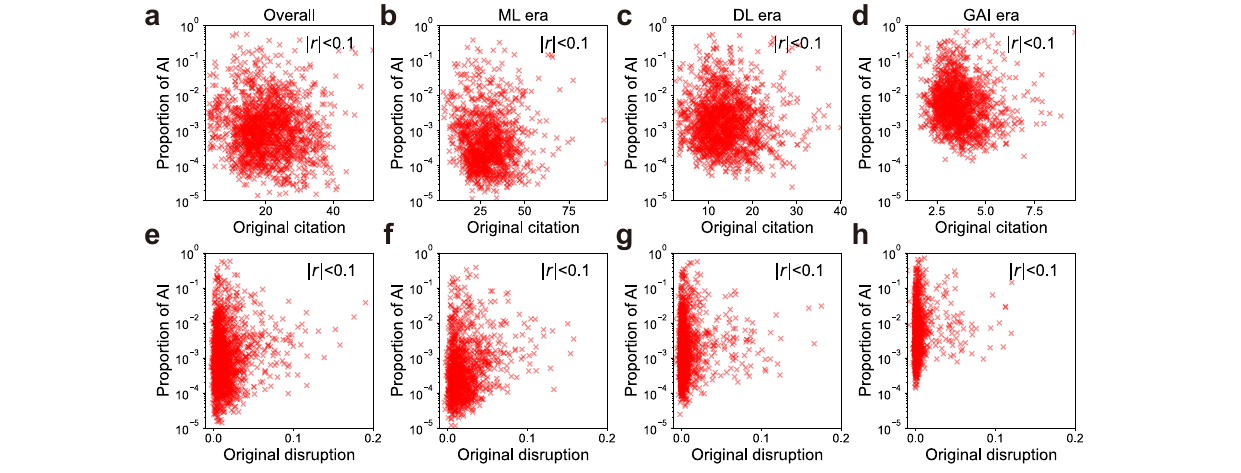}
\caption{
\textbf{Selectivity of AI adoption in different topics associated with the topics’ original impact.}
In all panels, each scatter shows the degree of AI penetration in a topic and the original citation count or disruption score within that topic.
The corresponding Pearson’s $r$ is indicated in each panel.
Results show that across topics and AI eras, neither the original citation nor the original disruption is obviously correlated with the proportion of AI adoption, suggesting that AI adoption does not selectively favor topics based on differential original impact.
}
\label{figsR1S15b}
\end{figure}

\clearpage
\newpage
\begin{figure}[ht]
\centering
\includegraphics[width=\textwidth]{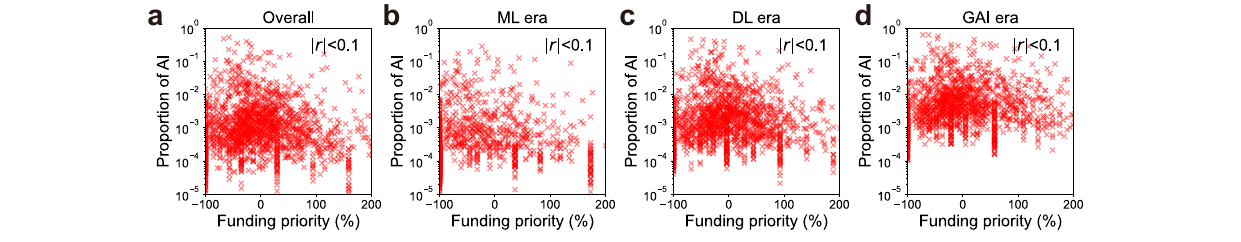}
\caption{
\textbf{Selectivity of AI adoption in different topics regarding funding priority across the topics.}
In all panels, each scatter shows the degree of AI penetration in a topic and funding priority to that topic.
The corresponding Pearson’s $r$ is indicated correspondingly in each panel. 
Results show that across topics and AI eras, there are heterogeneous funding priorities for different topics, but funding priority is not obviously correlated with the proportion of AI adoption, suggesting that AI adoption does not selectively favor topics prioritized by funding agencies.
}
\label{figsR1S15c}
\end{figure}

\clearpage
\newpage
\begin{figure}[ht]
\centering
\includegraphics[width=\textwidth]{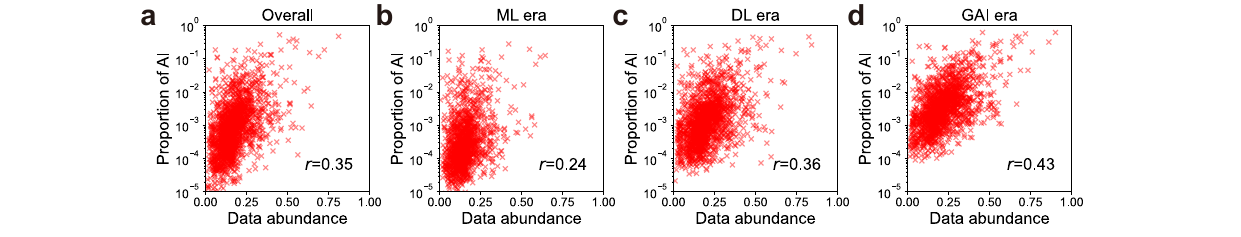}
\caption{
\textbf{Selectivity of AI adoption in different topics regarding data abundance in the topics.}
In all panels, each scatter shows the degree of AI penetration in a topic and the data abundance in that topic.
The corresponding Pearson’s $r$ is indicated correspondingly in each panel. The results show that across topics and AI eras, the data abundance is positively correlated with the proportion of AI adoption, suggesting that AI is more likely to be adopted in topics with abundant data resources.
Notably, the correlation of AI-use with frequency of mentioned data resources rises with the size and intensity of AI-models.
}
\label{figsR1S15d}
\end{figure}

\clearpage
\newpage
\begin{figure}[ht]
\centering
\includegraphics[width=\textwidth]{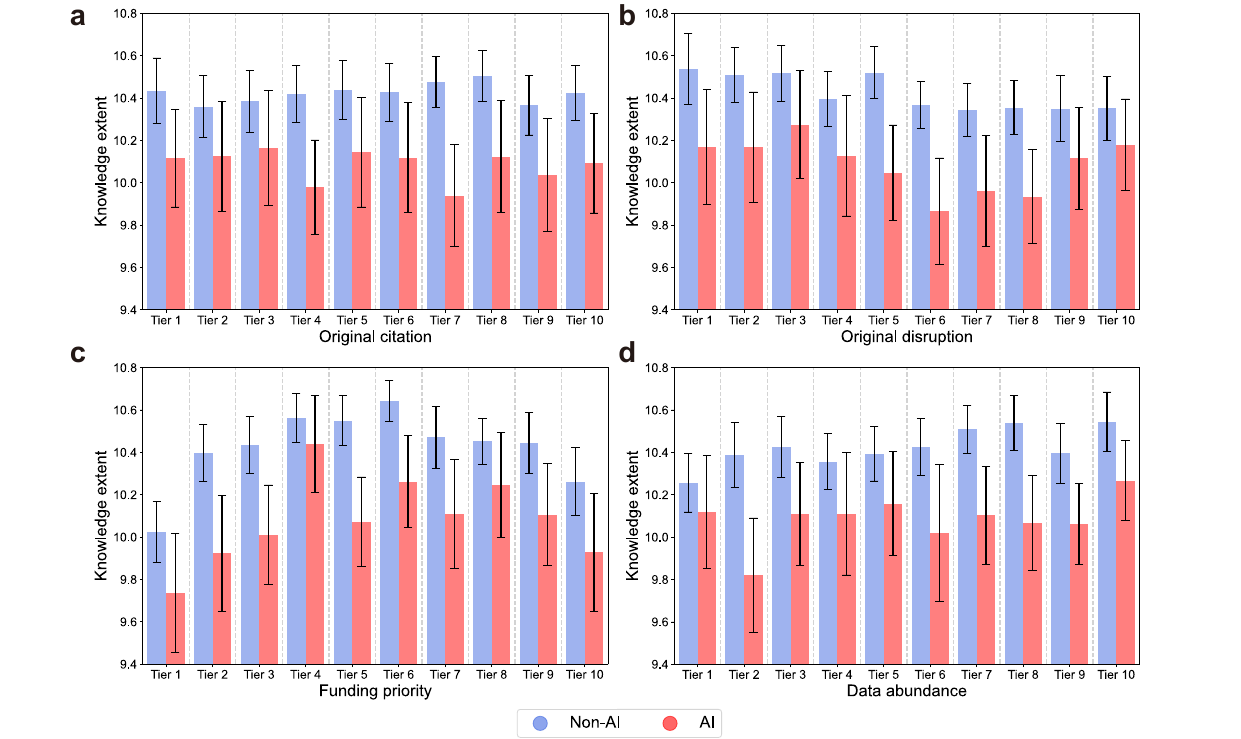}
\caption{
\textbf{Stratified analyses controlling for the effects of potential confounding variables on the contracted knowledge extent of AI-augmented research.}
The results show that when dividing each external factor into 10 tiers based on magnitude, the contraction in knowledge extent for AI-augmented research remains evident within each tier ($p<0.001, n=17,529,094$).
Therefore, regardless of whether a topic is more or less likely to be selected for AI adoption, existing AI-augmented research within the topic consistently covers a contracted knowledge space compared to non-AI research.
For all panels, 99\% CIs are shown as error bars centred at the mean.
All statistical tests use a two-sided t-test.
}
\label{figsR1S16}
\end{figure}

\clearpage
\newpage
\begin{figure}[ht]
\centering
\includegraphics[width=\textwidth]{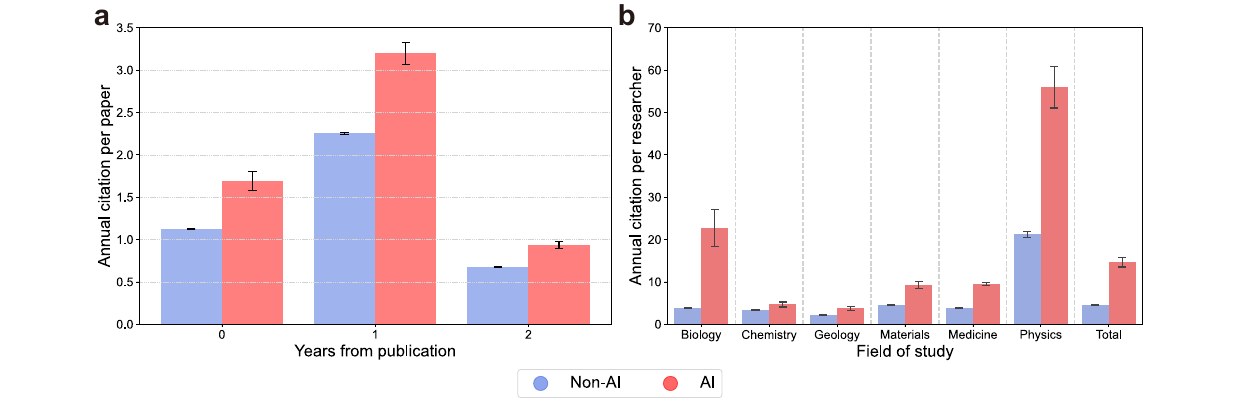}
\caption{
\textbf{Replication of “AI enlarges the impact of papers” with only the subset of papers published during the GAI era.}
\textbf{(a)} Annual citations after publication of AI and non-AI papers ($n=1,888,694$).
\textbf{(b)} Average annual citations of researchers adopting AI and their counterparts without AI ($p<0.001, n=2,888,737$).
For all panels, 99\% CIs are shown as error bars centred at the mean.
All statistical tests use a two-sided t-test.
Results in the GAI era are consistent with those obtained over the full time span featured in the main text.
}
\label{figLLMera2}
\end{figure}

\clearpage
\newpage
\begin{figure}[ht]
\centering
\includegraphics[width=\textwidth]{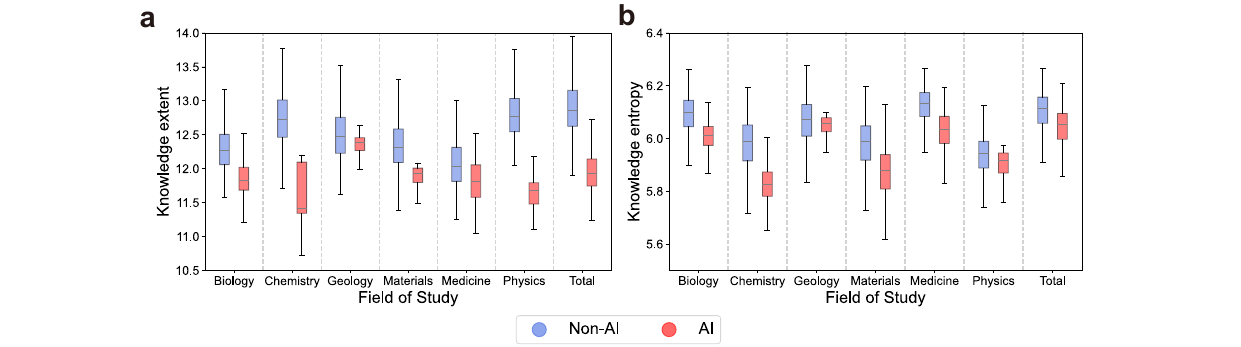}
\caption{
\textbf{Replication of “AI usage is associated with a contraction in knowledge extent within and across scientific fields” with only the subset of papers published during the GAI era.}
\textbf{(a)} Knowledge extent of AI and non-AI papers in each field ($n=1,000$ samples in each field).
\textbf{(b)} Knowledge entropy of AI and non-AI papers in each field ($n=1,000$ samples in each field).
Boxplots are centred at the median and bounded at the first and third quartile (Q1 and Q3), with 1.5 times of the inter-quartile range (IQR) shown as whiskers from the box.
Results in the GAI era are consistent with those obtained over the full time span featured in the main text.
}
\label{figLLMera3}
\end{figure}

\clearpage
\newpage
\begin{figure}[ht]
\centering
\includegraphics[width=\textwidth]{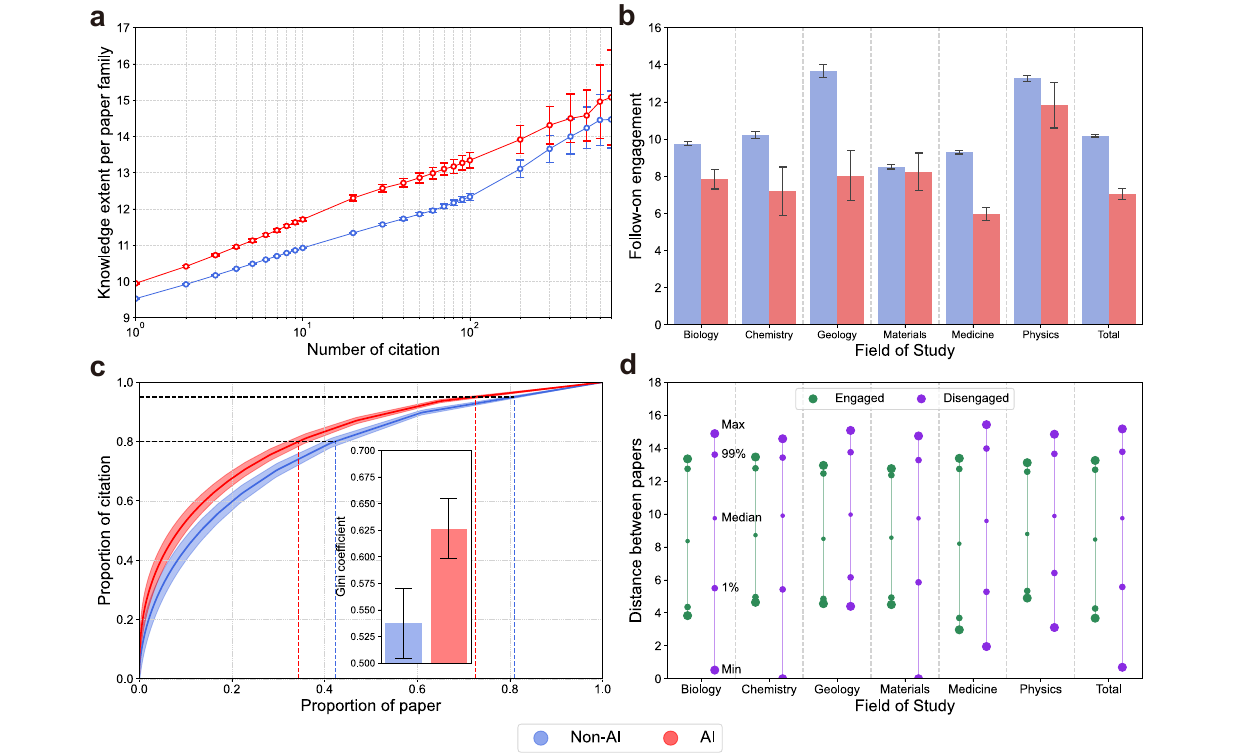}
\caption{
\textbf{Replication of “reduced follow-on engagement and more overlapped works in AI research” with only the subset of papers published during the GAI era.}
\textbf{(a)} Knowledge extent of individual AI and non-AI paper families ($n=1,888,694$).
\textbf{(b)} Engagement among papers that cite AI vs. non-AI papers ($p<0.001, n=518,156$).
\textbf{(c)} Distribution of citations to AI vs. non-AI papers ($p<0.001, n=100$ sampled paper groups).
\textbf{(d)} Distribution of distances between paper pairs that cite the same papers, with or without citing each other—engaged versus disengaged ($n=258,529$ sampled paper pairs).
For all panels, 99\% CIs are shown as error bars or error bands centred at the mean.
All statistical tests use a two-sided t-test.
Results in the GAI era are consistent with those on the full time span featured in the main text.
}
\label{figLLMera4}
\end{figure}

\clearpage
\newpage
\begin{figure}[ht]
\centering
\includegraphics[width=\textwidth]{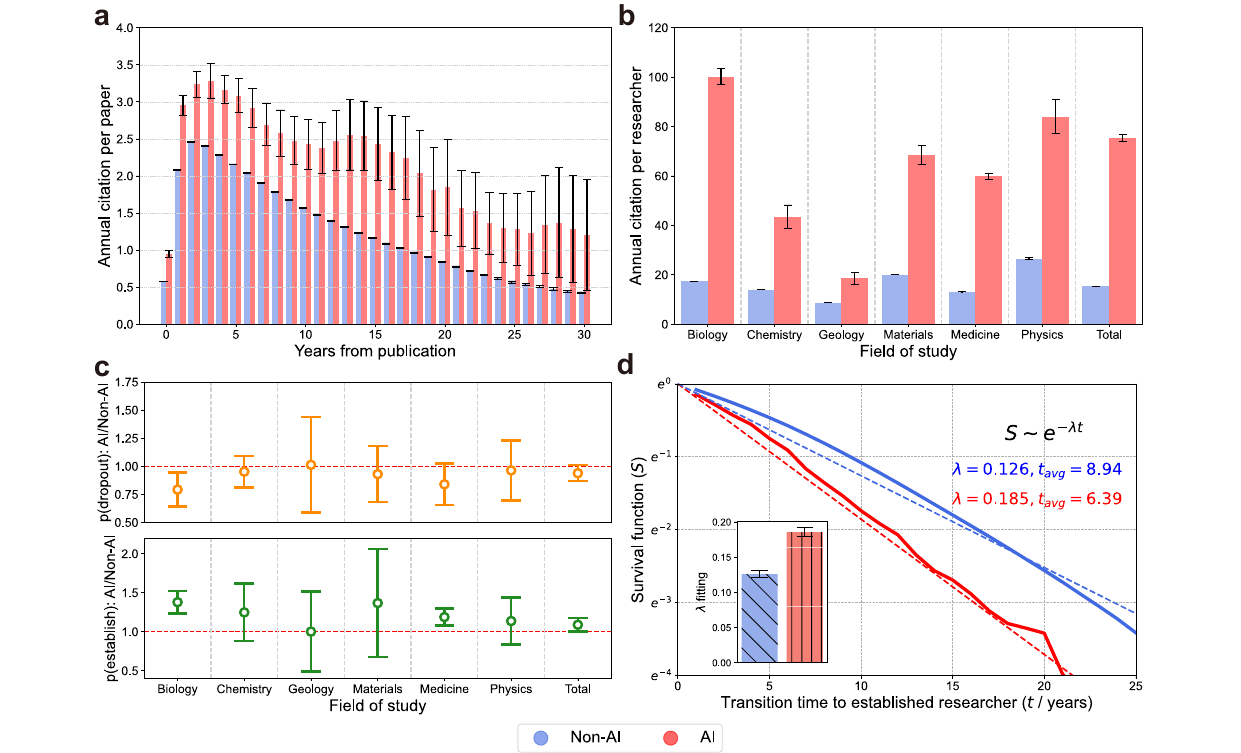}
\caption{
\textbf{Replication of “AI enlarges the impact of papers and enhances the career of researchers” on the WOS dataset.}
\textbf{(a)} Annual citations after publication of AI and non-AI papers ($n=16,706,988$).
\textbf{(b)} Average annual citations of researchers adopting AI and their counterparts without AI ($p<0.001, n=3,620,795$).
\textbf{(c)} The probability of two role transitions between junior scientists adopting AI and their counterparts without AI ($n=46$ year observations for each field).
\textbf{(d)} Survival functions for the transition from junior to established researcher ($p<0.001, n=1,556,338$).
For all panels, 99\% CIs are shown as error bars centred at the mean.
All statistical tests use a two-sided t-test.
Results on the WOS dataset are consistent with those from the OpenAlex dataset featured in the main text.
}
\label{figWoS2}
\end{figure}

\clearpage
\newpage
\begin{figure}[ht]
\centering
\includegraphics[width=\textwidth]{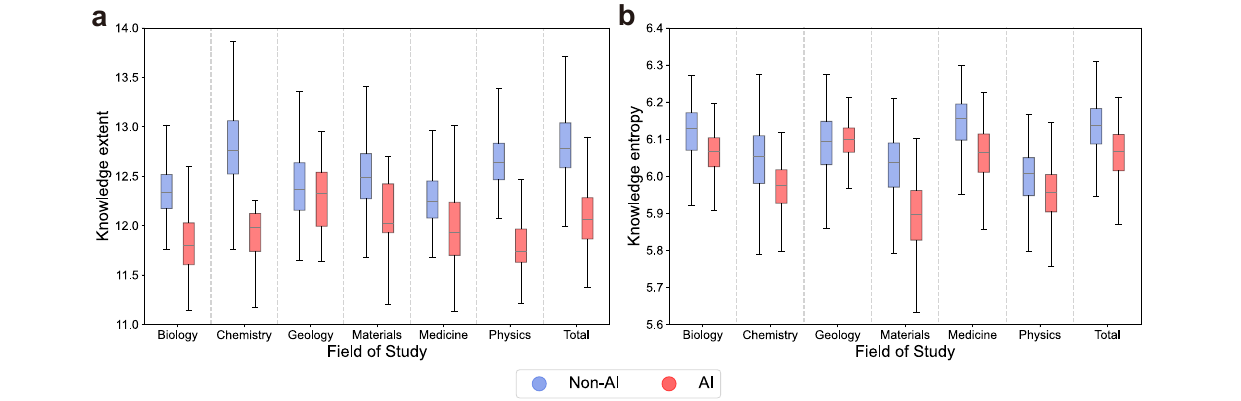}
\caption{
\textbf{Replication of “AI usage is associated with a contraction in knowledge extent within and across scientific fields” on the WOS dataset.}
\textbf{(a)} Knowledge extent of AI and non-AI papers in each field ($n=1,000$ samples in each field).
\textbf{(b)} Knowledge entropy of AI and non-AI papers in each field ($n=1,000$ samples in each field).
Boxplots are centred at the median and bounded at the first and third quartile (Q1 and Q3), with 1.5 times of the inter-quartile range (IQR) shown as whiskers from the box.
Results on the WOS dataset are consistent with the ones on OpenAlex featured in the main text.
}
\label{figWoS3}
\end{figure}

\clearpage
\newpage
\begin{figure}[ht]
\centering
\includegraphics[width=\textwidth]{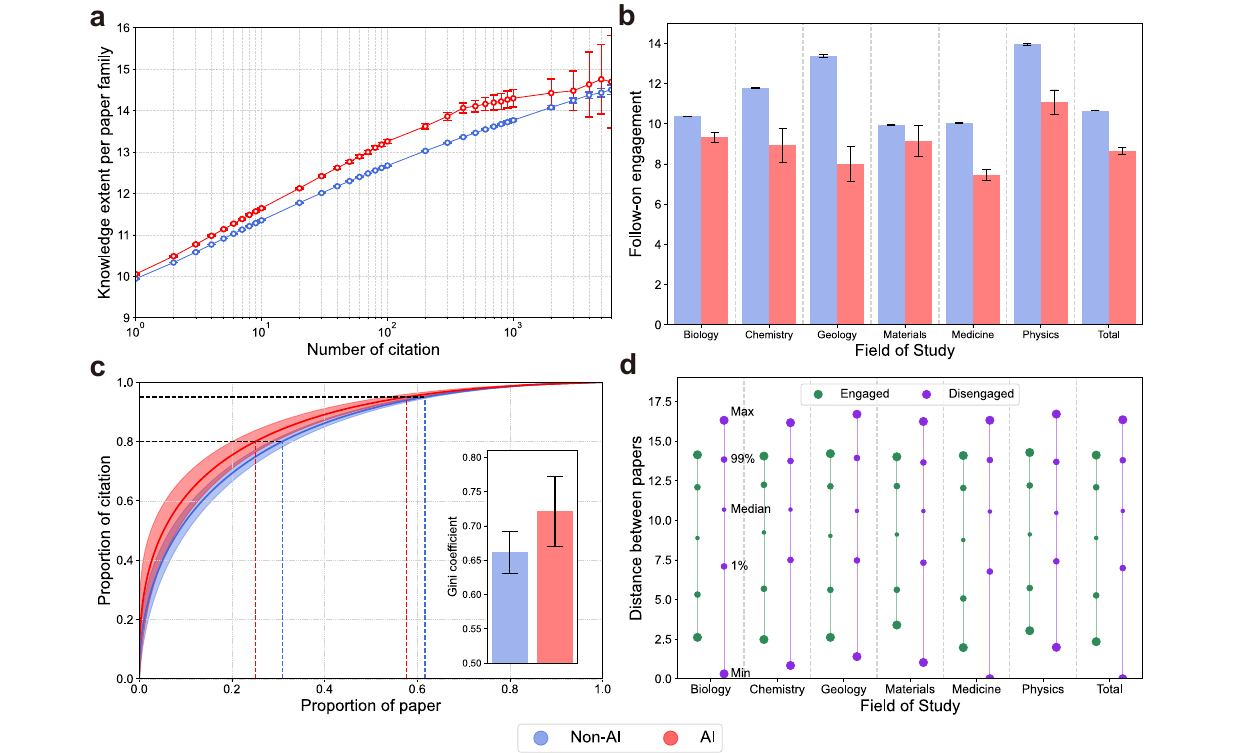}
\caption{
\textbf{Replication of “reduced follow-on engagement and more overlapped works in AI research” on the WOS dataset.}
\textbf{(a)} Knowledge extent of individual AI and non-AI paper families ($n=16,706,988$).
\textbf{(b)} Engagement among papers that cite AI vs. non-AI papers ($p<0.001, n=15,048,254$).
\textbf{(c)} The distribution of citations to AI vs. non-AI papers ($p<0.001, n=100$ sampled paper groups).
\textbf{(d)} The distribution of distances between paper pairs that cite the same papers, with or without citing each other—engaged versus disengaged ($n=674,132,352$ sampled paper pairs).
For all panels, 99\% CIs are shown as error bars or error bands centred at the mean.
All statistical tests use a two-sided t-test.
Results from the WOS dataset are consistent with those from the OpenAlex dataset featured in the main text.
}
\label{figWoS4}
\end{figure}

\clearpage
\newpage
\begin{figure}[ht]
\centering
\includegraphics[width=\textwidth]{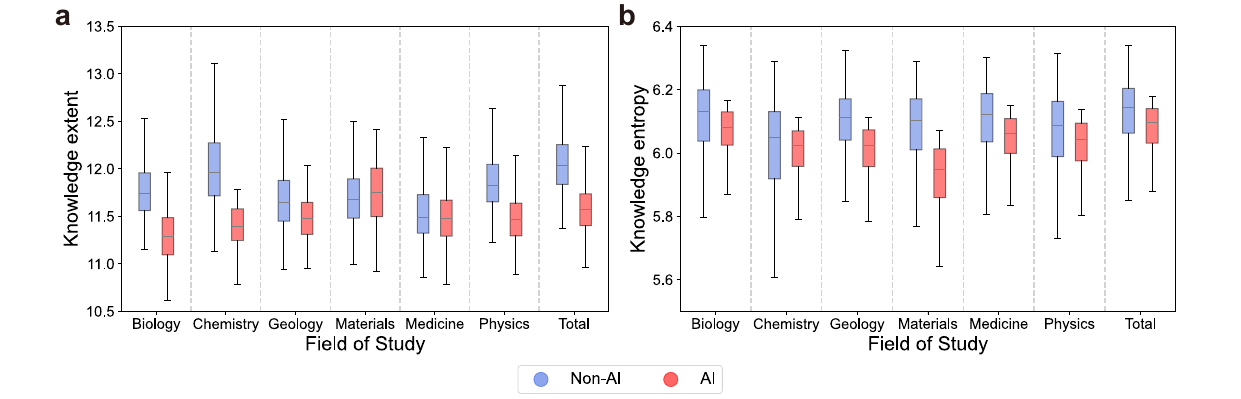}
\caption{
\textbf{Replication of “AI usage is associated with a contraction in knowledge extent within and across scientific fields” in an embedding space after dimensional reduction using PCA.}
\textbf{(a)} Knowledge extent of AI and non-AI papers in each field ($n=1,000$ samples in each field).
\textbf{(b)} Knowledge entropy of AI and non-AI papers in each field ($n=1,000$ samples in each field).
Boxplots are centred at the median and bounded at the first and third quartile (Q1 and Q3), with 1.5 times of the inter-quartile range (IQR) shown as whiskers from the box.
Results in the reduced-dimensional space are consistent with those from the original embedding space reported in the main text.
}
\label{figsR2S1a}
\end{figure}

\clearpage
\newpage
\begin{figure}[ht]
\centering
\includegraphics[width=\textwidth]{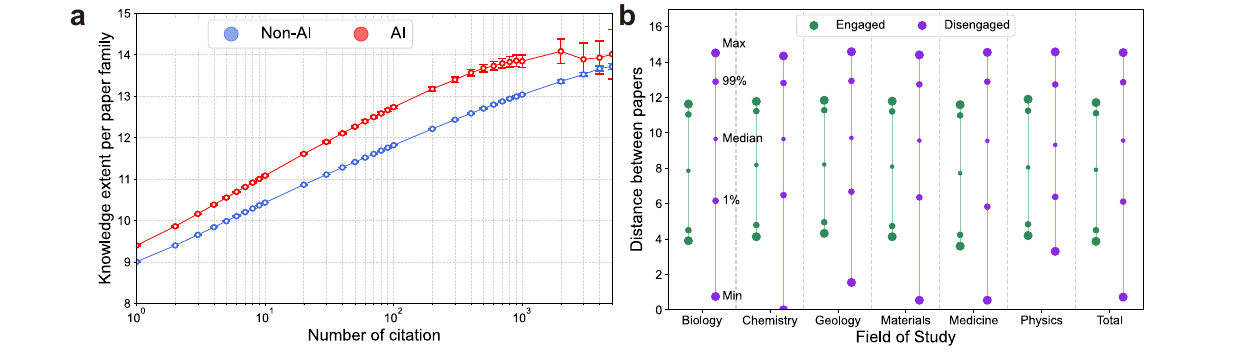}
\caption{
\textbf{Replication of “reduced follow-on engagement and more overlapped works in AI research” in an embedding space after dimensional reduction using PCA.}
\textbf{(a)} Knowledge extent of individual AI and non-AI paper families ($n=27,405,011$).
99\% CIs are shown as error bars centred at the mean.
\textbf{(b)} The distribution of distances between paper pairs that cite the same papers, with or without citing each other—engaged versus disengaged ($n=590,325,130$ sampled paper pairs).
Results in the reduced-dimensional space are consistent with those from the original embedding space reported in the main text.
}
\label{figsR2S1b}
\end{figure}

\clearpage
\newpage
\begin{figure}[ht]
\centering
\includegraphics[width=\textwidth]{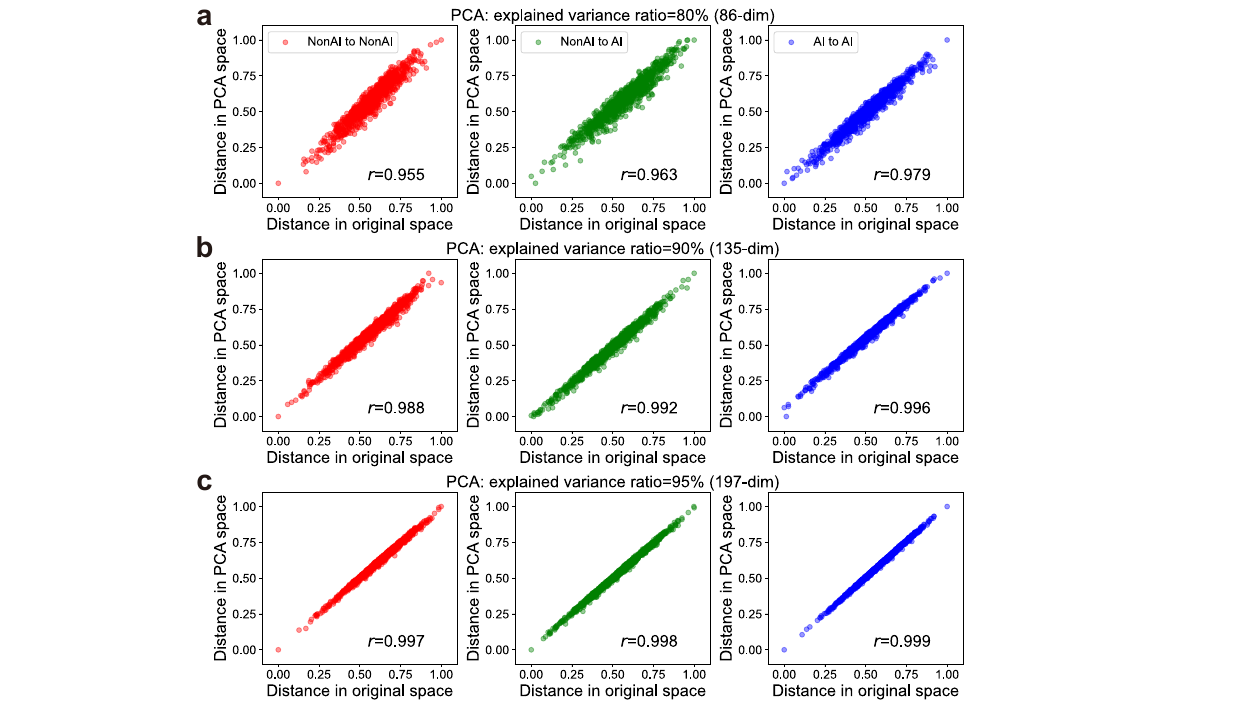}
\caption{
\textbf{General sensitivity check on the distance calculations in high-dimensional paper embedding space.}
We reduce the original embeddings to 86, 135, and 197 dimensions with PCA, which correspond to the principal components that explain \textbf{(a)} 80\%, \textbf{(b)} 90\%, and \textbf{(c)} 95\% of the total variance, respectively.
We then randomly sampled 1,000 NonAI-NonAI, 1,000 AI-AI, and 1,000 NonAI-AI paper pairs and calculated the distance for each pair within each of the reduced-dimensional spaces.
Results show that the distances in the various reduced-dimensional spaces are highly correlated with those in the original space, indicating the reliability of our paper embeddings for distance calculations.
}
\label{figsR2S2}
\end{figure}

\clearpage
\newpage
\begin{figure}[ht]
\centering
\includegraphics[width=\textwidth]{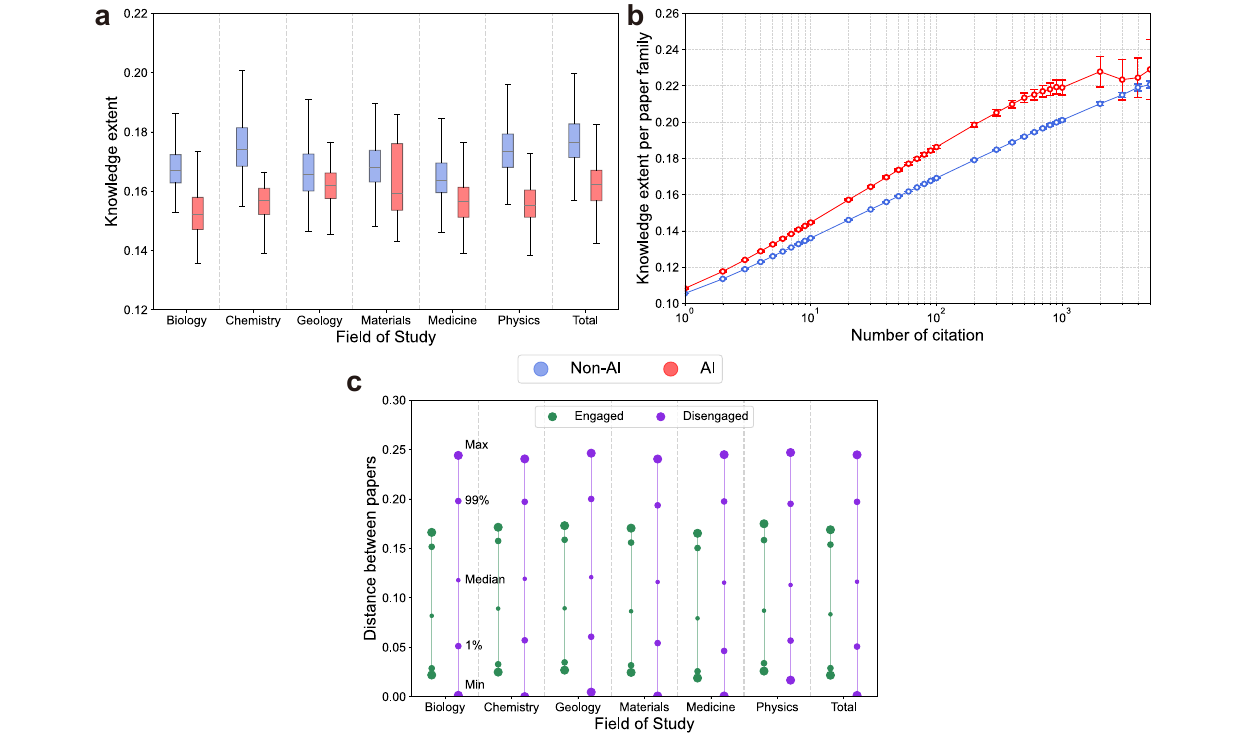}
\caption{
\textbf{Replication of results related to high-dimensional distance calculations with Cosine distance.}
\textbf{(a)} Knowledge extent of AI and non-AI papers in each field ($n=1,000$ samples in each field).
Boxplots are centred at the median and bounded at the first and third quartile (Q1 and Q3), with 1.5 times of the inter-quartile range (IQR) shown as whiskers from the box.
\textbf{(b)} Knowledge extent of individual AI and non-AI paper families ($n=27,405,011$).
99\% CIs are shown as error bars centred at the mean.
\textbf{(c)} The distribution of distances between paper pairs that cite the same papers, with or without citing each other-engaged versus disengaged ($n=590,325,130$ sampled paper pairs).
Results with Cosine distance are consistent with those with Euclidean distance reported in the main text.
}
\label{figsR2S5}
\end{figure}
\clearpage
\section*{Supplementary Tables}
\begin{table}[ht]
\caption{\textbf{Information of experts involved in scrutinizing our identification results}}
\centering
\begin{tabular}{@{}ccccc@{}}
\hline
ID & Majority      & Educational background & Number of publications & AI experience/y \\ \hline
1  & EECS          & Ph.D.                    & 31                     & 8                                 \\
2  & EECS          & Ph.D. student            & 4                      & 2                                 \\
3  & EECS          & Ph.D.                    & 7                      & 6                                 \\
4  & EECS          & Ph.D.                    & 11                     & 5                                 \\
5  & EECS          & Ph.D.                    & 21                     & 6                                 \\
6  & Physics, EECS & Ph.D.                    & 6                      & 3                                 \\
7  & EECS          & Ph.D.                    & 38                     & 7                                 \\
8  & EECS          & Ph.D.                    & 16                     & 5                                 \\
9  & EECS          & Ph.D.                    & 12                     & 6                                 \\
10 & Biology, EECS & Ph.D.                    & 15                     & 5                                 \\
11 & EECS          & Ph.D.                    & 40                     & 8                                 \\
12 & EECS          & Ph.D.                    & 21                     & 7                                 \\ \hline
\end{tabular}
\label{tab1}
\end{table}

\clearpage
\newpage
\begin{table}[ht]
\caption{\textbf{Number of papers sampled for expert annotation in identification accuracy evaluation}}
\centering
\begin{tabular}{@{}c|ccc|c@{}}
\toprule
Group               & ML era & DL era & GAI era & Overall \\ \midrule
Biology 1           & 59     & 33     & 18      & 110     \\
Biology 2           & 57     & 39     & 14      & 110     \\
Chemistry 1         & 60     & 32     & 18      & 110     \\
Chemistry 2         & 69     & 28     & 13      & 110     \\
Geology 1           & 69     & 25     & 16      & 110     \\
Geology 2           & 55     & 40     & 15      & 110     \\
Materials science 1 & 65     & 29     & 16      & 110     \\
Materials science 2 & 63     & 32     & 15      & 110     \\
Medicine 1          & 60     & 33     & 17      & 110     \\
Medicine 2          & 54     & 37     & 19      & 110     \\
Physics 1           & 66     & 29     & 15      & 110     \\
Physics 2           & 60     & 36     & 14      & 110     \\ \midrule
Overall             & 737    & 393    & 190     & 1320    \\ \bottomrule
\end{tabular}
\label{tab2}
\end{table}

\clearpage
\newpage
\begin{table}[ht]
\caption{\textbf{Fleiss’ Kappa of expert annotation results in identification accuracy evaluation}}
\centering
\begin{tabular}{@{}c|ccc|c@{}}
\toprule
Group               & ML era & DL era & GAI era & Overall \\ \midrule
Biology 1           & 0.955  & 0.943  & 0.926   & 0.950   \\
Biology 2           & 0.951  & 0.951  & 1.000   & 0.963   \\
Chemistry 1         & 1.000  & 0.952  & 0.924   & 0.976   \\
Chemistry 2         & 0.981  & 0.940  & 0.889   & 0.963   \\
Geology 1           & 0.921  & 0.941  & 1.000   & 0.939   \\
Geology 2           & 0.976  & 1.000  & 0.909   & 0.975   \\
Materials science 1 & 1.000  & 0.893  & 1.000   & 0.976   \\
Materials science 2 & 1.000  & 1.000  & 0.909   & 0.988   \\
Medicine 1          & 0.976  & 0.898  & 0.921   & 0.952   \\
Medicine 2          & 0.914  & 0.961  & 0.921   & 0.939   \\
Physics 1           & 0.960  & 1.000  & 0.909   & 0.964   \\
Physics 2           & 1.000  & 0.962  & 0.899   & 0.976   \\ \midrule
Overall             & 0.971  & 0.957  & 0.935   & 0.964   \\ \bottomrule
\end{tabular}
\label{tab3}
\end{table}

\begin{table}[ht]
\caption{\textbf{F1-score of BERT identification results against expert labels in identification accuracy evaluation}}
\centering
\begin{tabular}{@{}c|ccc|c@{}}
\toprule
Group               & ML era & DL era & GAI era & Overall \\ \midrule
Biology 1           & 0.923  & 0.920  & 0.889   & 0.917   \\
Biology 2           & 0.878  & 0.897  & 0.857   & 0.885   \\
Chemistry 1         & 0.809  & 0.857  & 0.900   & 0.844   \\
Chemistry 2         & 0.892  & 0.923  & 0.857   & 0.898   \\
Geology 1           & 0.800  & 0.812  & 0.941   & 0.826   \\
Geology 2           & 0.783  & 0.912  & 0.875   & 0.857   \\
Materials science 1 & 0.857  & 0.895  & 0.941   & 0.881   \\
Materials science 2 & 0.800  & 0.923  & 0.875   & 0.855   \\
Medicine 1          & 0.978  & 0.913  & 0.875   & 0.935   \\
Medicine 2          & 0.882  & 0.917  & 0.917   & 0.906   \\
Physics 1           & 0.825  & 0.833  & 0.875   & 0.835   \\
Physics 2           & 0.816  & 0.909  & 0.875   & 0.862   \\ \midrule
Overall             & 0.852  & 0.896  & 0.892   & 0.875   \\ \bottomrule
\end{tabular}
\label{tab4}
\end{table}

\clearpage
\newpage
\begin{sidewaystable}
\caption{The most frequently adopted AI methods in all selected disciplines in different AI development eras}
\centering
\small
%
\label{tab5}
\end{sidewaystable}

\clearpage
\newpage
\begin{sidewaystable}
\caption{The most frequently adopted AI methods in biology in different AI development eras}
\centering
\small
%
\label{tab6}
\end{sidewaystable}

\clearpage
\newpage
\begin{sidewaystable}
\caption{The most frequently adopted AI methods in chemistry in different AI development eras}
\centering
\small
%
\label{tab7}
\end{sidewaystable}

\clearpage
\newpage
\begin{sidewaystable}
\caption{The most frequently adopted AI methods in geology in different AI development eras}
\centering
\small
%
\label{tab8}
\end{sidewaystable}

\clearpage
\newpage
\begin{sidewaystable}
\caption{The most frequently adopted AI methods in materials science in different AI development eras}
\centering
\small
%
\label{tab9}
\end{sidewaystable}

\clearpage
\newpage
\begin{sidewaystable}
\caption{The most frequently adopted AI methods in medicine in different AI development eras}
\centering
\small
%
\label{tab10}
\end{sidewaystable}

\clearpage
\newpage
\begin{sidewaystable}
\caption{The most frequently adopted AI methods in physics in different AI development eras}
\centering
\small
%
\label{tab11}
\end{sidewaystable}

\clearpage
\newpage
\begin{table}[ht]
\caption{\textbf{Fraction of papers with alphabetically listed authors}}
\centering
\begin{tabular}{cccc}
\hline
Discipline        & OpenAlex author/\% & Random author ($\sigma$)/\% & Actual fraction/\% \\ \hline
Biology           & 16.59 & 15.03 (0.01) & 1.57  \\
Chemistry         & 17.80 & 16.70 (0.02) & 1.10  \\
Geology           & 22.15 & 20.28 (0.03) & 1.87  \\
Materials science & 14.87 & 13.82 (0.02) & 1.05 \\
Medicine          & 16.07 & 14.38 (0.01) & 1.69  \\
Physics           & 20.56 & 18.67 (0.01) & 1.89  \\ \hline
Total             & 16.89 & 15.31 (0.01) & 1.58  \\ \hline
\end{tabular}
\label{tab12}
\end{table}

\clearpage
\printbibliography[title={References}]

@inproceedings{krauss2024debunking,
  title={Debunking revolutionary paradigm shifts: evidence of cumulative scientific progress across science},
  author={Krauss, Alexander},
  booktitle={Proceedings A},
  volume={480},
  number={2302},
  pages={20240141},
  year={2024},
  organization={The Royal Society}
}

@article{king2009automation,
  title={The automation of science},
  author={King, Ross D and Rowland, Jem and Oliver, Stephen G and Young, Michael and Aubrey, Wayne and Byrne, Emma and Liakata, Maria and Markham, Magdalena and Pir, Pinar and Soldatova, Larisa N and others},
  journal={Science},
  volume={324},
  number={5923},
  pages={85--89},
  year={2009},
  publisher={American Association for the Advancement of Science}
}

@article{burger2020mobile,
  title={A mobile robotic chemist},
  author={Burger, Benjamin and Maffettone, Phillip M and Gusev, Vladimir V and Aitchison, Catherine M and Bai, Yang and Wang, Xiaoyan and Li, Xiaobo and Alston, Ben M and Li, Buyi and Clowes, Rob and others},
  journal={Nature},
  volume={583},
  number={7815},
  pages={237--241},
  year={2020},
  publisher={Nature Publishing Group UK London}
}

@inproceedings{wojtowicz2025undermining,
  title={Undermining mental proof: How ai can make cooperation harder by making thinking easier},
  author={Wojtowicz, Zachary and DeDeo, Simon},
  booktitle={Proceedings of the AAAI Conference on Artificial Intelligence},
  volume={39},
  number={2},
  pages={1592--1600},
  year={2025}
}

@article{salimi2023large,
  title={Large language models in ophthalmology scientific writing: ethical considerations blurred lines or not at all?},
  author={Salimi, Ali and Saheb, Hady},
  journal={American journal of ophthalmology},
  volume={254},
  pages={177--181},
  year={2023},
  publisher={Elsevier}
}

@inproceedings{liang2024mapping,
    title={Mapping the Increasing Use of {LLM}s in Scientific Papers},
    author={Weixin Liang and Yaohui Zhang and Zhengxuan Wu and Haley Lepp and Wenlong Ji and Xuandong Zhao and Hancheng Cao and Sheng Liu and Siyu He and Zhi Huang and Diyi Yang and Christopher Potts and Christopher D Manning and James Y. Zou},
    booktitle={First Conference on Language Modeling},
    year={2024}
}

@article{hwang2024can,
  title={Can ChatGPT assist authors with abstract writing in medical journals? Evaluating the quality of scientific abstracts generated by ChatGPT and original abstracts},
  author={Hwang, Taesoon and Aggarwal, Nishant and Khan, Pir Zarak and Roberts, Thomas and Mahmood, Amir and Griffiths, Madlen M and Parsons, Nick and Khan, Saboor},
  journal={PLoS One},
  volume={19},
  number={2},
  pages={e0297701},
  year={2024},
  publisher={Public Library of Science San Francisco, CA USA}
}

@article{kobak2025delving,
  title={Delving into LLM-assisted writing in biomedical publications through excess vocabulary},
  author={Kobak, Dmitry and Gonz{\'a}lez-M{\'a}rquez, Rita and Horv{\'a}t, Em{\H{o}}ke-{\'A}gnes and Lause, Jan},
  journal={Science Advances},
  volume={11},
  number={27},
  pages={eadt3813},
  year={2025},
  publisher={American Association for the Advancement of Science}
}

@article{evans2008electronic,
  title={Electronic publication and the narrowing of science and scholarship},
  author={Evans, James A},
  journal={science},
  volume={321},
  number={5887},
  pages={395--399},
  year={2008},
  publisher={American Association for the Advancement of Science}
}

@article{gao2024quantifying,
  title={Quantifying the use and potential benefits of artificial intelligence in scientific research},
  author={Gao, Jian and Wang, Dashun},
  journal={Nature Human Behaviour},
  pages={1--12},
  year={2024},
  publisher={Nature Publishing Group UK London}
}

@article{wang2023scientific,
  title={Scientific discovery in the age of artificial intelligence},
  author={Wang, Hanchen and Fu, Tianfan and Du, Yuanqi and Gao, Wenhao and Huang, Kexin and Liu, Ziming and Chandak, Payal and Liu, Shengchao and Van Katwyk, Peter and Deac, Andreea and others},
  journal={Nature},
  volume={620},
  number={7972},
  pages={47--60},
  year={2023},
  publisher={Nature Publishing Group UK London}
}

@article{kuhlman2003design,
  title={Design of a novel globular protein fold with atomic-level accuracy},
  author={Kuhlman, Brian and Dantas, Gautam and Ireton, Gregory C and Varani, Gabriele and Stoddard, Barry L and Baker, David},
  journal={science},
  volume={302},
  number={5649},
  pages={1364--1368},
  year={2003},
  publisher={American Association for the Advancement of Science}
}

@article{jumper2021highly,
  title={Highly accurate protein structure prediction with AlphaFold},
  author={Jumper, John and Evans, Richard and Pritzel, Alexander and Green, Tim and Figurnov, Michael and Ronneberger, Olaf and Tunyasuvunakool, Kathryn and Bates, Russ and {\v{Z}}{\'\i}dek, Augustin and Potapenko, Anna and others},
  journal={nature},
  volume={596},
  number={7873},
  pages={583--589},
  year={2021},
  publisher={Nature Publishing Group}
}

@article{hinton2002training,
  title={Training products of experts by minimizing contrastive divergence},
  author={Hinton, Geoffrey E},
  journal={Neural computation},
  volume={14},
  number={8},
  pages={1771--1800},
  year={2002},
  publisher={MIT Press}
}

@article{hinton2006reducing,
  title={Reducing the dimensionality of data with neural networks},
  author={Hinton, Geoffrey E and Salakhutdinov, Ruslan R},
  journal={science},
  volume={313},
  number={5786},
  pages={504--507},
  year={2006},
  publisher={American Association for the Advancement of Science}
}

@article{lecun2015deep,
  title={Deep learning},
  author={LeCun, Yann and Bengio, Yoshua and Hinton, Geoffrey},
  journal={nature},
  volume={521},
  number={7553},
  pages={436--444},
  year={2015},
  publisher={Nature Publishing Group UK London}
}

@article{krizhevsky2012imagenet,
  title={Imagenet classification with deep convolutional neural networks},
  author={Krizhevsky, Alex and Sutskever, Ilya and Hinton, Geoffrey E},
  journal={Advances in neural information processing systems},
  volume={25},
  year={2012}
}

@article{hopfield1984neurons,
  title={Neurons with graded response have collective computational properties like those of two-state neurons.},
  author={Hopfield, John J},
  journal={Proceedings of the national academy of sciences},
  volume={81},
  number={10},
  pages={3088--3092},
  year={1984},
  publisher={National Acad Sciences}
}

@article{hopfield1982neural,
  title={Neural networks and physical systems with emergent collective computational abilities.},
  author={Hopfield, John J},
  journal={Proceedings of the national academy of sciences},
  volume={79},
  number={8},
  pages={2554--2558},
  year={1982},
  publisher={National Acad Sciences}
}

@article{merton1968matthew,
  title={The Matthew effect in science: The reward and communication systems of science are considered.},
  author={Merton, Robert K},
  journal={Science},
  volume={159},
  number={3810},
  pages={56--63},
  year={1968},
  publisher={American Association for the Advancement of Science}
}

@article{mcmahan2018ambiguity,
  title={Ambiguity and engagement},
  author={McMahan, Peter and Evans, James},
  journal={American Journal of Sociology},
  volume={124},
  number={3},
  pages={860--912},
  year={2018},
  publisher={University of Chicago Press Chicago, IL}
}

@article{fawzi2022discovering,
  title={Discovering faster matrix multiplication algorithms with reinforcement learning},
  author={Fawzi, Alhussein and Balog, Matej and Huang, Aja and Hubert, Thomas and Romera-Paredes, Bernardino and Barekatain, Mohammadamin and Novikov, Alexander and R Ruiz, Francisco J and Schrittwieser, Julian and Swirszcz, Grzegorz and others},
  journal={Nature},
  volume={610},
  number={7930},
  pages={47--53},
  year={2022},
  publisher={Nature Publishing Group}
}

@article{degrave2022magnetic,
  title={Magnetic control of tokamak plasmas through deep reinforcement learning},
  author={Degrave, Jonas and Felici, Federico and Buchli, Jonas and Neunert, Michael and Tracey, Brendan and Carpanese, Francesco and Ewalds, Timo and Hafner, Roland and Abdolmaleki, Abbas and de Las Casas, Diego and others},
  journal={Nature},
  volume={602},
  number={7897},
  pages={414--419},
  year={2022},
  publisher={Nature Publishing Group}
}

@article{adiguzel2023revolutionizing,
  title={Revolutionizing education with AI: Exploring the transformative potential of ChatGPT},
  author={Ad{\i}g{\"u}zel, Tufan and Kaya, Mehmet Haldun and Cansu, Fatih K{\"u}r{\c{s}}at},
  journal={Contemporary Educational Technology},
  year={2023},
  publisher={Bastas}
}

@article{akgun2022artificial,
  title={Artificial intelligence in education: Addressing ethical challenges in K-12 settings},
  author={Akgun, Selin and Greenhow, Christine},
  journal={AI and Ethics},
  volume={2},
  number={3},
  pages={431--440},
  year={2022},
  publisher={Springer}
}

@article{mesko2023imperative,
  title={The imperative for regulatory oversight of large language models (or generative AI) in healthcare},
  author={Mesk{\'o}, Bertalan and Topol, Eric J},
  journal={NPJ digital medicine},
  volume={6},
  number={1},
  pages={120},
  year={2023},
  publisher={Nature Publishing Group UK London}
}

@article{loh2022application,
  title={Application of explainable artificial intelligence for healthcare: A systematic review of the last decade (2011--2022)},
  author={Loh, Hui Wen and Ooi, Chui Ping and Seoni, Silvia and Barua, Prabal Datta and Molinari, Filippo and Acharya, U Rajendra},
  journal={Computer Methods and Programs in Biomedicine},
  pages={107161},
  year={2022},
  publisher={Elsevier}
}

@article{ahmed2022artificial,
  title={From artificial intelligence to explainable artificial intelligence in industry 4.0: a survey on what, how, and where},
  author={Ahmed, Imran and Jeon, Gwanggil and Piccialli, Francesco},
  journal={IEEE Transactions on Industrial Informatics},
  volume={18},
  number={8},
  pages={5031--5042},
  year={2022},
  publisher={IEEE}
}

@article{boiko2023autonomous,
  title={Autonomous chemical research with large language models},
  author={Boiko, Daniil A and MacKnight, Robert and Kline, Ben and Gomes, Gabe},
  journal={Nature},
  volume={624},
  number={7992},
  pages={570--578},
  year={2023},
  publisher={Nature Publishing Group UK London}
}

@article{stokel2023chatgpt,
  title={What ChatGPT and generative AI mean for science},
  author={Stokel-Walker, Chris and Van Noorden, Richard},
  journal={Nature},
  volume={614},
  number={7947},
  pages={214--216},
  year={2023},
  publisher={Nature}
}

@article{gilson2023does,
  title={How does ChatGPT perform on the united states medical licensing examination? The implications of large language models for medical education and knowledge assessment},
  author={Gilson, Aidan and Safranek, Conrad W and Huang, Thomas and Socrates, Vimig and Chi, Ling and Taylor, Richard Andrew and Chartash, David and others},
  journal={JMIR Medical Education},
  volume={9},
  number={1},
  pages={e45312},
  year={2023},
  publisher={JMIR Publications Inc., Toronto, Canada}
}

@article{varadi2022alphafold,
  title={AlphaFold Protein Structure Database: massively expanding the structural coverage of protein-sequence space with high-accuracy models},
  author={Varadi, Mihaly and Anyango, Stephen and Deshpande, Mandar and Nair, Sreenath and Natassia, Cindy and Yordanova, Galabina and Yuan, David and Stroe, Oana and Wood, Gemma and Laydon, Agata and others},
  journal={Nucleic acids research},
  volume={50},
  number={D1},
  pages={D439--D444},
  year={2022},
  publisher={Oxford University Press}
}

@article{park2023papers,
  title={Papers and patents are becoming less disruptive over time},
  author={Park, Michael and Leahey, Erin and Funk, Russell J},
  journal={Nature},
  volume={613},
  number={7942},
  pages={138--144},
  year={2023},
  publisher={Nature Publishing Group UK London}
}

@article{milojevic2018changing,
  title={Changing demographics of scientific careers: The rise of the temporary workforce},
  author={Milojevi{\'c}, Sta{\v{s}}a and Radicchi, Filippo and Walsh, John P},
  journal={Proceedings of the National Academy of Sciences},
  volume={115},
  number={50},
  pages={12616--12623},
  year={2018},
  publisher={National Acad Sciences}
}

@article{frank2019evolution,
  title={The evolution of citation graphs in artificial intelligence research},
  author={Frank, Morgan R and Wang, Dashun and Cebrian, Manuel and Rahwan, Iyad},
  journal={Nature Machine Intelligence},
  volume={1},
  number={2},
  pages={79--85},
  year={2019},
  publisher={Nature Publishing Group UK London}
}

@Misc{mag,
    year = {2015},
    author = {Microsoft},
    howpublished = {\url{https://www.microsoft.com/en-us/research/project/microsoft-academic-graph}},
    title = {Microsoft Academic Graph}
}

@Misc{oag,
    year = {2020},
    author = {Aminer},
    howpublished = {\url{https://www.aminer.cn/oag-2-1}},
    title = {Open Academic Graph}
}

@Misc{openalex,
    year = {2025},
    author = {OpanAlex},
    howpublished = {\url{https://openalex.org/}},
    title = {OpenAlex}
}

@Misc{WoS,
    year = {2025},
    author = {Clarivate},
    howpublished = {\url{https://www.webofscience.com}},
    title = {Web of Science}
}

@article{mongeon2016journal,
  title={The journal coverage of Web of Science and Scopus: a comparative analysis},
  author={Mongeon, Philippe and Paul-Hus, Ad{\`e}le},
  journal={Scientometrics},
  volume={106},
  pages={213--228},
  year={2016},
  publisher={Springer}
}

@inproceedings{DBLP:conf/naacl/DevlinCLT19,
  author    = {Jacob Devlin and
               Ming{-}Wei Chang and
               Kenton Lee and
               Kristina Toutanova},
  editor    = {Jill Burstein and
               Christy Doran and
               Thamar Solorio},
  title     = {{BERT:} Pre-training of Deep Bidirectional Transformers for Language
               Understanding},
  booktitle = {Proceedings of the 2019 Conference of the North American Chapter of
               the Association for Computational Linguistics: Human Language Technologies,
               {NAACL-HLT} 2019, Minneapolis, MN, USA, June 2-7, 2019, Volume 1 (Long
               and Short Papers)},
  pages     = {4171--4186},
  publisher = {Association for Computational Linguistics},
  year      = {2019}
}

@inproceedings{wolf-etal-2020-transformers,
    title = "Transformers: State-of-the-Art Natural Language Processing",
    author = "Thomas Wolf and Lysandre Debut and Victor Sanh and Julien Chaumond and Clement Delangue and Anthony Moi and Pierric Cistac and Tim Rault and Rémi Louf and Morgan Funtowicz and Joe Davison and Sam Shleifer and Patrick von Platen and Clara Ma and Yacine Jernite and Julien Plu and Canwen Xu and Teven Le Scao and Sylvain Gugger and Mariama Drame and Quentin Lhoest and Alexander M. Rush",
    booktitle = "Proceedings of the 2020 Conference on Empirical Methods in Natural Language Processing: System Demonstrations",
    month = oct,
    year = "2020",
    address = "Online",
    publisher = "Association for Computational Linguistics",
    pages = "38--45"
}

@Misc{jcr,
    year = {2021},
    author = {Clarivate},
    howpublished = {\url{https://jcr.clarivate.com/jcr/home},
    title = {Journal Citation Reports.}},
}

@article{xu2022flat,
  title={Flat teams drive scientific innovation},
  author={Xu, Fengli and Wu, Lingfei and Evans, James},
  journal={Proceedings of the National Academy of Sciences},
  volume={119},
  number={23},
  pages={e2200927119},
  year={2022},
  publisher={National Acad Sciences}
}

@article{fleiss1971measuring,
  title={Measuring nominal scale agreement among many raters.},
  author={Fleiss, Joseph L},
  journal={Psychological bulletin},
  volume={76},
  number={5},
  pages={378},
  year={1971},
  publisher={American Psychological Association}
}

@article{landis1977measurement,
  title={The measurement of observer agreement for categorical data},
  author={Landis, J Richard and Koch, Gary G},
  journal={biometrics},
  pages={159--174},
  year={1977},
  publisher={JSTOR}
}

@article{segler2018planning,
  title={Planning chemical syntheses with deep neural networks and symbolic AI},
  author={Segler, Marwin HS and Preuss, Mike and Waller, Mark P},
  journal={Nature},
  volume={555},
  number={7698},
  pages={604--610},
  year={2018},
  publisher={Nature Publishing Group UK London}
}

@article{chu2021slowed,
  title={Slowed canonical progress in large fields of science},
  author={Chu, Johan SG and Evans, James A},
  journal={Proceedings of the National Academy of Sciences},
  volume={118},
  number={41},
  pages={e2021636118},
  year={2021},
  publisher={National Acad Sciences}
}

@article{wu2019large,
  title={Large teams develop and small teams disrupt science and technology},
  author={Wu, Lingfei and Wang, Dashun and Evans, James A},
  journal={Nature},
  volume={566},
  number={7744},
  pages={378--382},
  year={2019},
  publisher={Nature Publishing Group UK London}
}

@article{sekara2018chaperone,
  title={The chaperone effect in scientific publishing},
  author={Sekara, Vedran and Deville, Pierre and Ahnert, Sebastian E and Barab{\'a}si, Albert-L{\'a}szl{\'o} and Sinatra, Roberta and Lehmann, Sune},
  journal={Proceedings of the National Academy of Sciences},
  volume={115},
  number={50},
  pages={12603--12607},
  year={2018},
  publisher={National Acad Sciences}
}

@book{kingman1992poisson,
  title={Poisson processes},
  author={Kingman, John Frank Charles},
  volume={3},
  year={1992},
  publisher={Clarendon Press}
}

@article{kendall1960birth,
  title={Birth-and-death processes, and the theory of carcinogenesis},
  author={Kendall, David G},
  journal={Biometrika},
  volume={47},
  number={1/2},
  pages={13--21},
  year={1960},
  publisher={JSTOR}
}

@article{meisling1958discrete,
  title={Discrete-time queuing theory},
  author={Meisling, Torben},
  journal={Operations Research},
  volume={6},
  number={1},
  pages={96--105},
  year={1958},
  publisher={INFORMS}
}

@Misc{HuggingFace,
howpublished = {\url{https://huggingface.co/docs/transformers/model_doc/bert\#transformers.BertForSequenceClassification}},
title = {BERT pretrain model from Hugging Face.},
author = {Hugging Face}}

@article{lin2023remote,
  title={Remote collaboration fuses fewer breakthrough ideas},
  author={Lin, Yiling and Frey, Carl Benedikt and Wu, Lingfei},
  journal={Nature},
  volume={623},
  number={7989},
  pages={987--991},
  year={2023},
  publisher={Nature Publishing Group UK London}
}

@inproceedings{DBLP:conf/emnlp/SinghDCDF23,
  author       = {Amanpreet Singh and
                  Mike D'Arcy and
                  Arman Cohan and
                  Doug Downey and
                  Sergey Feldman},
  title        = {SciRepEval: {A} Multi-Format Benchmark for Scientific Document Representations},
  booktitle    = {{EMNLP}},
  pages        = {5548--5566},
  publisher    = {Association for Computational Linguistics},
  year         = {2023}
}

@inproceedings{he2016deep,
  title={Deep residual learning for image recognition},
  author={He, Kaiming and Zhang, Xiangyu and Ren, Shaoqing and Sun, Jian},
  booktitle={Proceedings of the IEEE conference on computer vision and pattern recognition},
  pages={770--778},
  year={2016}
}

@article{milojevic2015quantifying,
  title={Quantifying the cognitive extent of science},
  author={Milojevi{\'c}, Sta{\v{s}}a},
  journal={Journal of Informetrics},
  volume={9},
  number={4},
  pages={962--973},
  year={2015},
  publisher={Elsevier}
}

@article{porter2009science,
  title={Is science becoming more interdisciplinary? Measuring and mapping six research fields over time},
  author={Porter, Alan and Rafols, Ismael},
  journal={Scientometrics},
  volume={81},
  number={3},
  pages={719--745},
  year={2009},
  publisher={Akad{\'e}miai Kiad{\'o}, co-published with Springer Science+ Business Media BV~…}
}

@article{rumelhart1986learning,
  title={Learning representations by back-propagating errors},
  author={Rumelhart, David E and Hinton, Geoffrey E and Williams, Ronald J},
  journal={nature},
  volume={323},
  number={6088},
  pages={533--536},
  year={1986},
  publisher={Nature Publishing Group UK London}
}

@article{lecun1989backpropagation,
  title={Backpropagation applied to handwritten zip code recognition},
  author={LeCun, Yann and Boser, Bernhard and Denker, John S and Henderson, Donnie and Howard, Richard E and Hubbard, Wayne and Jackel, Lawrence D},
  journal={Neural computation},
  volume={1},
  number={4},
  pages={541--551},
  year={1989},
  publisher={MIT Press}
}

@article{borger2023artificial,
  title={Artificial intelligence takes center stage: exploring the capabilities and implications of ChatGPT and other AI-assisted technologies in scientific research and education},
  author={Borger, Jessica G and Ng, Ashley P and Anderton, Holly and Ashdown, George W and Auld, Megan and Blewitt, Marnie E and Brown, Daniel V and Call, Melissa J and Collins, Peter and Freytag, Saskia and others},
  journal={Immunology and cell biology},
  volume={101},
  number={10},
  pages={923--935},
  year={2023},
  publisher={Wiley Online Library}
}

@article{lawrence2024accelerating,
  title={Accelerating AI for science: open data science for science},
  author={Lawrence, Neil D and Montgomery, Jessica},
  journal={Royal Society Open Science},
  volume={11},
  number={8},
  pages={231130},
  year={2024},
  publisher={The Royal Society}
}

@article{fortunato2018science,
  title={Science of science},
  author={Fortunato, Santo and Bergstrom, Carl T and B{\"o}rner, Katy and Evans, James A and Helbing, Dirk and Milojevi{\'c}, Sta{\v{s}}a and Petersen, Alexander M and Radicchi, Filippo and Sinatra, Roberta and Uzzi, Brian and others},
  journal={Science},
  volume={359},
  number={6379},
  pages={eaao0185},
  year={2018},
  publisher={American Association for the Advancement of Science}
}

@article{ioannidis2014estimates,
  title={Estimates of the continuously publishing core in the scientific workforce},
  author={Ioannidis, John PA and Boyack, Kevin W and Klavans, Richard},
  journal={PloS one},
  volume={9},
  number={7},
  pages={e101698},
  year={2014},
  publisher={Public Library of Science San Francisco, USA}
}

@inproceedings{DBLP:conf/emnlp/BeltagyLC19,
  author       = {Iz Beltagy and
                  Kyle Lo and
                  Arman Cohan},
  title        = {SciBERT: {A} Pretrained Language Model for Scientific Text},
  booktitle    = {{EMNLP/IJCNLP} {(1)}},
  pages        = {3613--3618},
  publisher    = {Association for Computational Linguistics},
  year         = {2019}
}

@inproceedings{DBLP:conf/acl/CohanFBDW20,
  author       = {Arman Cohan and
                  Sergey Feldman and
                  Iz Beltagy and
                  Doug Downey and
                  Daniel S. Weld},
  title        = {{SPECTER:} Document-level Representation Learning using Citation-informed
                  Transformers},
  booktitle    = {{ACL}},
  pages        = {2270--2282},
  publisher    = {Association for Computational Linguistics},
  year         = {2020}
}

@inproceedings{chen2016xgboost,
  title={Xgboost: A scalable tree boosting system},
  author={Chen, Tianqi and Guestrin, Carlos},
  booktitle={Proceedings of the 22nd acm sigkdd international conference on knowledge discovery and data mining},
  pages={785--794},
  year={2016}
}

@article{hill2025pivot,
  title={The pivot penalty in research},
  author={Hill, Ryan and Yin, Yian and Stein, Carolyn and Wang, Xizhao and Wang, Dashun and Jones, Benjamin F},
  journal={Nature},
  pages={1--8},
  year={2025},
  publisher={Nature Publishing Group UK London}
}

@article{damoiseaux2006consistent,
  title={Consistent resting-state networks across healthy subjects},
  author={Damoiseaux, Jessica S and Rombouts, Serge ARB and Barkhof, Frederik and Scheltens, Philip and Stam, Cornelis J and Smith, Stephen M and Beckmann, Christian F},
  journal={Proceedings of the national academy of sciences},
  volume={103},
  number={37},
  pages={13848--13853},
  year={2006},
  publisher={National Academy of Sciences}
}

@article{lin2023evolutionary,
  title={Evolutionary-scale prediction of atomic-level protein structure with a language model},
  author={Lin, Zeming and Akin, Halil and Rao, Roshan and Hie, Brian and Zhu, Zhongkai and Lu, Wenting and Smetanin, Nikita and Verkuil, Robert and Kabeli, Ori and Shmueli, Yaniv and others},
  journal={Science},
  volume={379},
  number={6637},
  pages={1123--1130},
  year={2023},
  publisher={American Association for the Advancement of Science}
}

@article{kandlikar2005representing,
  title={Representing and communicating deep uncertainty in climate-change assessments},
  author={Kandlikar, Milind and Risbey, James and Dessai, Suraje},
  journal={Comptes Rendus. G{\'e}oscience},
  volume={337},
  number={4},
  pages={443--455},
  year={2005}
}

@article{redner1998popular,
  title={How popular is your paper? An empirical study of the citation distribution},
  author={Redner, Sidney},
  journal={The European Physical Journal B-Condensed Matter and Complex Systems},
  volume={4},
  number={2},
  pages={131--134},
  year={1998},
  publisher={Springer}
}

@article{seglen1992skewness,
  title={The skewness of science},
  author={Seglen, Per O},
  journal={Journal of the American society for information science},
  volume={43},
  number={9},
  pages={628--638},
  year={1992},
  publisher={Wiley Online Library}
}

@inproceedings{hao2024HLM,
  author       = {Qianyue Hao and
                  Jingyang Fan and
                  Fengli Xu and
                  Jian Yuan and
                  Yong Li},
  title        = {HLM-Cite: Hybrid Language Model Workflow for Text-based Scientific
                  Citation Prediction},
  booktitle    = {NeurIPS},
  year         = {2024}
}

@inproceedings{smith1988using,
  title={Using the ADAP learning algorithm to forecast the onset of diabetes mellitus},
  author={Smith, Jack W and Everhart, James E and Dickson, William C and Knowler, William C and Johannes, Robert Scott},
  booktitle={Proceedings of the annual symposium on computer application in medical care},
  pages={261},
  year={1988}
}

@article{hasan2020diabetes,
  title={Diabetes prediction using ensembling of different machine learning classifiers},
  author={Hasan, Md Kamrul and Alam, Md Ashraful and Das, Dola and Hossain, Eklas and Hasan, Mahmudul},
  journal={IEEE Access},
  volume={8},
  pages={76516--76531},
  year={2020},
  publisher={IEEE}
}

@article{kumari2021ensemble,
  title={An ensemble approach for classification and prediction of diabetes mellitus using soft voting classifier},
  author={Kumari, Saloni and Kumar, Deepika and Mittal, Mamta},
  journal={International Journal of Cognitive Computing in Engineering},
  volume={2},
  pages={40--46},
  year={2021},
  publisher={Elsevier}
}

@article{lockhart2023name,
  title={Name-based demographic inference and the unequal distribution of misrecognition},
  author={Lockhart, Jeffrey W and King, Molly M and Munsch, Christin},
  journal={Nature Human Behaviour},
  volume={7},
  number={7},
  pages={1084--1095},
  year={2023},
  publisher={Nature Publishing Group UK London}
}


\newpage
\section*{Data availability}
The OpenAlex dataset for research papers and researchers is available at \url{https://docs.openalex.org/download-all-data/openalex-snapshot}.
The Web of Science (WoS) dataset for research papers and researchers is available at \url{https://clarivate.com/academia-government/scientific-and-academic-research/research-discovery-and-referencing/web-of-science/web-of-science-core-collection}.
The Journal Citation Reports (JCR) dataset for the journal quantile is retrieved from \url{https://jcr.clarivate.com/jcr/browse-journals}.
The author contribution dataset is available at \url{https://zenodo.org/records/6569339}.

The pre-trained parameters for the BERT language model are available at \url{https://huggingface.co/docs/transformers}.
The pre-trained parameters for the SPECTER 2.0 text embedding model are available at \url{https://huggingface.co/allenai/specter2}.

\section*{Code availability}
This study used python 3.11.0 with software packages to conduct data analysis.
Required packages are numpy 1.26.4, pandas 2.2.3, scipy 1.15.2, sklearn 1.6.1, matplotlib 3.10.1.
The used t-SNE algorithm is imported from the sklearn package.
The codes developed in this study can be found at \url{https://github.com/tsinghua-fib-lab/AI-Impacts-Science}.

\section*{Author contributions}
F.X., Y.L. and J.E. jointly launched this research and designed the research outline.
Q.H. analyzed the data and prepared the figures.
All authors jointly participated in writing and revising the manuscript.

\end{document}